\documentclass[aps,prl,twocolumn,superscriptaddress,longbibliography]{revtex4-1} 
\usepackage{blindtext}
\usepackage{amssymb}
\usepackage{amsmath}
\usepackage{graphicx}
\usepackage[caption=false]{subfig}
\usepackage{floatrow}
\usepackage[dvipsnames]{xcolor}
\usepackage[version=3]{mhchem}
\usepackage{footnote}
\usepackage{bm}
\usepackage{dsfont}
\usepackage{xr-hyper}
\usepackage[hidelinks]{hyperref}
\usepackage[capitalise]{cleveref}
\usepackage{soul}
\usepackage[normalem]{ulem}
\usepackage{mathdots}
\usepackage{booktabs}
\usepackage{nicefrac}
\crefname{equation}{Eq.}{Eqs.}
\Crefname{equation}{Equation}{Equations}
\crefname{figure}{Fig.}{Figs.} 
\Crefname{figure}{Figure}{Figures}
\crefname{section}{Sect.}{Sects.}
\Crefname{section}{Section}{Sections}
\crefname{table}{Table}{Tables}
\crefname{appsec}{}{Appendices}

\newcommand{\abs}[1]{\ensuremath{\left|#1\right|}}

\newcommand{\comm}[2]{\ensuremath{\left[#1,#2\right]}}
\newcommand{\acomm}[2]{\ensuremath{\left\lbrace#1,#2\right\rbrace}}

\definecolor{catcolor}{rgb}{0.,0.5,0.5}
\definecolor{adrcolor}{rgb}{0.9,0.1,0.2}
\definecolor{cateditcolor}{rgb}{0.8,0.4,0.1}

\usepackage{tabularx}

\begin{document}

\floatsetup[figure]{style=plain,subcapbesideposition=top}

\title{Nonreciprocal devices based on voltage-tunable junctions}

\author{Catherine Leroux}
\email{Catherine.Leroux@USherbrooke.ca}
\affiliation{Institut quantique \& D\'epartement de Physique, Universit\'e de Sherbrooke, Sherbrooke J1K 2R1 QC, Canada}

\author{Adrian Parra-Rodriguez}
\affiliation{Institut quantique \& D\'epartement de Physique, Universit\'e de Sherbrooke, Sherbrooke J1K 2R1 QC, Canada}

\author{Ross Shillito}
\affiliation{Institut quantique \& D\'epartement de Physique, Universit\'e de Sherbrooke, Sherbrooke J1K 2R1 QC, Canada}

\author{Agustin Di Paolo}
\affiliation{Research Laboratory of Electronics, Massachusetts Institute of Technology, Cambridge, MA 02139, USA}

\author{William D. Oliver}
\affiliation{Research Laboratory of Electronics, Massachusetts Institute of Technology, Cambridge, MA 02139, USA}
\affiliation{Lincoln Laboratory, Massachusetts Institute of Technology, Lexington, MA 02421-6426, USA}
\affiliation{Department of Physics, Massachusetts Institute of Technology, Cambridge, MA 02139, USA}
\affiliation{Department of Electrical Engineering and Computer Science, Massachusetts Institute of Technology, Cambridge, MA 02139, USA}

\author{Charles M. Marcus} 
\affiliation{Center for Quantum Devices, Niels Bohr Institute, University of Copenhagen, 2100 Copenhagen, Denmark}

\author{Morten Kjaergaard} 
\affiliation{Center for Quantum Devices, Niels Bohr Institute, University of Copenhagen, 2100 Copenhagen, Denmark}

\author{Andr\'as Gyenis} 
\affiliation{Department of Electrical, Computer \& Energy Engineering, University of Colorado Boulder, Boulder, CO 80309, USA}

\author{Alexandre Blais}
\affiliation{Institut quantique \& D\'epartement de Physique, Universit\'e de Sherbrooke, Sherbrooke J1K 2R1 QC, Canada}
\affiliation{Canadian Institute for Advanced Research, Toronto, ON, Canada}

\date{\today}

\begin{abstract}
    We propose to couple the flux degree of freedom of one mode with the charge degree of freedom of a second mode in a hybrid superconducting-semiconducting architecture. Nonreciprocity can arise in this architecture in the presence of external static magnetic fields alone. We leverage this property to engineer a passive on-chip gyrator, the fundamental two-port nonreciprocal device which can be used to build other nonreciprocal devices such as circulators. We analytically and numerically investigate how the nonlinearity of the interaction, circuit disorder and parasitic couplings affect the scattering response of the gyrator. 
\end{abstract}
                 
\maketitle

\section{Introduction} 
Processing quantum information with high-fidelity requires interfaces for detecting, controlling and routing quantum signals and 
nonrecripocal devices are vital elements to realize these tasks~\cite{Naaman2021,Ranzani2014,Ranzani2015,Kamal2011}. 
At a fundamental level, nonreciprocity requires breaking time-reversal symmetry defined by the invariance of the system with respect to the transformation~$t\to-t$, where~$t$ is time. Equivalently, the Lagrangian of nonreciprocal devices is not conserved under the transformation~$\dot{\Phi} \to \dot{\Phi}$ and~$\Phi \to -\Phi$, where~$\Phi$ is the flux degree of freedom associated with a circuit mode. 

Under the usual capacitive or inductive interactions, modes in superconducting circuits typically couple through the same quadrature, for example charge-charge or flux-flux interactions. These couplings preserve time-reversal symmetry and lead to reciprocal two-body interactions. As a consequence, realizing circulators in Josephson junction-based quantum circuits 
often relies on parametric drives~\cite{Kerchkhoff2015,Sliwa2015,Kamal2011,Koch2010,Chapman2019,Dinc2017}, the Aharanov-Bohm effect, or its dual the Aharanov-Casher effect~\cite{Koch2010,Muller2018,Richman2021,Navarathna2022}. Other Josephson junction-based nonreciprocal devices include gyrators~\cite{Abdo2017}, isolators and directional amplifiers~\cite{Malz2018,Thorbeck2017,Abdo2014,Metelmann2015,Abdo2013,Macklin2015,HoEom2012,Vissers2016,Hover2012,Lecocq2017}. Optomechanical systems are also used in the design of nonreciprocal devices~\cite{Aspelmeyer2014, Ruesink2016,Bernier2017,Shen2016,Fang2017,Hafezi2012,Barzanjeh2017,Peterson2017}. Other proposals for nonreciprocity rely on the Hall effect~\cite{Viola2014,Mahoney2017,Bosco2017} and spatiotemporal modulation of conductivity in semiconductors~\cite{Dinc2017}.

Here, we propose to engineer a static coupling between two modes that involves distinct quadratures: one mode participates in the interaction via the flux operator, while the other mode via the charge operator.
This flux-charge interaction, which we refer to as FENNEC (Flux intErcoNNEcted with Charge), is realized by the use of weak-links with voltage-tunable potential energy~\cite{Weber2018,Larsen2015,Kringhoj2018,Casparis2019,Larsen2020,Vries2021,Lee2015,Haque2021}. We show how the FENNEC coupling, with the help of a static external magnetic field, can implement a gyrator, a building block of other nonreciprocal devices such as circulators.

The paper is organized as follows. 
In~\cref{sec: Flux-charge interaction}, we detail our proposal for a flux-charge interaction starting from the Andreev bound state energy spectrum of a weak-link. 
In~\cref{sec:Gyrator design}, we introduce a gyrator design based on the FENNEC interaction. We describe the system using mean-field calculations and provide numerical simulations in the presence of system nonidealities. As an application of this gyrator, we discuss a circulator design in~\cref{sec:Circulator design} before concluding in~\cref{sec: Conclusion}.

\section{Flux-charge interaction}
\label{sec: Flux-charge interaction}

\begin{figure}[t!]
    \centering
    \includegraphics[width=\textwidth]{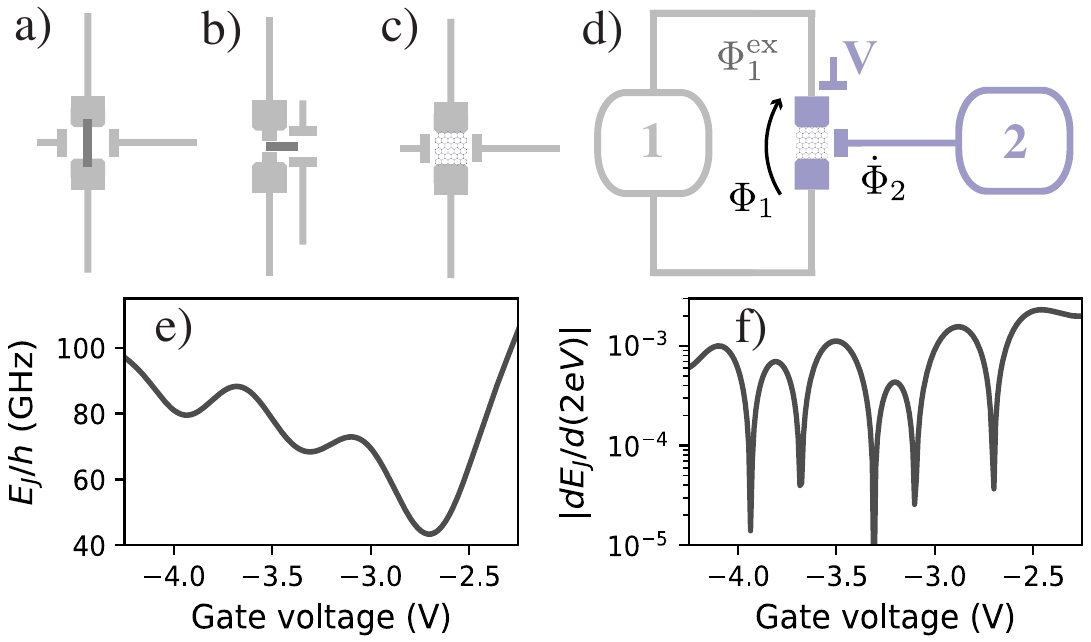}
    \caption{a) Nanowire junction, b) two-dimensional electron gas (2DEG) junction and c) graphene-based Josephson junction. Each semiconducting junction is gated by two voltage lines: one for the dc voltage bias and one to implement the flux-charge interaction. Panels a) and b) are inspired by Ref.~\cite{Weber2018}.  d) Circuit implementation of the flux-charge interaction. e) Approximate graphene junction energy taken from Ref.~\cite{Wang2019}. f) Magnitude of the first derivative of the Josephson energy in e) with respect to voltage in electronvolts, such that the derivative is dimensionless. (See panels e)-f) for nanowire and 2DEG junctions in~\cref{app:FENNEC Estimation of the interaction strength}.)}
    \label{fig:FENNEC_designs}
\end{figure}

Our approach to implement flux-charge coupling is based on voltage-tunable Josephson junctions~\cite{aguado2020}. These junctions can be realized by replacing the usual oxide separating the junction's superconductors by semiconducting nanowires~\cite{Larsen2015,de_lange_realization_2015,Kringhoj2018,Larsen2020,PhysRevApplied.14.064038}, two-dimensional electron gases (2DEGs)~\cite{Casparis2018,oconnell_yuan_epitaxial_2021,hertel2022gatetunable,hazard2022superconductingsemiconducting}, or van der Waals materials~\cite{Vries2021,Lee2015,lee_two-dimensional_2019,Wang2019,Haque2021}, see~\cref{fig:FENNEC_designs}a-c). The coupling between the superconductors separated by such barriers is governed by Andreev reflections. The total Andreev bound state (ABS) energy in a multichannel weak-link junction is~\cite{Larsen2015}
\begin{equation}
    \varepsilon_J(V,\Phi_1) = - \Delta\sum_i \sqrt{1 - T_i(V) \sin^2(\pi \Phi_1/\Phi_0)},
    \label{eq:ABS_energy}
\end{equation}
where~$\Delta$ is the superconducting gap,~$T_i(V)$ is the transmission probability of channel~$i$ which is controlled by the external gate voltage~$V$,~$\Phi_1$ is the gauge-invariant flux across the junction, and~$\Phi_0=h/(2e)$ is the flux quantum. Here, we focus on the weak transmission limit,~$T_i(V)\ll 1$, when
\begin{equation} \label{eq: EJ weak transmission}
    \varepsilon_J(V,\Phi_1) \approx - E_J(V) \cos(2\pi\Phi_1/\Phi_0),
\end{equation}
with the voltage-tunable Josephson coupling~$E_J(V) = \Delta\sum_i T_i(V)/4$. The large transmission limit is discussed further in~\cref{app:FENNEC Estimation of the interaction strength}.

In this work, we propose to couple the weak-link device (1) to a second mode (2) via the gate voltage [see~\cref{fig:FENNEC_designs}d)] such that the voltage~$V$ biasing the weak link is influenced by the voltage across the second mode,~$V \to V_0 + \dot{\Phi}_2$, where~$V_0$ is an external voltage bias and~$\dot{\Phi}_2$ is the time-derivative of the branch flux of the second mode. In the presence of an external flux~$\Phi_1^\mathrm{ex}$ threading a superconducting loop comprising the junction in~\cref{fig:FENNEC_designs}d), the charge-flux coupling is revealed by Taylor expanding~\cref{eq: EJ weak transmission} in~$\Phi_{1(2)}$ about the time-periodic field averages~$\langle \Phi_{1(2)}(t) \rangle$
\begin{equation}
    \varepsilon_J(V_0+\dot{\Phi}_2,\Phi_1-\Phi_1^\mathrm{ex}) 
    = \sum_{n,m=0}^\infty \frac{\partial^{n+m} \varepsilon_J}{\partial V^n\partial\Phi_1^m} \frac{\delta\dot{\Phi}_2^n\delta\Phi_1^m}{n!m!},
    \label{eq: charge phase deriv}
\end{equation}
where~$\delta \Phi_{1(2)} = \Phi_{1(2)} - \langle \Phi_{1(2)}(t) \rangle$. Equation~\ref{eq: charge phase deriv} leads to an interaction between the voltage of the second mode~$\dot{\Phi}_2$ and the flux of the first mode~$\Phi_1$, resulting in a flux-charge interaction in the Hamiltonian describing the device~\cite{Vool2017}. 

In the quantized model,~$\Phi_{1}$ and~$\dot{\Phi}_{2}$ have fluctuations~$\propto (\Phi_0/2\pi)\sqrt{\pi Z_1/R_Q}$ and~$\propto \omega_2 (\Phi_0/2\pi)\sqrt{\pi Z_2/R_Q}$ respectively, with~$R_Q = h/(2e)^2\simeq 6.5$ k$\Omega$ the resistance quantum,~$Z_{i=(1,2)}$ the impedance of mode~$i = (1,2)$, and ~$\omega_2$ the frequency of the second mode. In what follows we work in the limit $\abs{\partial^n E_J/\partial V^n} \ll n!\omega_2^{1-n}(\pi Z_2/R_Q)^{(1-n)/2}(\Phi_0/2\pi)^{1-n}\abs{\partial E_J/\partial V}~$ for~$n>1$ such that only the first derivative of~$E_J$ contributes to the interaction Lagrangian and~$\sqrt{\pi Z_1/R_Q}\ll 1$ which is appropriate for a low impedance mode. Under these conditions we truncate~\cref{eq: charge phase deriv} to its first derivative with respect to~$V$ and to first order in~$\Phi_1$ resulting in an interaction Lagrangian of the form [see~\cref{app:FENNEC Lagrangian} for details]
\begin{equation} \label{eq:L_int_main}
    \mathcal{L}_\mathrm{int} \approx \frac{G_{21}}{2} \dot{\Phi}_2\Phi_1, 
\end{equation}
where using~\cref{eq: EJ weak transmission,eq: charge phase deriv} the flux-charge coupling strength is
\begin{equation}\label{eq: G equation}
    G_{21} = \frac{4\pi}{R_Q}\frac{E_J'(V_0+\langle\dot{\Phi}_2\rangle)}{2e}\sin
    \left[\frac{2\pi}{\Phi_0}(\Phi_1^\mathrm{ex}-\langle\Phi_1(t)\rangle)\right], 
\end{equation}
where $E_J'(V) \equiv \partial E_J/\partial V$. The flux-charge interaction of~\cref{eq:L_int_main} breaks time-reversal symmetry since it is not conserved under the transformation~$\dot \Phi\to \dot \Phi$ and~$\Phi \to -\Phi$, and therefore has the form needed to implement nonreciprocal devices~\cite{ParraRodriguez2019,Rymarz2021}. 

The coupling~$G_{21}$, which is largest in magnitude at~$\Phi_1^\mathrm{ex}=\pm\Phi_0/4$, is generally smaller than~$1/R_Q=2e/\Phi_0$ as suggested by the derivative of the energy dispersion in~\cref{fig:FENNEC_designs}f) which is obtained from the experimental data of Ref.~\cite{Wang2019}. By optimizing the device geometry beyond what was done in Ref.~\cite{Wang2019}, it is possible to increase the electrostatic coupling between the gateline and the semiconducting region of the SNS junction, thereby making~$E_J'/2e$ larger than  reported in~\cref{fig:FENNEC_designs}d). In principle,~$G_{21}$ can also be increased using parametric amplification~\cite{Lemonde2016,Leroux2018,Groszkowski2020}. An analysis of the interaction strength based on spectroscopy data for different types of junctions can be found in~\cref{app:FENNEC Estimation of the interaction strength}. We also note that the leading order effects of the junction nonlinearity are captured by the mean-field approximation of~\cref{eq: G equation} where the field averages have to be solved for self-consistently. 

Here,~$E_J''(V_0)\langle \dot{\Phi}_2 \rangle$ is generally much smaller than~$E_J'(V_0)$ in magnitude, such that~$E_J'(V_0+\langle \dot{\Phi}_2 \rangle)\approx E_J'(V_0)$ in~\cref{eq: G equation}. However, for increasing photon numbers in the first mode, the time-average of~\cref{eq: G equation} decreases in magnitude. Indeed, the averaged flux field in the first mode is~$\langle \Phi_{1}(t) \rangle \approx (\Phi_0/2\pi) \sqrt{\pi Z_{1}/R_Q}\sum_{n=1}^\infty \alpha_{1}^{(n)} e^{in \omega t}/\sqrt{2}+\mathrm{h.c}$ with~$\alpha_{1}^{(n)}$ the displacement in the~$n$th harmonic of the flux field due to an input signal with frequency~$\omega$. To leading order in~$Z_1/R_Q \ll 1$, the time-averaged~\cref{eq: G equation} is then~$(4\pi E_J'/2e R_Q)\sin (2\pi \Phi_1^\mathrm{ex}/\Phi_0)(1 - \pi Z_1 N_1/2 R_Q)$ where~$N_1=\sum_{n=1}^\infty |\alpha_1^{(n)}|^2$ is the photon number in the first mode. As will be explained later, the impedance~$Z_1$ plays a key role in defining a maximum photon number,~$\sim R_Q/\pi Z_1$, which 
that can be allowed in the first mode before the FENNEC interaction is impacted by the junction's nonlinearity. A smaller impedance~$Z_1$ results in a larger maximum photon number before the interaction is suppressed.

Moreover, at external fluxes where~$\sin(2\pi\Phi_1^\mathrm{ex}/\Phi_0)=\pm 1$ and~$|G_{21}|$ maximized,~$G_{12}$ is to first order insensitive to flux noise but sensitive to charge noise proportionally to the second derivative of~$E_J$ with respect to voltage. However, because~$\partial G_{21}/\partial V_0$ is orders of magnitude smaller than~$G_{12}$, charge and flux noise have negligible effects on the FENNEC interaction strength at those external flux biases. (See~\cref{app:FENNEC Noise sensitivity} for details.)

\section{Gyrator design}
\label{sec:Gyrator design}

The simplest and most fundamental nonreciprocal device based on the flux-charge coupling of~\cref{eq:L_int_main} is the gyrator. An ideal gyrator is characterized by the scattering matrix
\begin{equation}
    \boldsymbol{\mathcal{S}} = \begin{pmatrix}  0 & 1 \\ -1 & 0\end{pmatrix},
    \label{eq:S gyr ideal}
\end{equation}
which relates the amplitude of the incoming ($\boldsymbol{a}$) and outgoing ($\boldsymbol{b}$) fields, at each port of the device via~$\boldsymbol{b}=\boldsymbol{S}\cdot\boldsymbol{a}$.
The circuit Lagrangian of an ideal gyrator takes the form of~\cite{ParraRodriguez2019,Rymarz2021}
\begin{equation}
    \mathcal{L}_{\mathrm{gyr}} = \frac{G}{2} \left(\dot{\Phi}_2 \Phi_1 - \dot{\Phi}_1\Phi_2\right), \label{eq:L gyr}
\end{equation}
where~$G$ is the conductance of the gyrator,~$\Phi_{1(2)}$ is the branch flux and~$\dot{\Phi}_{1(2)}$ is the voltage at port 1(2). 

\begin{figure}[t!]
    \centering
    \includegraphics[width=\textwidth]{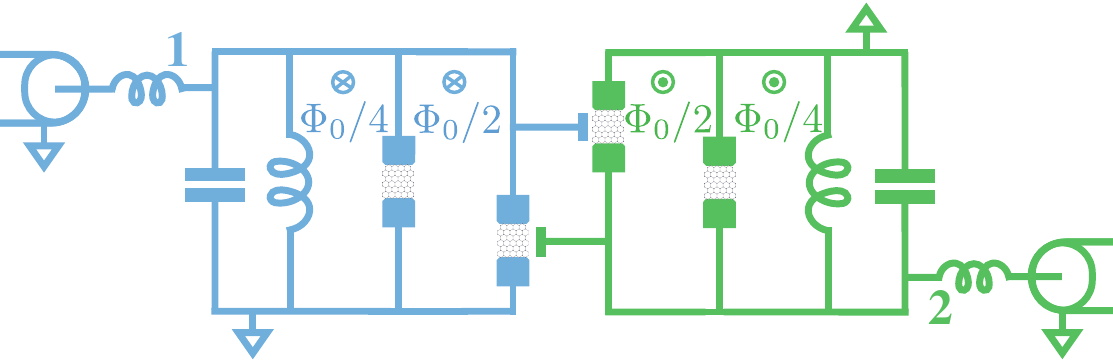}
    \caption{Proposed lumped-element gyrator design. The gyrator is divided in two symmetric subcircuits, shown in blue (left) and green (right). Each half comprises two superconducting loops, where a SQUID of semiconducting junctions is shunted by an inductance~$L_0$ and a capactiance~$C_0$. The two junctions have equal superconducting gap~$\Delta$ and transmission probabilities~$T_i(V)$, and are biased at the same dc voltage~$V_0$. The superconducting loops are threaded by~$\pm \Phi_0/4$ and~$\pm \Phi_0/2$, with~$+$ ($-$) for the blue (green) half. The two parts are connected through the FENNEC interaction: the branch flux of one subcircuit is coupled to the voltage port of neighboring junction of the other subcircuit. Each subcircuit is connected to an input-output transmission line.} 
    \label{fig:gyrator circuit}
\end{figure}

To realize~$\mathcal{L}_{\mathrm{gyr}}$ using the FENNEC interaction, we consider the lumped-element circuit of~\cref{fig:gyrator circuit} comprising two identical internal modes~$1$ (blue) and~$2$ (green). Each mode contains a symmetric SQUID loop of semiconducting junctions biased at half quantum flux. The FENNEC interaction is realized by capacitively coupling each mode to the voltage port of the other mode's voltage-tunable junction. The presence of semiconducting junctions in half-quantum-flux-biased SQUIDs results only in the flux-charge interaction without any additional nonlinearity in the inductance of the internal modes of the gyrator. Both modes are also shunted by LC circuits with resonance frequencies setting the central frequency of the device. The flux bias between the LC circuit and the SQUID loop is set to one quarter quantum flux to render the FENNEC interaction quadratic as needed for gyration in~\cref{eq:L gyr}. Finally, each internal gyrator mode is coupled to an external port via an inductance. Stray capacitive coupling between the two modes will be mostly present in a realistic implementation and will be briefly analyzed later on when we discuss circuit disorder. We stress that the device, which involves only two modes, is both compact and passive.

Using~\cref{eq:L_int_main,eq: G equation} with~$\Phi_1^\mathrm{ex} = \Phi_0/4$ and~$\Phi_2^\mathrm{ex} = -\Phi_0/4$, we find that the circuit of~\cref{fig:gyrator circuit} results in an effective interaction Lagrangian of the form of~\cref{eq:L gyr} with a time-dependent conductance 
\begin{equation} \label{eq: G gyrator}
    G = \frac{G_{21}-G_{12}}{2} \approx G_\mathrm{max}\left[
1- \frac{\pi^2}{\Phi_0^2}(\langle \Phi_1(t) \rangle^2+\langle  \Phi_2(t)\rangle^2)\right], 
\end{equation}
with~$G_{21(12)}$ defined in~\cref{eq: G equation} and
\begin{equation} \label{eq: G max}
    G_\mathrm{max} = \frac{4\pi }{R_Q}\frac{E_J'(V_0)}{2e}.
\end{equation}
The approximation in~\cref{eq: G gyrator} results from Taylor expanding in the field averages~$\langle \Phi_{1(2)}(t) \rangle$ to second order and neglecting the second derivative of~$E_J$ (see~\cref{app:Gyrator Mean-field theory}). Typical device parameters (see~\cref{fig:FENNEC_designs}) result in~$G_\mathrm{max}\ll 1/R_Q$ and therefore a narrow bandwidth. In~\cref{eq: G gyrator} the averaged flux field is~$\langle \Phi_{1(2)}(t) \rangle \approx (\Phi_0/2\pi) \sqrt{\pi Z_0/R_Q}\sum_{n=1}^\infty \alpha_{1(2)}^{(n)} e^{in \omega t}/\sqrt{2}+\mathrm{h.c}$ with~$\alpha_{1(2)}^{(n)}$ the displacement in the~$n$th harmonic of the flux field due to an input signal with frequency~$\omega$, and~$Z_0 = \sqrt{L_0/C_0}$ the characteristic impedance of the shunting LC. The total photon number in mode~$1(2)$ is therefore~$N_{1(2)}=\sum_{n=1}^\infty |\alpha_{1(2)}^{(n)}|^2$. The time-dependent contributions in~\cref{eq: G gyrator} result in frequency mixing and, as a consequence, a time-averaged conductance~\cref{eq: G gyrator} that decreases from its optimal value with increasing input power. Within the rotating-wave approximation, it is useful to approximate~\cref{eq: G gyrator} by its time-average 
\begin{equation}
\label{eq:compressed G}
    G \approx G_\mathrm{max}\left[1-\frac{\pi Z_0 N}{2R_Q}\right]
\end{equation}
with~$N= (N_1+N_2)/2$ the average photon number in the gyrator which is proportional to the input power. As discussed in further detail below, a reduced conductance leads to increased reflection. The effects of frequency-mixing due to the counter-rotating terms that are dropped in~\cref{eq: G gyrator} are analyzed in~\cref{app:Gyrator Effective linear response theory}.

\begin{figure*}[t!]
    \centering
    \includegraphics[width=0.85\textwidth]{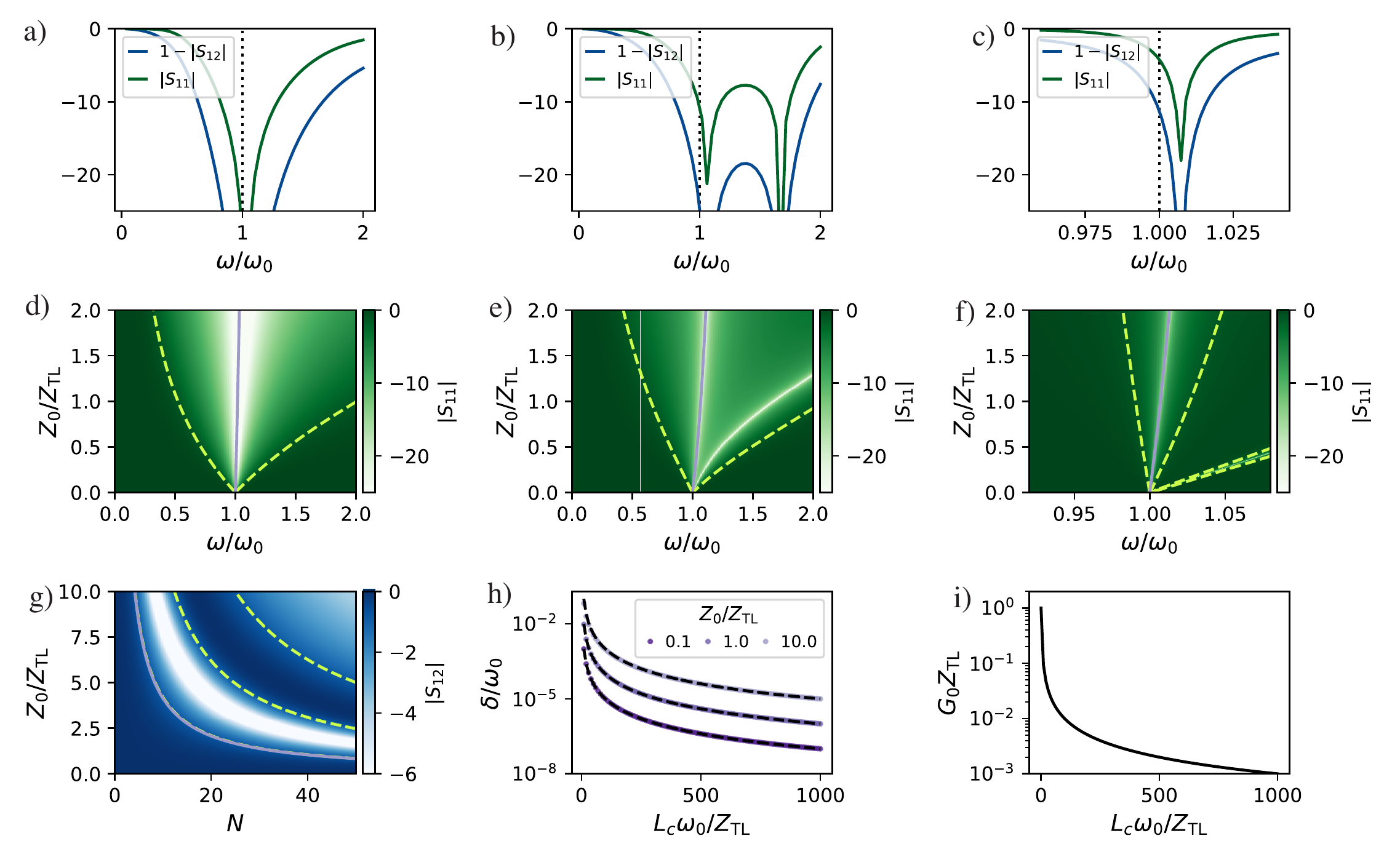}
    \caption{a-c) Reflection~$|S_{11}|$ and transmission~$|S_{12}|$ in dB computed analytically at the impedance matching condition~$G=G_0$ as a function of the frequency~$\omega$ renormalized by the resonance frequency~$\omega_0=1/\sqrt{C_0L_0}$, for different coupling inductances~$L_c=Z_\mathrm{TL}/\omega_0 \times \{0.05,0.50,5.00\}$ respectively, and for fixed load impedance~$Z_0=\sqrt{L_0/C_0}=10 Z_\mathrm{TL}$ and in the~$N=0$ limit. d-f) Two-dimensional version of a-c) where~$Z_0$ is varied. The dashed lines are the~$|\tan(2\theta_\omega)| = 1$ boundaries, where reflection starts to dominate over transmission. Two such boundaries closest to~$\omega=\omega_0$ are highlighted in light green. The central frequency is identified with a purple line. g) Transmission~$|S_{12}|$ versus the photon number~$N$ in the gyrator and the load impedance~$Z_0$. The 1~dB-compression level is highlighted in light green and corresponds to the maximum photon number that can be tolerated in the gyrator. The analytical estimate from~\cref{eq:N_max} is the purple line which shows agreement with the 1~dB level. h) Numerically computed frequency bandwidth near~$\omega=\omega_0$ using the boundaries highlighted in d-f) versus the coupling inductance~$L_c$, for different load impedances~$Z_0$. The black lines correspond to~\cref{eq:Delta}. Details of the fitting algorithm can be found in~\cref{app:scattering_matrix_numerics}. i) Impedance-matched conductance as a function of coupling inductance~$L_c$, see~\cref{eq:G0}.}
    \label{fig:s_matrix_gyrator}
\end{figure*}

\emph{Scattering matrix.} Starting from the equations of motion of the mean-field Lagrangian, we find that the linear scattering response can be expressed as
\begin{equation}
\label{eq:S matrix}
    \boldsymbol{S}(\omega)  = \left(\boldsymbol{1}-\frac{\boldsymbol{Z_\omega}^{-1}}{Z_\mathrm{TL}^{-1}}\right)^{-1}\cdot\left(\boldsymbol{1}+\frac{\boldsymbol{Z_\omega}^{-1}}{Z_\mathrm{TL}^{-1}}\right), 
\end{equation}
(see~\cref{app:Gyrator Scattering matrix of the linearized system}) where~$Z_\mathrm{TL}$ is the characteristic impedance of the input-output transmission lines, and~$\boldsymbol{Z_\omega} =  i \omega \boldsymbol{L_c}+ [i\omega \boldsymbol{C} + (i\omega\boldsymbol{L})^{-1} +i  G\boldsymbol{\sigma_y}]^{-1}$ encodes the total impedance of the gyrator modes. Here,~$\boldsymbol{C}$ and~$\boldsymbol{L}$ are the~$2\times 2$ capacitance and inductance matrices of the gyrator modes respectively,~$\boldsymbol{\sigma_y}$ is the Pauli matrix, and~$\boldsymbol{L_c}$ is the coupling inductance matrix between the transmission lines and the gyrator modes. In the ideal case where~$\boldsymbol{L_c}=L_c \boldsymbol{1}$,~$\boldsymbol{C}=C_0 \boldsymbol{1}$ and~$\boldsymbol{L}=L_0 \boldsymbol{1}$, the scattering matrix reduces to the simple form
\begin{equation}\label{eq: s mat pauli}
    \boldsymbol{S}(\omega) = \cos(2\theta_\omega)\boldsymbol{1} +i \sin(2\theta_\omega)\boldsymbol{\sigma_y},
\end{equation}
where
\begin{equation}\label{eq:tan(2theta)}
    \tan(2\theta_\omega) = \frac{2G\overline{Z}_\mathrm{TL}(\omega)}{1-\overline{Z}_\mathrm{TL}^2(\omega)/\overline{Z}_0^2(\omega)-G^2\overline{Z}_\mathrm{TL}^2(\omega)},
\end{equation}
and
\begin{align}
    & \overline{Z}_\mathrm{TL}(\omega) = \frac{Z_\mathrm{TL}}{\left[1+Z_c(\omega)/Z_0(\omega)\right]^2+ G^2Z_c^2(\omega)}, \\
    & \overline{Z}_0(\omega) = \frac{Z_0(\omega)}{1 + [Z_c(\omega)/Z_0(\omega)]\left[1 +  G^2Z_0^2(\omega)\right]},
\end{align}
are the frequency-dependent characteristic impedance of the lines and renormalized load impedance due to the coupling inductance, respectively. Here~$Z_0(\omega) = [i\omega C_0 + (i\omega L_0)^{-1}]^{-1}$ is the impedance of the load whereas~$Z_c(\omega) = i \omega L_c$ is the impedance of the coupling inductance. $\boldsymbol{S}$ approaches the ideal scattering matrix of a gyrator~\cref{eq:S gyr ideal} for~$|\tan(2\theta_{\omega})| \to \infty$ or, equivalently, when the circuit is perfectly impedance-matched such that transmission is maximal. 

\emph{Central frequency.} The central frequency of the device corresponds to the frequency for which the denominator in~\cref{eq:tan(2theta)} vanishes with the smallest~$G$ possible. As discussed in further details in~\cref{app:Gyrator Effective linear response theory}, the central frequency is close to the resonance frequency of the internal gyrator modes~$\omega_0= 1/\sqrt{L_0 C_0}$. 

\emph{Impedance-matched conductance.} The conductance for which the scattering matrix approaches that of an ideal gyrator at~$\omega=\omega_0$ is approximately
\begin{equation}
    G_0 = \overline{Z}_\mathrm{TL}(\omega_0)^{-1}=  \frac{\sqrt{1+2 \left(\sqrt{2}L_c\omega_0/Z_\mathrm{TL}\right)^2}-1}{Z_\mathrm{TL} \left(\sqrt{2}L_c\omega_0/Z_\mathrm{TL}\right)^2},\label{eq:G0}
\end{equation}
which becomes~$G_0=Z_\mathrm{TL}^{-1}$ as~$L_c\rightarrow 0$. To maximize transmission, we set~$G_\mathrm{max}$ in~\cref{eq: G max} equal to~$G_0$ in~\cref{eq:G0}. We also note that~$G_0$ decreases with increasing~$L_c$. As will be shown below, we ideally want~$L_c=0$ such as to maximize the frequency bandwidth of the device leaving us with the constraint~$G_\mathrm{max}=Z_\mathrm{TL}^{-1}$. In cases where the transmission lines have a characteristic impedance~$Z_\mathrm{TL}\ll G_\mathrm{max}^{-1}$, which is most likely for typical circuit parameters, we can nonetheless use a matching circuit between the lines and the gyrator~\cite{Pozar2009,Naaman2021}.

\emph{Frequency bandwidth.} We also introduce the frequency bandwidth~$\delta = \omega_+-\omega_-$ for gyration with~$\omega_\pm$ the cut-off frequencies for which reflection equals transmission, where~$|\tan(2\theta_\omega)| = 1$. At large~$L_c$, where~$G_0 \approx (L_c\omega_0)^{-1}$, we find (see~\cref{app:Gyrator Effective linear response theory})
\begin{equation}
    \delta \approx \frac{Z_0 Z_\mathrm{TL}}{ L_c^2\omega_0}. \label{eq:Delta}
\end{equation}
The same expression for zero~$L_c$ is instead~$\delta = 2\omega_0 \sqrt{1+\beta\left( Z_0/Z_\mathrm{TL}\right)^2}$ where~$4\beta=G^2Z_\mathrm{TL}+2|G|Z_\mathrm{TL}-1$.

\emph{Compression point.} As discussed above, frequency mixing can lead to reduced transmission and here we define the compression level as the maximum average photon number~$N$ for which the scattering-matrix components deviate by 1~dB from the expected values in the zero-photon linear limit. Near the central frequency~$\omega_0$ we find that~$|\tan (2\theta_{\omega_0})| \approx 2 (1-x)/[1 - (1-x)^2]$, where~$x=\pi Z_0 N/2R_Q$ using the mean-field expression for the conductance in~\cref{eq:compressed G}, see~\cref{app:Gyrator Scattering matrix of the linearized system}. From this expression, we find a maximum average photon number 
\begin{equation}
    N_\mathrm{max} \approx \frac{R_Q}{\pi Z_0} \label{eq:N_max}
\end{equation}
by setting~$|\tan(2\theta_{\omega_0})|\approx 1.31$ and with~$\theta_{\omega_0}$ the angle at which transmission drops by 1~dB in~\cref{eq: s mat pauli} at the central frequency~$\omega_0$ with nonzero average photon number~$N$. That the maximal photon number~$N_\mathrm{max}$ decreases with increasing~$Z_0$ is a signature that the system dynamics is more affected by the junctions nonlinearity for large zero-point fluctuations of the internal gyrator modes. 
Assuming a typical mode impedance~$Z_0=50\,\Omega$,~\cref{eq:N_max} leads to~$N_\mathrm{max}\approx 41$ photons.

\emph{Numerical results.} The reflection and transmission coefficients of the scattering matrix in the linear regime (i.e. $N \ll R_Q/\pi Z_0$) for different~$L_c$ and~$Z_0$ are shown in~\cref{fig:s_matrix_gyrator}a-f). The frequency bandwidth is shown in~\cref{fig:s_matrix_gyrator}h) and the optimal conductance~\cref{eq:G0} is shown in~\cref{fig:s_matrix_gyrator}i). In panels a-f), we observe that the central frequency (purple line near~$\omega_0$) slightly deviates from~$\omega_0$ as a function of~$Z_0/Z_\mathrm{TL}$ for non-zero~$L_c$ with our choice of conductance~$G_0$ (see~\cref{app:Gyrator Scattering matrix of the linearized system} for analytical estimates). The dashed light green contours in panels d-f) about~$\omega=\omega_0$ correspond to~$\omega_\pm$. We note that the frequency bandwidth near~$\omega = \omega_0$ also quickly decreases with increasing~$L_c$, which is clearly illustrated in panel h) where we see excellent agreement with~\cref{eq:Delta} for large~$L_c$ values. Panel i) illustrates that the optimal conductance~$G_0$ is inversely proportional to~$L_c$. Compression is also shown within mean-field theory in~\cref{fig:s_matrix_gyrator}g), with the purple line corresponding to~\cref{eq:N_max}. 

\emph{Noise sensitivity}. The gyrator interaction in~\cref{eq:L gyr} is akin to a Jaynes-Cummings interaction between two resonant LC oscillators that are the internal gyrator modes. This quadratic model, with energy splitting~$2G$, is insensitive to both charge and flux noise. Nevertheless, for the design of~\cref{fig:gyrator circuit}, and within mean-field theory, the interaction strength given by~\cref{eq: G gyrator} is sensitive to both charge noise~$\dot{\Phi}_{1(2)}\to \dot{\Phi}_{1(2)}+\delta \dot{\Phi}_{1(2)}$ and flux noise~$\Phi_{1(2)}\to \Phi_{1(2)}+\delta \Phi_{1(2)}$. To leading order in the noise, we find that~$G \to G + (\partial G_\mathrm{max}/\partial V_0) [\delta \dot{\Phi}_1+\delta\dot{\Phi}_2]/2$ is insensitive to flux noise but sensitive to charge noise. $\partial G_\mathrm{max}/\partial V_0 \propto E_J''(V_0)$ is however orders of magnitude smaller than~$G_\mathrm{max}$ and consequently charge noise is negligible. Derivations and full analysis for both flux and charge noise can be found in~\cref{app:Gyrator Noise sensitivity}.

\emph{Circuit disorder.} Gyration is fragile to frequency mismatches and stray couplings, both unavoidable in realistic circuit implementations and resulting in~$\boldsymbol{\sigma_z}$ and~$\boldsymbol{\sigma}_x$ components in the scattering matrix~\cref{eq: s mat pauli}. We consider~$\boldsymbol{L_c}=L_c \boldsymbol{1} + d L_c \boldsymbol{\sigma_z}$,~$\boldsymbol{C}= C_0\boldsymbol{1} + dC_0 \boldsymbol{\sigma_z} - C_{12}\boldsymbol{\sigma_x}$ and~$\boldsymbol{L} = L_0\boldsymbol{1} + dL_0\boldsymbol{\sigma_z} - L_{12}\boldsymbol{\sigma_x}$ with~$dL_c$,~$d C_0$,~$d L_0$ the disorder in~$L_c$,~$C_0$,~$L_0$, respectively, and~$C_{12}$,~$L_{12}$ the parasitic capacitive and inductive couplings between active nodes and loops, respectively. As shown in~\cref{app:Circuit disorder}, deviations in the scattering matrix elements, proportional to~$\boldsymbol{\sigma_z}$ and~$\boldsymbol{\sigma_x}$, are much smaller than unity for~$d L_c\ll Z_\mathrm{TL}/\omega_0$,~$d C_0 \ll Z_\mathrm{TL}G_0^2/\omega_0$,~$d L_0 \ll L_0^2\omega_0 Z_\mathrm{TL} G_0^2$,~$C_{12} \ll Z_\mathrm{TL} G_0^2/\omega_0$ and~$L_{12}\ll L_0^2\omega_0 Z_\mathrm{TL} G_0^2$. These constraints are all realizable in superconducting circuits. We note that a larger optimal conductance~$G_0$ [i.e. a smaller~$L_c$ in~\cref{eq:G0}] renders the device less sensitive to circuit disorder, which is also a direct consequence of a larger frequency bandwidth, see ~\cref{eq:Delta}. Further discussions regarding circuit disorder can be found in~\cref{app:Circuit disorder}. 

\emph{Optimal circuit parameters.} Important circuit parameters are the coupling inductance~$L_c$, the conductance~$G_\mathrm{max}$ in~\cref{eq: G max} which must be set to the optimal conductance value~$G_0$, and the characteristic impedance~$Z_0$ of the shunting LC resonators. For typical semiconducting junctions,~$G_\mathrm{max}\ll Z_\mathrm{TL}^{-1}$ which forces~$L_c$ to be large such that the condition that ~$G_\mathrm{max}=G_0$ can be satisfied accordingly to~\cref{eq:G0} unless we use a matching circuit between the transmission lines and the gyrator. A larger~$L_c$ (i.e.~smaller~$G_0$) results in a smaller frequency bandwidth [see~\cref{eq:Delta}] and increased sensitivity to circuit disorder as noted in the previous paragraph. We also require~$Z_0\ll R_Q$ to maximize~\cref{eq:N_max} which equally contributes in reducing the frequency bandwidth. Overall the larger~$G_\mathrm{max}$ can be made the larger the frequency bandwidth and the smallest the sensitivity to circuit disorder. 

\emph{Beyond mean-field theory.} So far we have used the mean-field approach to capture the leading order effects of the circuit nonlinearity in the scattering matrix.  However, this approach does not take into account the impact of quantum fluctuations. The time evolution of the full circuit under a dissipative master equation is analyzed in~\cref{app:Lindblad Master equation}, where we show that quantum fluctuations are indeed negligible when comparing reflection against mean-field theory for different input powers and load impedances. In~\cref{app: generic lagrangian}, we also follow the circuit-quantization procedure~\cite{Vool2017} on a generic circuit with the FENNEC interaction which is nonlinear in both the phase and charge quadratures due to the higher derivatives of~$E_J$. To this end, we introduce a perturbative expansion for the canonical charges with respect to the voltages~$\dot{\Phi}_1$ and~$\dot{\Phi}_2$ to take into account the higher derivatives of~$E_J$. The perturbative expansion yields nonlinear corrections to the quantized circuit Hamiltonian. The leading order effect resulting from the second derivative of~$E_J$ is a nonlinear capacitive energy that depends on the phase of the other mode. 

\section{Circulator design}
\label{sec:Circulator design}

\begin{figure}[t!]
    \centering
    \includegraphics[width=\textwidth]{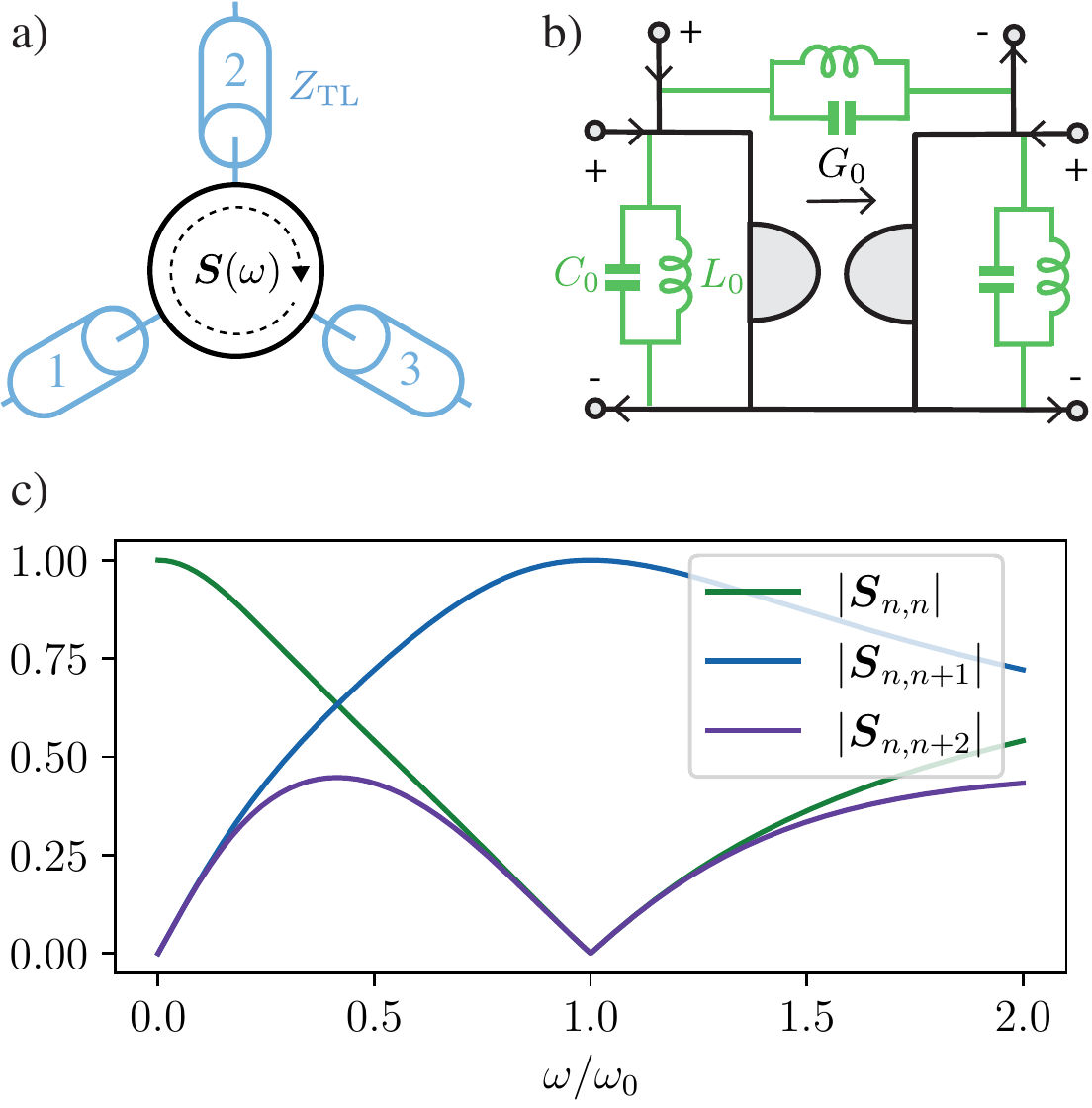}
    \caption{a) Three-port circulator characterized by a scattering matrix~$\boldsymbol{S}(\omega)$ with input-output transmission lines of impedance~$Z_\mathrm{TL}$. b) Lumped-element design for a circulator based on a single gyrator with its circuit symbol shown in gray and corresponding to the implementation in~\cref{fig:gyrator circuit}. The three equal loads shown in green have a characteristic impedance~$Z_0$. The gyrator here can be implemented with the FENNEC interaction. c) Absolute values of the full scattering matrix elements versus the frequency for~$n=1,2,3$.
   }
    \label{fig:Circulator_sketch}
\end{figure}

Having demonstrated that the flux-charge interaction leads to the fundamental two-port nonreciprocal element, we use first principles of circuit theory to build more general multi-port devices. As an example,~\cref{fig:Circulator_sketch}b) shows a symmetric version of a circulator built from the gyrator design of~\cref{fig:gyrator circuit}. 

The limitations and imperfections of our gyrator design imparted by either the junction nonlinearity or circuit disorder, as discussed in the previous section, will be the same for the circulator design in~\cref{fig:Circulator_sketch}b). For simplicity, here we therefore consider the gyrator to be ideal. The circulator in~\cref{fig:Circulator_sketch}b) was already analyzed in Ref.~\cite{Carlin:1964} also considering an ideal gyrator.

In the linear semi-classical regime, i.e. within mean-field theory and for small input photon numbers where FENNEC acts as an ideal gyrator, this device is described by the scattering matrix
\begin{align}
    \boldsymbol{S}=\begin{pmatrix}
    0&1&0\\0&0&-1\\1&0&0
    \end{pmatrix},
\end{align}
when the system is probed at resonance ($\omega=\omega_{0}$) and is impedance matched ($Z_\mathrm{TL}=Z_{0}=1/G_0$). The absolute values of the full scattering matrix elements are shown in~\cref{fig:Circulator_sketch}c) while details of analytical expressions can be found in Appendix~\ref{app:Circulator_from_gyrator}.

\section{Conclusion}
\label{sec: Conclusion}

We proposed a flux-charge interaction that breaks time-reversal symmetry in the presence of static external magnetic fields and which can be used as a building block for passive nonreciprocal devices such as  gyrators and circulators. We analytically and numerically investigated the scattering matrix of a gyrator based on the this interaction. The strength of the FENNEC interaction, which we wish to maximize, will determine both the frequency bandwidth of the device and the sensitivity to circuit disorder. The nonlinearity of the junctions will also result in compression similarly to other proposals for circulators~\cite{Koch2010,Muller2018,Richman2021}. Despite its narrow bandwidth, the advantages of our gyrator are both its compactness and passiveness. 

Beyond applications to nonreciprocal devices, the FENNEC interaction yields either quadratic or nonlinear two-body interactions opening up new possibilities for engineering two-qubit gates and next-generation superconducting qubits~\cite{Gyenis2021}. Indeed, based on the recent by proposal by~\textcite{Rymarz2021}, it can be shown that GKP states~\cite{Gottesman2001} can be stabilized with this interaction~\cite{Leroux2022}.

\section*{Acknowledgments}
We thank Ilan Rosen for insightful discussions. This research was funded in part by NSERC, the Canada First Research Excellence Fund, and the U.S. Army Research Office grants No.~W911NF2210042, W911NF18S0116, and W911NF2210023.

\bibliography{refs.bib}

\newpage

\onecolumngrid

\appendix

\section{FENNEC interaction properties}
\label{app:FENNEC interaction properties}

\subsection{Time-reversal symmetry}
\label{app:FENNEC Time-reversal symmetry}

Voltages and currents are typically considered even and odd variables with respect to time inversion, i.e. $V(-t)=V(t)$ and~$I(-t)=-I(t)$. Given that the fluxes and charges are their time integrals respectively, i.e. $\Phi(t)=\int_{0}^t V(\tau) d\tau$ and~$Q(t)=\int_{0}^t I(\tau) d\tau$, we would then define~$\Phi\rightarrow -\Phi$ and~$\dot\Phi\rightarrow \dot \Phi$ under time inversion. 

\subsection{FENNEC Lagrangian}
\label{app:FENNEC Lagrangian}

We consider a generic circuit Lagrangian of the form~$\mathcal L = \mathcal L_0 + \mathcal L_{\mathrm int}$, where 
\begin{equation} \label{eq:L_0}
    \mathcal L_0 =  \dot{\boldsymbol{\Phi}}^T\cdot \frac{\boldsymbol{C_0}}{2}\cdot \dot{\boldsymbol{\Phi}} +\dot{\boldsymbol{\Phi}}^T\cdot \frac{\boldsymbol{C_c}}{2}\cdot \dot{\boldsymbol{\Phi}} + \left(\dot{\boldsymbol{\Phi}}-\boldsymbol{V}\right)^T\cdot \frac{\boldsymbol{C_J}}{2}\cdot \left(\dot{\boldsymbol{\Phi}}-\boldsymbol{V}\right)  - U\left(\boldsymbol{\varphi}\right),
\end{equation}
is the Lagrangian due to all standard superconducting circuit elements, 
\begin{equation} \label{eq:L_int}
\begin{split}
    \mathcal L_{\mathrm int} & =- \varepsilon_J(\Delta_1,\boldsymbol{T_1},V_1+\dot{\Phi}_2,\varphi_1^{\mathrm{ex}},\varphi_1)  =-\sum_{n,m=0}^\infty \frac{\dot{\Phi}_2^n}{n!}\frac{\Phi_1^m}{m!} \left(\frac{2\pi}{\Phi_0}\right)^m \frac{\partial^{n+m} \varepsilon_J(\Delta_1,\boldsymbol{T_1},V_1,\varphi_1^{\mathrm{ex}},0)}{\partial V^n\partial \varphi^m} 
\end{split}
\end{equation}
results from the FENNEC interaction alone. Here~$\boldsymbol{\Phi} = \left(\Phi_1, \ \Phi_2\right)$ is a vector comprising the branch flux~$\Phi_1$ ($\Phi_2$) of the first (second) mode,~$\boldsymbol{\varphi} = 2\pi\boldsymbol{\Phi}/\Phi_0$ are the associated branch phases,~$\boldsymbol{C_0}$ and~$\boldsymbol{C_c}$ are capacitance matrices due to the shunt capacitors and the coupling capacitors respectively,~$\boldsymbol{C_J}$ is the capacitance matrix associated with the coupling to the control voltage lines~$\boldsymbol{V}$,~$U(\boldsymbol{\varphi})$ is any additional potential energy of the two modes,
\begin{equation}
    \varepsilon_J(\Delta,\boldsymbol{T},V,\varphi^{\mathrm{ex}},\varphi) = -\Delta \sum_i\sqrt{1-[\boldsymbol{T}(V)]_i\sin^2\left(\frac{ \varphi-\varphi^{\mathrm ex}}{2}\right)},
\end{equation}
is the form of the Andreev bound-state energy of any semiconducting junction in the circuit,~$\Delta_k$ is the gap energy of the~$k$th junction with transmissions~$[\boldsymbol{T_k}]_i$,~$\Phi_k^{\mathrm ex}$ is an external flux threading the~$k$th loop. In this work we focus on the leading order contribution of the interaction Lagrangian
\begin{equation} \label{eq:L_int^2}
    \mathcal L_{\textit{fennec}} = G  \dot{\Phi}_2\Phi_1/2,
\end{equation}
where we defined the amplitudes 
\begin{equation}
    G =- \frac{4\pi}{\Phi_0}\frac{\partial^2 \varepsilon_J(\Delta_1,\boldsymbol{T_1},V_1,\varphi_1^{\mathrm{ex}},0)}{\partial V\partial \varphi}=\frac{4\pi}{\Phi_0} \frac{\Delta_1}{4}\sum_i [\boldsymbol{T_1}(V_1)]_i' \sin\left(\varphi_1^{\mathrm ex}\right)\frac{1+[\boldsymbol{T_1}(V_1)]_i\sin^2\left(\varphi_1^{\mathrm ex}/2\right)/2}{\sqrt{1-[\boldsymbol{T_1}(V_1)]_i\sin^2\left(\varphi_1^{\mathrm ex}/2\right)}^3}.
\end{equation}
In what follows we truncate the interaction Lagrangian to quadratic order, 
\begin{equation}
    \mathcal{L}_{\mathrm{int}}\approx  \frac{ c_2}{2}\dot{\Phi}_2^2 - \frac{\Phi_1^2}{2\ell_1} + \alpha_2 \dot{\Phi}_2 + \beta_1 \Phi_1 + \frac{G}{2}  \dot{\Phi}_2\Phi_1
\end{equation}
where we defined the charge offset 
\begin{align} 
    & \alpha_2 = -\frac{\partial \varepsilon_J(\Delta_1,\boldsymbol{T_1},V_1,\varphi_1^\mathrm{ex},0)}{\partial V} =- \sum_i\frac{\Delta_1 [\boldsymbol{T_1}(V_1)]_i'\sin^2\left(\varphi_1^{\mathrm ex}/2\right)/2}{ \sqrt{1-[\boldsymbol{T_1}(V_1)]_i\sin^2\left(\varphi_1^{\mathrm ex}/2\right)}^3}, 
\end{align}
the phase offset 
\begin{align}
    & \beta_1 =- \frac{2\pi}{\Phi_0} \frac{\partial \varepsilon_J(\Delta_1,\boldsymbol{T_1},V_1,\varphi_1^\mathrm{ex},0)}{\partial \varphi} = \frac{2\pi}{\Phi_0}\sum_i\frac{\Delta_1[\boldsymbol{T_1}(V_1)]_i\sin(\varphi_1^{\mathrm ex})/4}{ \sqrt{1-[\boldsymbol{T_1}(V_1)]_i\sin^2\left( \varphi_1^{\mathrm ex}/2\right)}^3}, 
\end{align}
the shift in the capacitance 
\begin{align}
    & c_2 = -\frac{\partial^2 \varepsilon_J(\Delta_1,\boldsymbol{T_1},V_1,\varphi_1^\mathrm{ex},0)}{\partial V^2} = -\sum_i\frac{\Delta_1\sin^2\left(\varphi_1^{\mathrm ex}/2\right)/2}{ \sqrt{1-[\boldsymbol{T_1}(V_1)]_i\sin^2\left(\varphi_1^{\mathrm ex}/2\right)}} \left[[\boldsymbol{T_1}(V_1)]_i'' - \frac{([\boldsymbol{T_1}(V_1)]_i)^2\sin^2\left(\varphi_1^{\mathrm ex}/2\right)/2}{1-[\boldsymbol{T_1}(V_1)]_i\sin^2\left(\varphi_1^{\mathrm ex}/2\right)}\right], 
\end{align}
and the shift in the inductance 
\begin{align}
    & \frac{1}{\ell_1} =\left(\frac{2\pi}{\Phi_0}\right)^2 \frac{\partial^2 \varepsilon_J(\Delta_1,\boldsymbol{T_1},V_1,\varphi_1^\mathrm{ex},0)}{\partial \varphi^2} =  \left(\frac{2\pi}{\Phi_0}\right)^2\sum_i\frac{\Delta_1 [\boldsymbol{T_1}(V_1)]_i/4}{ \sqrt{1-[\boldsymbol{T_1}(V_1)]_i\sin^2\left(\varphi_1^{\mathrm ex}/2\right)}^3}\left[  \cos(\varphi_1^{\mathrm ex}) + [\boldsymbol{T_1}(V_1)]_i\sin^4\left(\varphi_1^{\mathrm ex}/2\right)\right].
\end{align}
We already advertise that at~$\varphi_1^\mathrm{ex}=\pm \pi/2$, in the weak transmission limit~$[\boldsymbol{T_1}(V_1)]_i \ll 1$ and more generally~$|[\boldsymbol{T_1}(V_1)]_i'|, |[\boldsymbol{T_1}(V_1)]_i''|\ll 1$, the shifts~$c_2$ and~$1/\ell_1$ are negligible contributions to the capacitance and inductance of modes 2 and 1 respectively. 

\subsection{Weak transmission limit}
\label{app:FENNEC Weak transmission limit}

We further simplify the system Lagrangian by considering the weak transmission limit~$[\boldsymbol{T_1}(V_1)]_i\ll 1$ where we find that  
\begin{equation}
    \varepsilon_J(\Delta,\boldsymbol{T},V,\varphi^{\mathrm{ex}},\varphi) \approx \Delta + \frac{\Delta \sum_i[\boldsymbol{T_1}(V_1)]_i}{4} - \frac{\Delta \sum_i[\boldsymbol{T_1}(V_1)]_i}{4}\cos\left(\varphi-\varphi^{\mathrm ex}\right) = \Delta + E_J(\Delta,\boldsymbol{T},V) - E_J(\Delta,\boldsymbol{T},V)\cos\left(\varphi-\varphi^{\mathrm ex}\right),
\end{equation}
where we defined the effective Josephson energy~$E_J(\Delta,\boldsymbol{T},V) = \Delta \sum_i[\boldsymbol{T_1}(V_1)]_i/4$. We consequently find that 
\begin{align}
    & G \approx \frac{4\pi}{\Phi_0} \frac{\partial E_J(\Delta_1,\boldsymbol{T_1},V_1)}{\partial V}\sin(\varphi_1^\mathrm{ex}), \\ 
    & \alpha_2 = -\frac{\partial E_J(\Delta_1,\boldsymbol{T_1},V_1)}{\partial V} + \frac{\partial E_J(\Delta_1,\boldsymbol{T_1},V_1)}{\partial V}\cos\left(\varphi^{\mathrm ex}\right), \\
    & \beta_1 = \frac{2\pi}{\Phi_0} E_J(\Delta_1,\boldsymbol{T_1},V_1)\sin(\varphi_1^\mathrm{ex}), \\
    & c_2 = -\frac{\partial^2 E_J(\Delta_1,\boldsymbol{T_1},V_1)}{\partial V^2} + \frac{\partial^2 E_J(\Delta_1,\boldsymbol{T_1},V_1)}{\partial V^2}\cos\left(\varphi^{\mathrm ex}\right), \\ 
    & \frac{1}{\ell_1} =\left(\frac{2\pi}{\Phi_0}\right)^2 E_J(\Delta_1,\boldsymbol{T_1},V_1)\cos(\varphi_1^\mathrm{ex}).
\end{align}
In what follows we drop the small shifts~$c_2$ and~$1/\ell_1$ for compactness.

\subsection{Noise sensitivity}
\label{app:FENNEC Noise sensitivity}

In this section we analyze the noise sensitivity of the device. Notice that given the term~$\varepsilon_J(\Delta_1,\boldsymbol{T_1},V_1+\dot{\Phi}_2,\varphi_1^{\mathrm{ex}},\varphi_1)$, charge noise in the second mode, such that~$\dot{\Phi}_2 \to \dot{\Phi}_2+\delta \dot{\Phi}_2$, is equivalent to~$V_1 \to V_1 +\delta \dot{\Phi}_2$. Similarly, flux noise in the first mode, such that~$\varphi_1\to \varphi_1 + \delta \varphi_1$, is equivalent to~$\varphi_1^\mathrm{ex}\to \varphi_1^\mathrm{ex} - \delta\varphi_1$.

\emph{Charge noise.} In presence of charge noise, which amounts to~$V_1\to V_1 + \delta \dot{\Phi}_2$ in the FENNEC interaction strength~$G$, we find that~$G\to G+\delta G$ where 
\begin{equation}
    \delta G \approx  \frac{4\pi}{\Phi_0}\frac{\partial^2 E_J(\Delta_1,\boldsymbol{T_1},V_1)}{\partial V^2} \sin(\varphi_1^\mathrm{ex})\delta \dot{\Phi}_2
\end{equation}
to leading order in the noise. 

\emph{Flux noise.} In presence of flux noise, which can be implemented with~$\varphi_1^\mathrm{ex}\to \varphi_1^\mathrm{ex}- \delta \varphi_1$ in the FENNEC interaction strength~$G$, we find that~$G\to G+\delta G$ where 
\begin{equation}
    \delta G \approx - \left(\frac{4\pi}{\Phi_0}\right)^2 \frac{\partial E_J(\Delta_1,\boldsymbol{T_1},V_1)}{\partial V}\cos(\varphi_1^\mathrm{ex})\delta \Phi_1.
\end{equation}

\emph{Resolution of the strength.} Another important point is the resolution of the DC gate voltage bias,~$\delta V$, which must satisfy 
\begin{equation}
   \delta V \ll \abs{ \frac{4\pi}{\Phi_0}\frac{\partial^2 E_J(\Delta_1,\boldsymbol{T_1},V_1)}{\partial V^2} \sin(\varphi_1^\mathrm{ex})}^{-1}.
\end{equation}

\subsection{Mean-field theory}
\label{app:FENNEC Mean-field theory}

In this section we linearized the FENNEC interaction within a mean-field theory approximation: 
\begin{equation} \label{eq:L_int^mf}
\begin{split}
    \mathcal L_{\mathrm int}^\mathrm{mf} & =-\sum_{n,m=0}^\infty \frac{\delta\dot{\Phi}_2^n}{n!}\frac{\delta\Phi_1^m}{m!} \left(\frac{2\pi}{\Phi_0}\right)^m \frac{\partial^{n+m} \varepsilon_J(\Delta_1,\boldsymbol{T_1},V_1+\langle \dot{\Phi}_2 \rangle,\varphi_1^{\mathrm{ex}}-\langle \varphi_1 \rangle,0)}{\partial V^n\partial \varphi^m},
\end{split}
\end{equation}
where~$\delta \Phi_k = \Phi_k - \langle \Phi_k \rangle$. The field averages have to be solved self-consistently. To second order in the fluctuations we arrive at the effective interaction Lagrangian
\begin{equation}
    \mathcal{L}_{\mathrm{int}}\approx \alpha_2(t) \dot{\Phi}_2 + \beta_1(t) \Phi_1 + G(t)\dot{\Phi}_2\Phi_1/2,
\end{equation}
where we defined 
\begin{align}
    & G(t) =- \frac{4\pi}{\Phi_0}\frac{\partial^2 \varepsilon_J(\Delta_1,\boldsymbol{T_1},V_1+\langle \dot{\Phi}_2 \rangle,\varphi_1^{\mathrm{ex}}-\langle \varphi_1 \rangle,0)}{\partial V\partial \varphi}, \\
    & \alpha_2(t) =- \frac{\partial \varepsilon_J(\Delta_1,\boldsymbol{T_1},V_1+\langle \dot{\Phi}_2 \rangle,\varphi_1^\mathrm{ex}-\langle \varphi_1 \rangle,0)}{\partial V} - \langle \Phi_1\rangle G(t)/2 \\
    & \beta_1(t) =- \frac{2\pi}{\Phi_0} \frac{\partial \varepsilon_J(\Delta_1,\boldsymbol{T_1},V_1+\langle \dot{\Phi}_2 \rangle,\varphi_1^\mathrm{ex}-\langle \varphi_1 \rangle,0)}{\partial \varphi} - \langle \dot{\Phi}_2\rangle G(t)/2.
\end{align}
To quartic order in the flux we find the approximate interaction Lagrangian~$\mathcal{L}_{\mathrm{int}}^\mathrm{mf} = G(t)\dot{\Phi}_2 \Phi_1$ where 
\begin{equation}
\begin{split}
    G(t) = \frac{4\pi }{\Phi_0}\frac{\partial E_J(\Delta_1,\boldsymbol{T_1},V_1)}{\partial V}\sin\left(\varphi_1^\mathrm{ex}\right)\left(1-\frac{\langle \varphi_1 \rangle^2}{2}\right) + \frac{4\pi }{\Phi_0}\frac{\partial^3E_J(\Delta_1,\boldsymbol{T_1},V_1)}{\partial V^3}\sin\left(\varphi_1^\mathrm{ex}\right) \frac{\langle \dot{\Phi}_2 \rangle^2}{2} 
   \\ -  \frac{4\pi}{\Phi_0} \frac{\partial^2E_J(\Delta_1,\boldsymbol{T_1},V_1)}{\partial V^2}\cos\left(\varphi_1^\mathrm{ex}\right)\langle \dot{\Phi}_2 \rangle\langle \varphi_1 \rangle. 
\end{split}
\end{equation}
We consider the second and third derivatives of~$E_J$ to be negligible. $R_Q = \Phi_0/(2e) = h/(2e)^2\simeq 6.5$ k$\Omega$ the resistance quantum.

\subsection{Estimation of the interaction strength}
\label{app:FENNEC Estimation of the interaction strength}

We remark that, following standard circuit quantization, the FENNEC interaction yields a Hamiltonian term~$(g/\hbar) \hat q_2 \hat \Phi_1$ where~$g =8E_{C_2} E_J'(V_0)/2e$ and~$E_{C_2}$ isthe charging energy of the second mode.

\begin{figure}[h!]
    \centering
    \includegraphics[width=0.75\textwidth]{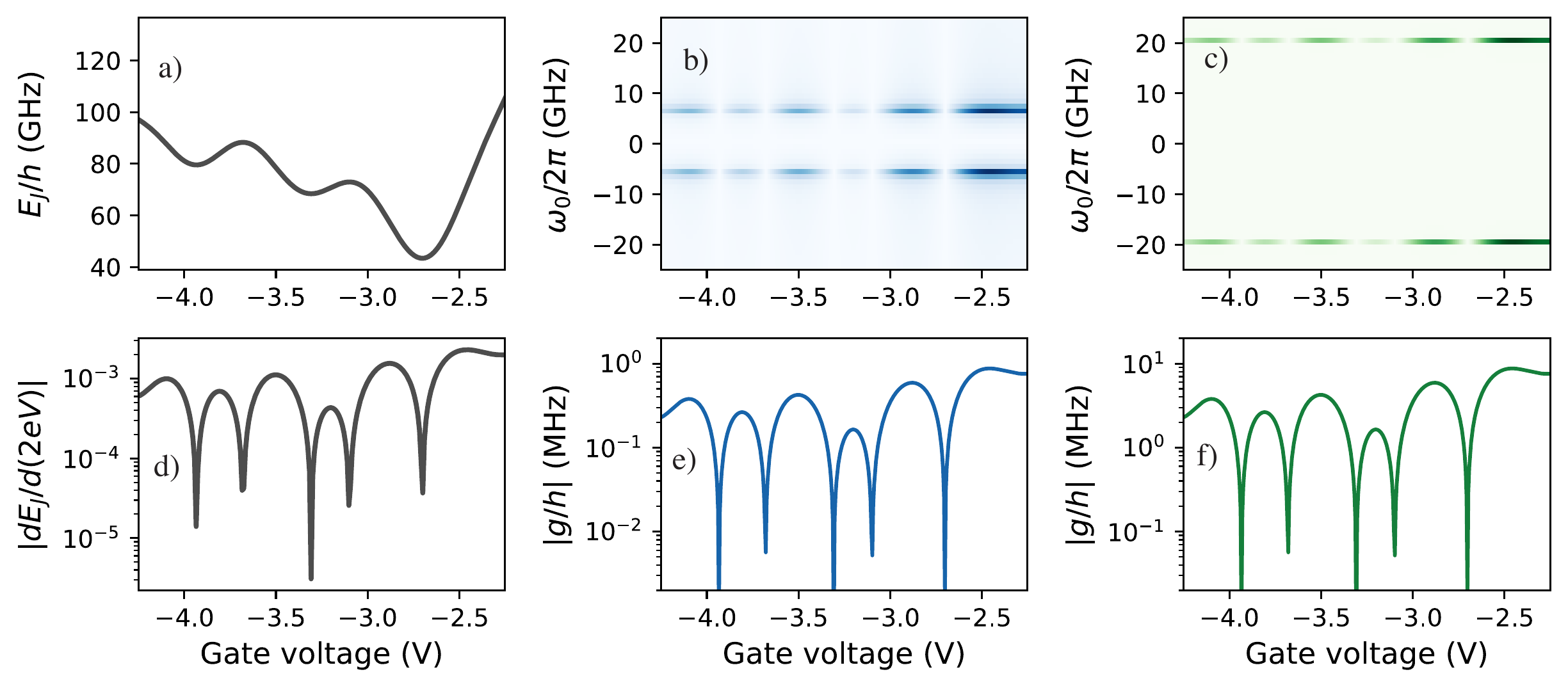}
    \caption{a) Josephson energy estimated from~\cite{Wang2019}. b)-c) Numerical discrete Fourier transform for different DC gate voltages. Here we add a small AC voltage with frequency and amplitude both determined by the capacitive and inductive energies of a fictitious second mode. The inductive energy is~$50/h$ GHz and the capactivie energy is~$0.1/h$ GHz in b) and~$1.0/h$ GHz in c). d) First derivative of the Josephson energy in a). e)-f) line-cut of b)-c) respectively at the frequency of the AC voltage. e) and f) follow the pattern of the first derivative in d).}
    \label{fig:FENNEC_interaction_graphene_fit}
\end{figure}

\begin{figure}[h!]
    \centering
    \includegraphics[width=0.75\textwidth]{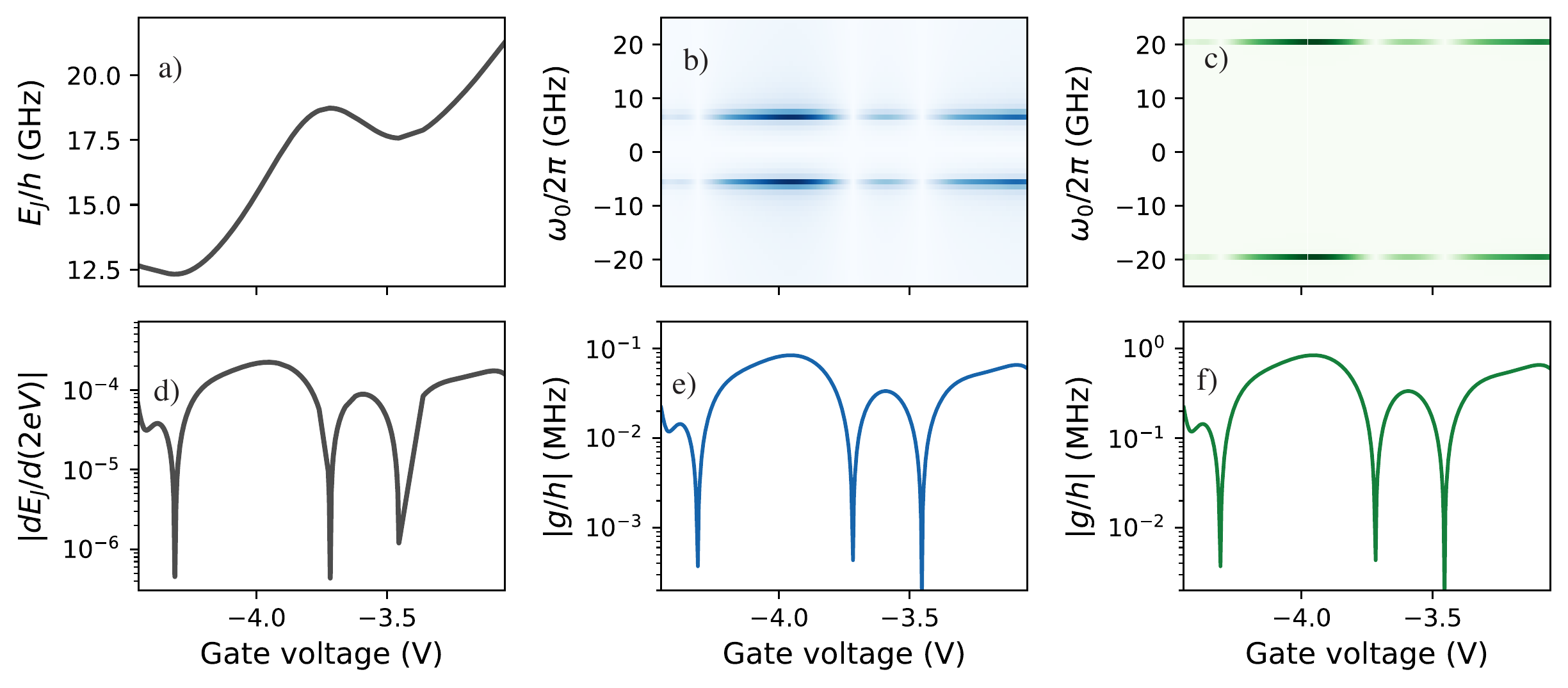}
    \caption{See caption of~\cref{fig:FENNEC_interaction_graphene_fit}. Based on the spectroscopy data in~\cite{Casparis2018}.}
    \label{fig:FENNEC_interaction_2deg_fit}
\end{figure}

\begin{figure}[h!]
    \centering
    \includegraphics[width=0.75\textwidth]{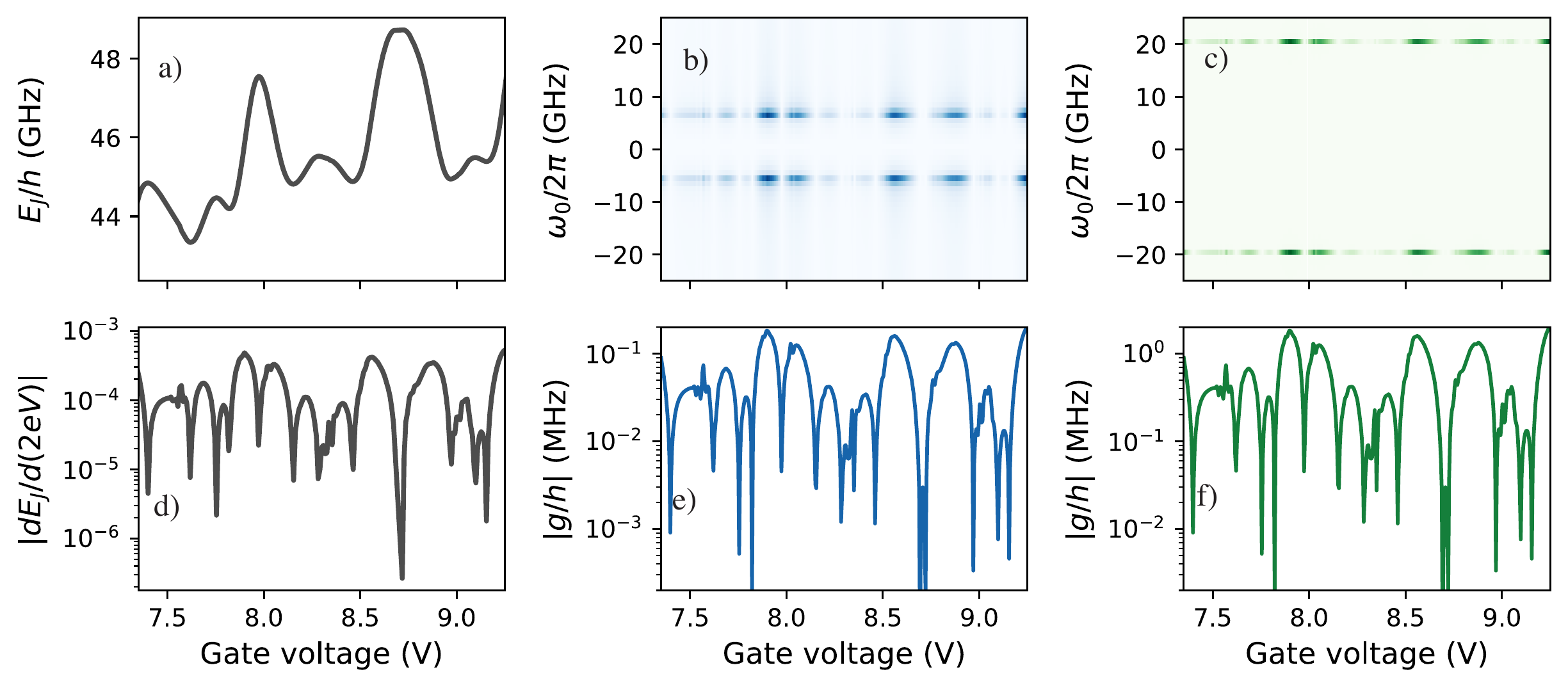}
    \caption{See caption of~\cref{fig:FENNEC_interaction_graphene_fit}. Based on the measured gatemon frequency in~\cite{Larsen2015}.}
    \label{fig:FENNEC_interaction_nanowire_fit}
\end{figure}

The Josephson energy~$E_J$ in the weak transmission limit is estimated from the approximate Gatemon transition energy formula 
\begin{equation}
    f_Q \approx \left(\sqrt{8 E_C E_J}-E_C\right)/h, \label{eq:f_Q}
\end{equation}
where~$E_C$ is the measured charging energy provided in~\cite{Larsen2015,Casparis2018,Wang2019}. We numerically compute the derivative using an interpolated spline that fits the~$f_Q$ that was experimentally measured. We also numerically confirm that the FENNEC interaction strength is indeed proportional to this derivative in~\cref{fig:FENNEC_interaction_graphene_fit,fig:FENNEC_interaction_2deg_fit,fig:FENNEC_interaction_nanowire_fit}. Two-dimensional electron gas junction have smoother energy with respect to the gate voltage (see~\cref{fig:FENNEC_interaction_2deg_fit}) but generally weaker first derivative. Nanowire junctions can in principle yield larger first derivatives (see~\cref{fig:FENNEC_interaction_nanowire_fit}) but appear more noisy. Graphene junctions result in both large first derivatives and smooth profiles (see~\cref{fig:FENNEC_interaction_graphene_fit}).

We also note that in the regime of a single channel with large transmission~$T(V)$ we instead find ~$\varepsilon_J(V,\Phi_1) \approx - \Delta |\cos(\pi\Phi_1/\Phi_0)| +  (\Delta/2)(T(V)-1) \sin^2(\pi \Phi_1/\Phi_0)|\sec(\pi\Phi_1/\Phi_0)|$, which is more sensitive to the external voltage~$V$ near half flux quantum. In other words, it is possible to find larger FENNEC interaction strengths by working in the large transmission limit.

\section{Gyrator implementation}
\label{app:Gyrator implementation}

\subsection{System Lagrangian}
\label{app:Gyrator System Lagrangian}

We consider a generic circuit Lagrangian of the form
\begin{equation}\label{eq: app full gyrator Lagrangian}
\mathcal L = \mathcal L_0 + \mathcal L_{\mathrm int} + \mathcal L_{\mathrm cor}, 
\end{equation}
where 
\begin{equation} \label{eq:L_0}
    \mathcal L_0 =  \dot{\boldsymbol{\Phi}}^T\cdot \frac{\boldsymbol{C_0}}{2}\cdot \dot{\boldsymbol{\Phi}} +\dot{\boldsymbol{\Phi}}^T\cdot \frac{\boldsymbol{C_c}}{2}\cdot \dot{\boldsymbol{\Phi}} + \left(\dot{\boldsymbol{\Phi}}-\boldsymbol{V}\right)^T\cdot \frac{\boldsymbol{C_J}}{2}\cdot \left(\dot{\boldsymbol{\Phi}}-\boldsymbol{V}\right)  - \boldsymbol{\Phi}^T \cdot \frac{\boldsymbol{L_0}^{-1}}{2}\cdot \boldsymbol{\Phi},
\end{equation}
is the Lagrangian due to all standard superconducting circuit elements, 
\begin{equation} \label{eq:L_int}
\begin{split}
    \mathcal L_{\mathrm int} & =- \varepsilon_J(\Delta_1,\boldsymbol{T_1},V_1+\dot{\Phi}_2,\varphi_1^{\mathrm{ex}},\varphi_1) - \varepsilon_J(\Delta_2,\boldsymbol{T_2},V_2+\dot{\Phi}_1,\varphi_2^{\mathrm{ex}},\varphi_2) \\
    & =-\sum_{n,m=0}^\infty \frac{\dot{\Phi}_2^n}{n!}\frac{\Phi_1^m}{m!} \left(\frac{2\pi}{\Phi_0}\right)^m \frac{\partial^{n+m} \varepsilon_J(\Delta_1,\boldsymbol{T_1},V_1,\varphi_1^{\mathrm{ex}},0)}{\partial V^n\partial \varphi^m} - \sum_{n,m=0}^\infty \frac{\dot{\Phi}_1^n}{n!}\frac{\Phi_2^m}{m!} \left(\frac{2\pi}{\Phi_0}\right)^m \frac{\partial^{n+m} \varepsilon_J(\Delta_2,\boldsymbol{T_2},V_2,\varphi_2^{\mathrm{ex}},0)}{\partial V^n\partial \varphi^m}
\end{split}
\end{equation}
results from the FENNEC interaction alone, and
\begin{equation}
\begin{split}
    \mathcal L_{\mathrm cor} & = - \varepsilon_J(\Delta_1,\boldsymbol{T_1},V_1,\varphi_1^{\mathrm{ex}}-\pi,\varphi_1) - \varepsilon_J(\Delta_2,\boldsymbol{T_2},V_2,\varphi_2^{\mathrm{ex}}+\pi,\varphi_2) \\ 
    & =-\sum_{m=0}^\infty \frac{\Phi_1^m}{m!} \left(\frac{2\pi}{\Phi_0}\right)^m \frac{\partial^{m} \varepsilon_J(\Delta_1,\boldsymbol{T_1},V_1,\varphi_1^{\mathrm{ex}}-\pi,0)}{\partial \varphi^m} - \sum_{m=0}^\infty \frac{\Phi_2^m}{m!} \left(\frac{2\pi}{\Phi_0}\right)^m \frac{\partial^{m} \varepsilon_J(\Delta_2,\boldsymbol{T_2},V_2,\varphi_2^{\mathrm{ex}}+\pi,0)}{\partial \varphi^m} .
\end{split}
\end{equation}
will be used to cancel the potentially large interaction-free part of~$\mathcal{L}_{\mathrm{int}}$ since~$\varepsilon_J(\Delta,\boldsymbol{T},V,\varphi^{\mathrm{ex}}\pm\pi,0)=-\varepsilon_J(\Delta,\boldsymbol{T},V,\varphi^{\mathrm{ex}},0)$ in the weak transmission limit~$[\boldsymbol{T}(V_1)]_i\ll 1$, as will be clear below. Here~$\boldsymbol{\Phi} = \left(\Phi_1, \ \Phi_2\right)$ is a vector comprising the branch flux~$\Phi_1$ ($\Phi_2$) of the first (second) mode,~$\boldsymbol{\varphi} = 2\pi\boldsymbol{\Phi}/\Phi_0$ are the associated branch phases,~$\boldsymbol{C_0}$ and~$\boldsymbol{C_c}$ are capacitance matrices due to the shunt capacitors and the coupling capacitors respectively,~$\boldsymbol{C_J}$ is the capacitance matrix associated with the coupling to the control voltage lines~$\boldsymbol{V}$,~$\boldsymbol{L_0}$ is an inductance matrix,
\begin{equation}
    \varepsilon_J(\Delta,\boldsymbol{T},V,\varphi^{\mathrm{ex}},\varphi) = -\Delta \sum_i \sqrt{1-[\boldsymbol{T}(V)]_i\sin^2\left(\frac{ \varphi-\varphi^{\mathrm ex}}{2}\right)},
\end{equation}
is the form of the Andreev bound-state energy of any semiconducting junction in the circuit,~$\Delta_k$ is the gap energy of the~$k$th junction with transmission probability~$[\boldsymbol{T_k}(V_k)]_i$,~$\Phi_k^{\mathrm ex}$ is an external flux threading the~$k$th loop. In this work we focus on the leading order contribution of the interaction Lagrangian
\begin{equation} \label{eq:L_int^2}
    \mathcal L_{\mathrm{target}} = G_+(t) \left(\dot{\Phi}_2\Phi_1+\dot{\Phi}_1\Phi_2\right)/2 + G_-(t) \left(\dot{\Phi}_2\Phi_1-\dot{\Phi}_1\Phi_2\right)/2,
\end{equation}
where we defined the amplitudes 
\begin{equation}
    G_\pm =- \frac{2\pi}{\Phi_0}\frac{\partial^2 \varepsilon_J(\Delta_1,\boldsymbol{T_1},V_1,\varphi_1^{\mathrm{ex}},0)}{\partial V\partial \varphi} \mp \frac{2\pi}{\Phi_0}\frac{\partial^2 \varepsilon_J(\Delta_2,\boldsymbol{T_2},V_2,\varphi_2^{\mathrm{ex}},0)}{\partial V\partial \varphi}.
\end{equation}
Here 
\begin{equation}
    \frac{\partial^2 \varepsilon_J(\Delta,\boldsymbol{T},V,\varphi^{\mathrm{ex}},0)}{\partial V\partial \varphi} =- \frac{\Delta}{4}\sum_i\frac{\partial [\boldsymbol{T}(V)]_i}{\partial V} \sin\left(\varphi^{\mathrm ex}\right)\frac{1-[\boldsymbol{T}(V)]_i\sin^2\left(\varphi^{\mathrm ex}/2\right)/2}{\sqrt{1-[\boldsymbol{T}(V)]_i\sin^2\left(\varphi^{\mathrm ex}/2\right)}^3}.
\end{equation}
Moreover~$G_-(t) = 1/(2R)$, where~$R$ is the resistance of a gyrator. Overall we truncate the interaction Lagrangian to 
\begin{equation}
    \mathcal{L}_{\mathrm{int}}+\mathcal L_{\mathrm cor}\approx \sum_{k=1}^2\left[\frac{ c_k}{2}\dot{\Phi}_k^2 + \frac{\Phi_k^2}{2\ell_k} + \alpha_k \dot{\Phi}_k + \beta_k \Phi_k\right]+ G_+(t) \left(\dot{\Phi}_2\Phi_1+\dot{\Phi}_1\Phi_2\right)/2 + G_-(t) \left(\dot{\Phi}_2\Phi_1-\dot{\Phi}_1\Phi_2\right)/2
\end{equation}
where we defined 
\begin{align}
    & c_k = -\frac{\partial^2 \varepsilon_J(\Delta_\ell,\boldsymbol{T_\ell},V_\ell,\varphi_\ell^\mathrm{ex},0)}{\partial V^2}, \\ 
    & \frac{1}{\ell_k} =-\left(\frac{2\pi}{\Phi_0}\right)^2 \frac{\partial^2 \varepsilon_J(\Delta_k,\boldsymbol{T_k},V_k,\varphi_k^\mathrm{ex},0)}{\partial \varphi^2}-\left(\frac{2\pi}{\Phi_0}\right)^2 \frac{\partial^2 \varepsilon_J(\Delta_k,\boldsymbol{T_k},V_k,\varphi_k^\mathrm{ex}+(-1)^k\pi,0)}{\partial \varphi^2}, \\ 
    & \alpha_k = -\frac{\partial \varepsilon_J(\Delta_\ell,\boldsymbol{T_\ell},V_\ell,\varphi_\ell^\mathrm{ex},0)}{\partial V}, \\
    & \beta_k =- \frac{2\pi}{\Phi_0} \frac{\partial \varepsilon_J(\Delta_k,\boldsymbol{T_k},V_k,\varphi_k^\mathrm{ex},0)}{\partial \varphi} - \frac{2\pi}{\Phi_0} \frac{\partial \varepsilon_J(\Delta_k,\boldsymbol{T_k},V_k,\varphi_k^\mathrm{ex}+(-1)^k\pi,0)}{\partial \varphi}.
\end{align}
For optimal gyration we wish for~$G_-(t)$ ($G_+(t)$) to be maximized (minimized). $G_-(t)$ leads to a resonant Jaynes-Cummings-type interaction with a~$\pi/2$ relative phase ($i\hat a^\dagger \hat b + \mathrm{h.c}$) whereas~$G_+(t)$ leads to a off-resonant two-mode-squeezing-type interaction ($i\hat a^\dagger \hat b^\dagger + \mathrm{h.c}$). 

\subsection{Weak transmission limit}
\label{app:Gyrator Weak transmission limit}

In the weak transmission ($[\boldsymbol{T_k}(V_k)]_i\ll 1$) limit we find that  
\begin{equation}
    \varepsilon_J(\Delta,\boldsymbol{T},V,\varphi^{\mathrm{ex}},\varphi) \approx \Delta + \frac{\Delta \sum_i [\boldsymbol{T}(V)]_i}{4} - \frac{\Delta \sum_i [\boldsymbol{T}(V)]_i}{4}\cos\left(\varphi-\varphi^{\mathrm ex}\right) = \Delta + E_J(\Delta,\boldsymbol{T},V) - E_J(\Delta,\boldsymbol{T},V)\cos\left(\varphi-\varphi^{\mathrm ex}\right).
\end{equation}
Notice that 
\begin{align}
    & c_k = -\left(1-\cos\left(\varphi_\ell^\mathrm{ex}\right)\right)\frac{\partial^2 E_J(\Delta_\ell,\boldsymbol{T_\ell},V_\ell)}{\partial V^2}, \\ 
    & \frac{1}{\ell_k} =0, \\ 
    & \alpha_k = -\left(1-\cos\left(\varphi_\ell^\mathrm{ex}\right)\right)\frac{\partial E_J(\Delta_\ell,\boldsymbol{T_\ell},V_\ell)}{\partial V}, \\
    & \beta_k =0.
\end{align}
From now on we will drop~$c_k$ and~$1/\ell_k$ in the assumption that they are negligible contributions to the capacitance and inductance of the modes.

\paragraph{Gyration.} 

If the FENNEC interaction can prove useful for two-qubit gates the main application is the realization of nonreciprocal devices. It follows that 
\begin{equation}
    G_-(t) \approx \frac{\pi}{R_Q}  \sin\left(\varphi_1^{\mathrm ex}\right) \frac{1}{2e} \frac{\partial E_J(\Delta_1,\boldsymbol{T_1},V_1)}{\partial V}  - \frac{\pi}{R_Q} \sin\left(\varphi_2^{\mathrm ex}\right)\frac{1}{2e} \frac{\partial E_J(\Delta_2,\boldsymbol{T_2},V_2)}{\partial V}
\end{equation}
where we defined~$R_Q = \Phi_0/(2e) = h/(2e)^2\simeq 6.5$ k$\Omega$ the resistance quantum. Importantly, this implies that the resistance of the gyrator is 
\begin{equation}
    R = \frac{R_Q}{2\pi}\left(\sin\left(\varphi_1^{\mathrm ex}\right) \frac{1}{2e} \frac{\partial E_J(\Delta_1,\boldsymbol{T_1},V_1)}{\partial V}  -  \sin\left(\varphi_2^{\mathrm ex}\right)\frac{1}{2e} \frac{\partial E_J(\Delta_2,\boldsymbol{T_2},V_2)}{\partial V}\right)^{-1}.
\end{equation}
Typically~$|\partial E_J(\Delta_k,\boldsymbol{T_k},V_k)/\partial V|\ll 1$ and it is therefore clear that the resistance of the gyrator~$R$ is mostly likely larger than the resistance quantum~$R_Q$.  

\subsection{Noise sensitivity}
\label{app:Gyrator Noise sensitivity}

In this section we analyze the noise sensitivity of the device. Notice that given the term~$\varepsilon_J(\Delta_1,\boldsymbol{T_1},V_1+\dot{\Phi}_2,\varphi_1^{\mathrm{ex}},\varphi_1)$, charge noise in the second mode, such that~$\dot{\Phi}_2 \to \dot{\Phi}_2+\delta \dot{\Phi}_2$, is equivalent to~$V_1 \to V_1 +\delta \dot{\Phi}_2$. Similarly, flux noise in the first mode, such that~$\varphi_1\to \varphi_1 + \delta \varphi_1$, is equivalent to~$\varphi_1^\mathrm{ex}\to \varphi_1^\mathrm{ex} - \delta\varphi_1$.

\emph{Charge noise.} In presence of charge noise, which amounts to~$V_1\to V_1 + \delta \dot{\Phi}_2$ and ~$V_2\to V_2 + \delta \dot{\Phi}_1$ in the FENNEC interaction strength~$G_-(t)$, we find that~$G_-(t)\to G_-(t)+\delta G_-(t)$ where 
\begin{equation}
    \delta G_-(t) \approx \frac{\pi}{R_Q}  \sin\left(\varphi_1^{\mathrm ex}\right) \frac{1}{2e} \frac{\partial^2 E_J(\Delta_1,\boldsymbol{T_1},V_1)}{\partial V^2} \delta \dot{\Phi}_2   - \frac{\pi}{R_Q} \sin\left(\varphi_2^{\mathrm ex}\right)\frac{1}{2e} \frac{\partial^2 E_J(\Delta_2,\boldsymbol{T_2},V_2)}{\partial V^2}\delta \dot{\Phi}_1
\end{equation}
to leading order in the noise.

The frequencies of the normal modes of gyrator become~$\omega_\pm = \omega_0 \pm G_-(t) \pm \delta G_-(t)$. We observe that the dispersion is linear in charge noise and determined by the second derivative of~$E_J$.

\emph{Flux noise.} In presence of flux noise, which can be implemented with~$\varphi_1^\mathrm{ex}\to \varphi_1^\mathrm{ex}- \delta \varphi_1$ and ~$\varphi_2^\mathrm{ex}\to \varphi_2^\mathrm{ex}- \delta \varphi_2$ in the FENNEC interaction strength~$G_-(t)$, we find that~$G_-(t)\to G_-(t)+\delta G_-(t)$ where 
\begin{equation}
    \delta G_-(t) \approx -\frac{\pi}{R_Q}  \cos\left(\varphi_1^{\mathrm ex}\right)  \frac{1}{2e} \frac{\partial E_J(\Delta_1,\boldsymbol{T_1},V_1)}{\partial V} \delta \varphi_1  + \frac{\pi}{R_Q} \cos\left(\varphi_2^{\mathrm ex}\right) \frac{1}{2e} \frac{\partial E_J(\Delta_2,\boldsymbol{T_2},V_2)}{\partial V}\delta \varphi_2.
\end{equation}
We observe that the system is insensitive to flux noise to leading order at the optimal gyration point~$|\varphi_k^\mathrm{ex}|=\pi/2$. 

\subsection{Mean-field theory}
\label{app:Gyrator Mean-field theory}

In this section we linearized the FENNEC interaction within a mean-field theory approximation:
\begin{equation} \label{eq:L_int^mf}
\begin{split}
    \mathcal L_{\mathrm int}^\mathrm{mf} & = -\varepsilon_J(\Delta_1,\boldsymbol{T_1},V_1+\dot{\Phi}_2,\varphi_1^{\mathrm{ex}},\varphi_1) - \varepsilon_J(\Delta_2,\boldsymbol{T_2},V_2+\dot{\Phi}_1,\varphi_2^{\mathrm{ex}},\varphi_2) \\
    & =-\sum_{n,m=0}^\infty \frac{\delta\dot{\Phi}_2^n}{n!}\frac{\delta\Phi_1^m}{m!} \left(\frac{2\pi}{\Phi_0}\right)^m \frac{\partial^{n+m} \varepsilon_J(\Delta_1,\boldsymbol{T_1},V_1+\langle \dot{\Phi}_2 \rangle,\varphi_1^{\mathrm{ex}}-\langle \varphi_1 \rangle,0)}{\partial V^n\partial \varphi^m} \\
    & - \sum_{n,m=0}^\infty \frac{\delta\dot{\Phi}_1^n}{n!}\frac{\delta\Phi_2^m}{m!} \left(\frac{2\pi}{\Phi_0}\right)^m \frac{\partial^{n+m} \varepsilon_J(\Delta_2,\boldsymbol{T_2},V_2+\langle \dot{\Phi}_1 \rangle ,\varphi_2^{\mathrm{ex}}-\langle \varphi_2 \rangle,0)}{\partial V^n\partial \varphi^m},
\end{split}
\end{equation}
where~$\delta \Phi_k = \Phi_k - \langle \Phi_k \rangle$. Similarly,
\begin{equation} \label{eq:L_cor^mf}
\begin{split}
    \mathcal L_{\mathrm cor}^\mathrm{mf} & = -\varepsilon_J(\Delta_1,\boldsymbol{T_1},V_1+\dot{\Phi}_2,\varphi_1^{\mathrm{ex}}-\pi,\varphi_1) - \varepsilon_J(\Delta_2,\boldsymbol{T_2},V_2+\dot{\Phi}_1,\varphi_2^{\mathrm{ex}}+\pi,\varphi_2) \\
    & =-\sum_{m=0}^\infty \frac{\delta\Phi_1^m}{m!} \left(\frac{2\pi}{\Phi_0}\right)^m \frac{\partial^{m} \varepsilon_J(\Delta_1,\boldsymbol{T_1},V_1,\varphi_1^{\mathrm{ex}}-\pi-\langle \varphi_1 \rangle,0)}{\partial \varphi^m} \\
    & - \sum_{m=0}^\infty \frac{\delta\Phi_2^m}{m!} \left(\frac{2\pi}{\Phi_0}\right)^m \frac{\partial^{m} \varepsilon_J(\Delta_2,\boldsymbol{T_2},V_2 ,\varphi_2^{\mathrm{ex}}+\pi-\langle \varphi_2 \rangle,0)}{\partial \varphi^m}.
\end{split}
\end{equation}

To second order we therefore arrive at the effective interaction Lagrangian
\begin{equation}
    \mathcal{L}_{\mathrm{int}}^\mathrm{mf}+\mathcal{L}_{\mathrm{cor}}^\mathrm{mf}\approx \sum_{k=1}^2\left[ \alpha_k(t) \dot{\Phi}_k + \beta_k(t) \Phi_k\right]+ G_+(t) \left(\dot{\Phi}_2\Phi_1+\dot{\Phi}_1\Phi_2\right)/2 + G_-(t) \left(\dot{\Phi}_2\Phi_1-\dot{\Phi}_1\Phi_2\right)/2,
\end{equation}
where we defined 
\begin{align}
    & G_\pm(t) =- \frac{2\pi}{\Phi_0}\frac{\partial^2 \varepsilon_J(\Delta_1,\boldsymbol{T_1},V_1+\langle \dot{\Phi}_2 \rangle,\varphi_1^{\mathrm{ex}}-\langle \varphi_1 \rangle,0)}{\partial V\partial \varphi} \mp  \frac{2\pi}{\Phi_0}\frac{\partial^2 \varepsilon_J(\Delta_2,\boldsymbol{T_2},V_2+\langle \dot{\Phi}_1 \rangle,\varphi_2^{\mathrm{ex}}-\langle \varphi_2 \rangle,0)}{\partial V\partial \varphi}, \\
    & \alpha_k(t) =- \frac{\partial \varepsilon_J(\Delta_\ell,\boldsymbol{T_\ell},V_\ell+\langle \dot{\Phi}_k \rangle,\varphi_\ell^\mathrm{ex}-\langle \varphi_\ell \rangle,0)}{\partial V} - \langle \Phi_\ell\rangle \frac{G_+(t)+(-1)^\ell G_-(t)}{4}, \\
    \nonumber & \beta_k(t) =- \frac{2\pi}{\Phi_0} \frac{\partial \varepsilon_J(\Delta_k,\boldsymbol{T_k},V_k+\langle \dot{\Phi}_\ell \rangle,\varphi_k^\mathrm{ex}-\langle \varphi_k \rangle,0)}{\partial \varphi} - \frac{2\pi}{\Phi_0} \frac{\partial \varepsilon_J(\Delta_k,\boldsymbol{T_k},V_k,\varphi_k^\mathrm{ex}+(-1)^k\pi-\langle \varphi_k \rangle,0)}{\partial \varphi} \\
    & \phantom{\beta_k(t) =} - \langle \dot{\Phi}_\ell\rangle \frac{G_+(t)+(-1)^\ell G_-(t)}{4},
\end{align}
where~$\ell \neq k$. The field averages have to be solved self-consistently. To quartic order in the flux while neglecting higher order derivatives in either~$V$ or~$\varphi$, we find that
\begin{align}
   & G_\pm (t) = \frac{2\pi }{\Phi_0}\frac{\partial E_J(\Delta_1,\boldsymbol{T_1},V_1)}{\partial V}\sin\left(\varphi_1^\mathrm{ex}\right)\left(1-\frac{\langle \varphi_1 \rangle^2}{2}\right) \pm \frac{2\pi }{\Phi_0}\frac{\partial E_J(\Delta_2,\boldsymbol{T_2},V_2)}{\partial V}\sin\left(\varphi_2^\mathrm{ex}\right)\left(1-\frac{\langle \varphi_2 \rangle^2}{2}\right) \\
    & \alpha_k(t) = - \left(1-\cos(\varphi_\ell^\mathrm{ex})\right)\frac{\partial E_J(\Delta_\ell,\boldsymbol{T_\ell},V_\ell)}{\partial V},
\end{align}
and~$\beta_k(t)=0$.

\subsection{Scattering matrix of the linearized system}
\label{app:Gyrator Scattering matrix of the linearized system}

In this section, we focus on the linear mean-field Lagrangian~\cite{Hazard2019}:
\begin{equation} \label{eq: app mean field full gyrator Lagrangian}
\begin{split}
     \mathcal L^\mathrm{mf} =   \sum_{i=1}^2\int_{-\infty}^0 dx \left[\frac{c}{2}\left(\partial_t \tilde{\Phi}_i(x,t)\right)^2-\frac{1}{2\ell}\left(\partial_x \tilde{\Phi}_i(x,t)\right)^2\right] - \left(\boldsymbol{\tilde{\Phi}}(0,t)-\boldsymbol{\Phi}\right)^T\cdot \frac{\boldsymbol{L_c}^{-1}}{2}\cdot \left(\boldsymbol{\tilde{\Phi}}(0,t)-\boldsymbol{\Phi}\right) + \boldsymbol{\alpha}^T\cdot\dot{\boldsymbol{\Phi}} \\
     +  \dot{\boldsymbol{\Phi}}^T\cdot \frac{\boldsymbol{C}}{2}\cdot \dot{\boldsymbol{\Phi}} - \boldsymbol{\Phi}^T\cdot \frac{\boldsymbol{L}^{-1}}{2}\cdot \boldsymbol{\Phi} + \dot{\boldsymbol{\Phi}}^T \cdot \frac{G_+(t)\boldsymbol{\sigma_x}-iG_-(t)\boldsymbol{\sigma_y}}{4} \cdot \boldsymbol{\Phi} + \boldsymbol{\Phi}^T \cdot \frac{G_+(t)\boldsymbol{\sigma_x}+iG_-(t)\boldsymbol{\sigma_y}}{4} \cdot \dot{\boldsymbol{\Phi}},
\end{split}
\end{equation}
where~$\boldsymbol{L_c}$ is assumed to be diagonal. Here~$\boldsymbol{\sigma_x}$ and~$\boldsymbol{\sigma_y}$ are the Pauli matrices.

\subsubsection{Equations of motion}
\label{app:Gyrator Equations of motion}

The equations of motion for the effectively linearized Lagrangian are given by
\begin{equation}
    0 = \partial_t\frac{\partial \mathcal{L}^\mathrm{mf}}{\partial (\partial_t \tilde{\Phi}_i)} + \partial_x\frac{\partial \mathcal{L}^\mathrm{mf}}{\partial (\partial_x \tilde{\Phi}_i)}  - \frac{\partial \mathcal{L}^\mathrm{mf}}{\partial \tilde{\Phi}_i}= \frac{d}{dt}\frac{\partial \mathcal{L}^\mathrm{mf}}{\partial \dot{\Phi}_i} - \frac{\partial \mathcal{L}^\mathrm{mf}}{\partial \Phi_i}, \quad i = 1,2,
\end{equation}
which explicitly take the form 
\begin{align}
& c \partial_t^2 \boldsymbol{\tilde{\Phi}}(x,t) = \frac{1}{\ell} \partial_x^2 \boldsymbol{\tilde{\Phi}}(x,t), \label{eq:TL wave eq}\\
&\frac{1}{\ell} \partial_x\boldsymbol{\tilde{\Phi}}(0,t) =\boldsymbol{L_c}^{-1}\cdot \left(\boldsymbol{\tilde{\Phi}}(0,t)-\boldsymbol{\Phi}\right), \label{eq:boundary eq} \\
& 0 = \boldsymbol{C}\cdot \ddot{\boldsymbol{\Phi}} + \boldsymbol{L}^{-1}\cdot \boldsymbol{\Phi} -i G_-(t)\boldsymbol{\sigma_y}\cdot \dot{\boldsymbol{\Phi}}-\frac{1}{\ell} \partial_x\boldsymbol{\tilde{\Phi}}(0,t). \label{eq:gyrator wave eq}
\end{align}
Given the wave-equation~\cref{eq:TL wave eq} we find that the quantized field in the transmission line~$j$ has the form
\begin{align}
    & \boldsymbol{\tilde{\Phi}}(x,t) =  \sqrt{\frac{\hbar}{4\pi c}}\int_0^\infty \frac{d\omega}{\sqrt{\omega}}\left(e^{i\omega t+ik_\omega x} \boldsymbol{a_{\omega}}^{\dagger} + e^{i\omega t- ik_\omega x} \boldsymbol{b}_{\omega}^{\dagger}+\mathrm{h.c.}\right), \label{eq:tilde phi x t}
\end{align}
with the dispersion relation~$k_\omega = \omega\sqrt{c \hspace{0.1em} \ell}=\omega/\omega_\mathrm{TL}$ and commutation relations~$\comm{\hat a_{i,\omega}}{\hat a_{j,\omega'}^\dagger}=\delta_{ij}\delta(\omega-\omega')$ and~$\comm{\hat b_{i,\omega}}{\hat b_{j,\omega'}^\dagger}=\delta_{ij}\delta(\omega-\omega')$. Here~$\boldsymbol{a_\omega}$ ($\boldsymbol{b_\omega}$) are the annihilation operators associated with the ingoing (outgoing) fields at frequency~$\omega$. 

\subsubsection{Fourier transform}
\label{app:Gyrator Fourier transform}

We apply a Fourier transform on~\cref{eq:boundary eq,eq:gyrator wave eq} (with the definition~$\hat y(\omega) = \int_{-\infty}^{\infty} dt \   y(t) e^{i\omega t}/\sqrt{2\pi}$ and property~$y(-\omega) = [y(\omega)]^*$):
\begin{align}
&\boldsymbol{\hat \Phi}(\omega) =  \boldsymbol{\hat{\tilde{\Phi}}}(0,\omega)- \frac{\boldsymbol{L_c}}{\ell} \cdot \partial_x\boldsymbol{\hat{\tilde{\Phi}}}(0,\omega), \label{eq:boundary Fourier eq} \\
& 0 = \left(-\omega^2 \boldsymbol{C} + \boldsymbol{L}^{-1}\right)\cdot \boldsymbol{\hat \Phi}(\omega) -\int_{-\infty}^\infty d\omega' \  \omega'  \hat{ G}_-(\omega-\omega')\boldsymbol{\sigma_y}\cdot \boldsymbol{\hat \Phi}(\omega')-\frac{1}{\ell} \partial_x\boldsymbol{\hat{\tilde{\Phi}}}(0,\omega). \label{eq:gyrator wave Fourier eq}
\end{align}

\subsubsection{Expansion in the amplitude of the flux fields}
\label{app:Gyrator Expanion in the amplitude of the flux fields}

$G_-(t)$ depends on the average of the flux fields which we assume to have small amplitude. We write~$G_-(t) = \overline{G}_- + \lambda dG_-(t)$ where  
\begin{equation}
    \overline{G}_- = \frac{2\pi }{\Phi_0}\frac{\partial E_J(\Delta_1,\boldsymbol{T_}1,V_1)}{\partial V}\sin\left(\varphi_1^\mathrm{ex}\right)- \frac{2\pi }{\Phi_0}\frac{\partial E_J(\Delta_2,\boldsymbol{T_2},V_2)}{\partial V}\sin\left(\varphi_2^\mathrm{ex}\right)
\end{equation}
is the contribution that is independent of the flux fields, and \begin{equation}
\begin{split}
    \lambda d G_- (t) &= \frac{2\pi }{\Phi_0}\frac{\partial E_J(\Delta_2,\boldsymbol{T_2},V_2)}{\partial V}\sin\left(\varphi_2^\mathrm{ex}\right) \frac{\langle \varphi_2(t) \rangle^2}{2}-\frac{2\pi }{\Phi_0}\frac{\partial E_J(\Delta_1,\boldsymbol{T_1},V_1)}{\partial V}\sin\left(\varphi_1^\mathrm{ex}\right) \frac{\langle \varphi_1(t) \rangle^2}{2} 
\end{split},
\end{equation}
depends on the flux fields following the mean-field approximation. We do a perturbative expansion in~$\lambda$, i.e. $\hat{\boldsymbol{\tilde{\Phi}}}(0,\omega) = \sum_{k=0}^\infty \lambda ^k\hat{\boldsymbol{\tilde{\Phi}}}^{(k)}(0,\omega)$ and~$\lambda dG_-(t) = \sum_{k=0}^\infty \lambda ^{k+1}dG_-^{(k)}(t)$, and solve~\cref{eq:gyrator wave Fourier eq} in each order of~$\lambda$. For conciseness we stop at first order. 

\paragraph{Order 0:}

\begin{align}
& 0 = \left(-\omega^2 \boldsymbol{C} + \boldsymbol{L}^{-1}- \omega  \overline{G}_-\boldsymbol{\sigma_y}\right)\cdot \left(\boldsymbol{\hat{\tilde{\Phi}}}^{(0)}(0,\omega)- \frac{\boldsymbol{L_c}}{\ell} \cdot \partial_x\boldsymbol{\hat{\tilde{\Phi}}}^{(0)}(0,\omega)\right) -\frac{1}{\ell} \partial_x\boldsymbol{\hat{\tilde{\Phi}}}^{(0)}(0,\omega). \label{eq: order 0}
\end{align}

\paragraph{Order 1:}

\begin{align}
\begin{split}
    0 = \left(-\omega^2 \boldsymbol{C} + \boldsymbol{L}^{-1}- \omega  \overline{G}_-\boldsymbol{\sigma_y}\right)\cdot \left(\boldsymbol{\hat{\tilde{\Phi}}}^{(1)}(0,\omega)- \frac{\boldsymbol{L_c}}{\ell} \cdot \partial_x\boldsymbol{\hat{\tilde{\Phi}}}^{(1)}(0,\omega)\right)-\frac{1}{\ell} \partial_x\boldsymbol{\hat{\tilde{\Phi}}}^{(1)}(0,\omega) \\
    -\int_{-\infty}^\infty d\omega' \  \omega'  \hat{d G}_-^{(0)}(\omega-\omega')\boldsymbol{\sigma_y}\cdot \left(\boldsymbol{\hat{\tilde{\Phi}}}^{(0)}(0,\omega')- \frac{\boldsymbol{L_c}}{\ell} \cdot \partial_x\boldsymbol{\hat{\tilde{\Phi}}}^{(0)}(0,\omega')\right). \label{eq: order 1}
\end{split}
\end{align}

Consistently with the perturbative expansion we have that
\begin{equation}
\begin{split}
    \lambda d G_-^0 (t) & =\frac{2\pi }{\Phi_0}\frac{\partial E_J(\Delta_2,\boldsymbol{T_2},V_2)}{\partial V}\sin\left(\varphi_2^\mathrm{ex}\right) \frac{\langle \varphi_2^{(0)}(t) \rangle^2}{2}- \frac{2\pi }{\Phi_0}\frac{\partial E_J(\Delta_1,\boldsymbol{T_1},V_1)}{\partial V}\sin\left(\varphi_1^\mathrm{ex}\right) \frac{\langle \varphi_1^{(0)}(t) \rangle^2}{2}
\end{split}.
\end{equation}

\subsubsection{Input/output equations}
\label{app:Gyrator Input/output equations}

Accordingly to~\cref{eq:tilde phi x t} we have in the case~$\omega>0$ 
\begin{equation}
    \boldsymbol{\hat{\tilde{\Phi}}}^{(k)}(x,\omega>0) = \sqrt{\frac{\hbar}{4\pi c \omega}} \left(e^{- ik_\omega x}\boldsymbol{a_\omega}\delta_{k,0}+e^{ik_\omega x}\boldsymbol{b_\omega}^{(k)}\right),
\end{equation}
where~$\delta_{ij} = \begin{cases}1, & i=j \\ 0, & i\neq j \end{cases}$ is the discrete delta function. \cref{eq: order 0} then takes the form 
\begin{align}
    0 =  \boldsymbol{a_\omega}+\boldsymbol{b_\omega}^{(0)} + \frac{\boldsymbol{Z}(\omega)}{Z_\mathrm{TL}}\cdot  \left(\boldsymbol{a_\omega}-\boldsymbol{b_\omega}^{(0)}\right)
\end{align}
where~$Z_\mathrm{TL}=\sqrt{\ell/c}$ is the characteristic impedance of the transmission lines and 
\begin{equation}
    \boldsymbol{Z}(\omega) = i \omega \boldsymbol{L_c} + \left(i\omega \boldsymbol{C}  + (i\omega\boldsymbol{L})^{-1}  + i   \overline{G}_-\boldsymbol{\sigma_y}\right)^{-1} = i \omega \boldsymbol{L_c} + \boldsymbol{Z_0}(\omega).
\end{equation}
At zeroth order the scattering matrix is then 
\begin{equation}
    \boldsymbol{b_\omega}^{(0)} =\boldsymbol{S}^{(0)}(\omega) \cdot \boldsymbol{a_\omega} =\left(\frac{\boldsymbol{Z}(\omega)}{Z_\mathrm{TL}}-1\right)^{-1}\cdot\left(\frac{\boldsymbol{Z}(\omega)}{Z_\mathrm{TL}}+1\right)\cdot\boldsymbol{a_\omega}.
\end{equation}
Similarly we observe that~\cref{eq: order 1} reduces to
\begin{align}
    \boldsymbol{b_\omega}^{(1)}=\int_{-\infty}^\infty d\omega' \  \frac{\omega'}{\omega}  Z_\mathrm{TL}\hat{d G}_-^{(0)}(\omega-\omega')
    \left(\frac{\boldsymbol{Z}(\omega')}{Z_\mathrm{TL}}-1\right)^{-1}\cdot  \frac{\boldsymbol{Z_0}(\omega')}{Z_\mathrm{TL}} \cdot i2\boldsymbol{\sigma_y}\cdot  \left(\frac{\boldsymbol{Z}(\omega')}{Z_\mathrm{TL}}-1\right)^{-1}\cdot  \frac{\boldsymbol{Z_0}(\omega')}{Z_\mathrm{TL}} \cdot \boldsymbol{a_{\omega'}}, \label{eq:b omega 1}
\end{align}
using the identities 
\begin{align}
    1+\boldsymbol{S}^{(0)}(\omega) =2  \left(\frac{\boldsymbol{Z}(\omega)}{Z_\mathrm{TL}}-1\right)^{-1}\cdot \frac{\boldsymbol{Z}(\omega)}{Z_\mathrm{TL}} \quad \textrm{and} \quad   
    1-\boldsymbol{S}^{(0)}(\omega) =-2  \left(\frac{\boldsymbol{Z}(\omega)}{Z_\mathrm{TL}}-1\right)^{-1}.
\end{align}
Next we must solve for~$dG_-^{(0)}(t)$. \cref{eq:boundary Fourier eq} yields, for~$\omega>0$,
\begin{equation}
    \boldsymbol{\hat \Phi}^{(0)}(\omega) = \sqrt{\frac{\hbar}{4\pi c \omega}} \left(  \boldsymbol{a_\omega}+\boldsymbol{b_\omega}^{(0)}  +\frac{i\omega \boldsymbol{L_c}}{Z_\mathrm{TL}}\cdot \left(\boldsymbol{a_\omega}-\boldsymbol{b_\omega}^{(0)}\right) \right) = \sqrt{\frac{\hbar}{\pi c \omega}}   \left(\frac{\boldsymbol{Z}(\omega)}{Z_\mathrm{TL}}-1\right)^{-1}\frac{\boldsymbol{Z_0}(\omega)}{Z_\mathrm{TL}}\cdot \boldsymbol{a_\omega}, \label{eq:boundary Fourier ab eq}
\end{equation}

\subsubsection{Monotonic incoming field}
\label{app:Gyrator Monotonic incoming field}

In what follows we assume that the incoming field is monotonic with frequency~$\omega_0$. As a result we find that 
\begin{equation}
\begin{split}
    \langle \boldsymbol{\varphi}(t) \rangle^2 &= \chi  \left(\frac{\boldsymbol{Z}^\dagger(\omega_0)}{Z_\mathrm{TL}}-1\right)^{-1}\frac{\boldsymbol{Z_0}^\dagger(\omega_0)}{Z_\mathrm{TL}}\cdot |\langle\boldsymbol{a_{\omega_0}}\rangle|^2\cdot \frac{\boldsymbol{Z_0}(\omega_0)}{Z_\mathrm{TL}}\cdot \left(\frac{\boldsymbol{Z}(\omega_0)}{Z_\mathrm{TL}}-1\right)^{-1}\\
    &+\frac{\chi}{2} \left(\frac{\boldsymbol{Z}^T(\omega_0)}{Z_\mathrm{TL}}-1\right)^{-1}\frac{\boldsymbol{Z_0}^T(\omega_0)}{Z_\mathrm{TL}}\cdot \langle\boldsymbol{a_{\omega_0}} \rangle^2\cdot \frac{\boldsymbol{Z_0}(\omega_0)}{Z_\mathrm{TL}}\cdot \left(\frac{\boldsymbol{Z}(\omega_0)}{Z_\mathrm{TL}}-1\right)^{-1} e^{-i2\omega_0 t}+ \mathrm{h.c.},
\end{split}
\end{equation}
where we defined the quantity
\begin{equation}
    \chi = \left(\frac{2\pi}{\Phi_0}\right)^2\frac{2\hbar}{\pi c \omega} = \frac{4}{R_Q c \omega}
\end{equation}
Now we observe that 
\begin{align}
    \hat{dG}_-^{(0)}(\omega) = \hat{dG}_-^{(0)}(0) \delta(\omega) +  \hat{dG}_-^{(0)}(2\omega_0) \delta(\omega-\omega_0) +  \hat{dG}_-^{(0)}(-2\omega_0) \delta(\omega+\omega_0), 
\end{align}
in other words,~$\hat{dG}_-^{(0)}(\omega)$ is sharply peaked at three frequencies. The~$\omega=0$ component leads to compression, and under conservation of total exctations,~$\omega=\pm 2\omega_0$ lead to frequency mixing. Indeed, the outgoing fields are no longer monotonic as they oscillate at both~$\omega_0$ and~$3\omega_0$, the latter having much smaller amplitude. Importantly we see that the scattering matrix is rectangular:
\begin{align}
    \begin{pmatrix}
    \boldsymbol{b_{-3\omega_0}} \\
    \boldsymbol{b_{-\omega_0}} \\
    \boldsymbol{b_{+\omega_0}} \\
    \boldsymbol{b_{+3\omega_0}}
    \end{pmatrix}
     = 
     \begin{pmatrix}
     \boldsymbol{M}(-3\omega_0;-\omega_0) & \boldsymbol{0} \\
     \boldsymbol{S}^{(0)}(-\omega_0) & \boldsymbol{M}(-\omega_0;+\omega_0) \\
     \boldsymbol{M}(+\omega_0;-\omega_0) & \boldsymbol{S}^{(0)}(+\omega_0) \\
     \boldsymbol{0} & \boldsymbol{M}(+3\omega_0;+\omega_0) 
     \end{pmatrix}
     \cdot 
    \begin{pmatrix}
    \boldsymbol{a_{-\omega_0}} \\
    \boldsymbol{a_{+\omega_0}} 
    \end{pmatrix}
\end{align}
where we defined the frequency-mixing matrices 
\begin{align}
    \boldsymbol{M}(\omega;\omega') = \frac{\omega'}{\omega}  Z_\mathrm{TL}\hat{d G}_-^{(0)}(\omega-\omega')
    \left(\frac{\boldsymbol{Z}(\omega')}{Z_\mathrm{TL}}-1\right)^{-1}\cdot  \frac{\boldsymbol{Z_0}(\omega')}{Z_\mathrm{TL}} \cdot i2\boldsymbol{\sigma_y}\cdot  \left(\frac{\boldsymbol{Z}(\omega')}{Z_\mathrm{TL}}-1\right)^{-1}\cdot  \frac{\boldsymbol{Z_0}(\omega')}{Z_\mathrm{TL}},
\end{align}
which are ultimately proportional to~$\chi$.

\subsubsection{Effective linear response theory}
\label{app:Gyrator Effective linear response theory}

The leading order effect of the nonlinearity is compression: frequency mixing can be thought as its direct consequence. Compression is associated with the static component of~$dG_-(t)$ only. We therefore propose an effective linear response theory that captures compression:
\begin{equation}
    \boldsymbol{b_\omega} \simeq \boldsymbol{S}(\omega) \cdot \boldsymbol{a_\omega} =\left(\frac{\boldsymbol{Z}(\omega)}{Z_\mathrm{TL}}-1\right)^{-1}\cdot\left(\frac{\boldsymbol{Z}(\omega)}{Z_\mathrm{TL}}+1\right)\cdot\boldsymbol{a_\omega}, \label{eq: final scattering matrix}
\end{equation}
where 
\begin{align}
    &\boldsymbol{Z}(\omega) = i \omega \boldsymbol{L_c} + \left(i\omega \boldsymbol{C}  + (i\omega\boldsymbol{L})^{-1}  + i  G\boldsymbol{\sigma_y}\right)^{-1}, \\
    &G = \overline{G}_-+\lambda \hat{dG}_-^0(0) = \frac{2\pi }{\Phi_0}\frac{\partial E_J(\Delta_1,\boldsymbol{T_1},V_1)}{\partial V}\sin\left(\varphi_1^\mathrm{ex}\right)\left(1-\frac{\pi Z_0 N_1 }{2R_Q}\right)- \frac{2\pi }{\Phi_0}\frac{\partial E_J(\Delta_2,\boldsymbol{T_2},V_2)}{\partial V}\sin\left(\varphi_2^\mathrm{ex}\right)\left(1-\frac{\pi Z_0 N_2 }{2R_Q}\right), \label{eq:G_mf}
\end{align}
with~$N_{1(2)}$ the average photon number in the internal gyrator mode~$1(2)$ with characteristic impedance~$Z_0$. It can be verified that Taylor expanding~\cref{eq: final scattering matrix} to leading order in~$\lambda \hat{dG}_-^0(0)$ returns the linear part of~\cref{eq:b omega 1}. 

\paragraph{Ideal case.} We consider the limiting case~$\boldsymbol{L_c}=L_c \boldsymbol{1}$,~$\boldsymbol{C}=C_0 \boldsymbol{1}$ and~$\boldsymbol{L}=L_0 \boldsymbol{1}$. In this case we find that 
\begin{equation}
    \frac{\boldsymbol{Z}(\omega)}{Z_\mathrm{TL}} = \frac{Z_c(\omega)}{Z_\mathrm{TL}}\boldsymbol{1} + \frac{\left(Z_0^{-1}(\omega) \boldsymbol{1} + i G \boldsymbol{\sigma_y}\right)^{-1}}{Z_\mathrm{TL}} = \frac{Z_c(\omega)}{Z_\mathrm{TL}}\boldsymbol{1} + \frac{Z_\mathrm{TL}/Z_0(\omega) }{(Z_\mathrm{TL}/Z_0(\omega))^2+G^2Z_\mathrm{TL}^2}\boldsymbol{1} - \frac{i G Z_\mathrm{TL} }{(Z_\mathrm{TL}/Z_0(\omega))^2+G^2Z_\mathrm{TL}^2}\boldsymbol{\sigma_y},
\end{equation}
where~$Z_0(\omega) = \left(i\omega C_0 + (i\omega L_0)^{-1}\right)^{-1}$ is the load impedance and~$Z_c(\omega) = i \omega L_c$ is the impedance associated with the coupling inductance. Our goal will be to redefine~$Z_0(\omega)$ and~$Z_\mathrm{TL}$ such as to include~$L_c$. To this end consider 
\begin{equation}
    \frac{\boldsymbol{Z}(\omega)}{Z_\mathrm{TL}} =  \frac{\left(\overline{Z}_0^{-1}(\omega) \boldsymbol{1} + i G \boldsymbol{\sigma_y}\right)^{-1}}{\overline{Z}_\mathrm{TL}(\omega)} =  \frac{\overline{Z}_\mathrm{TL}(\omega)/\overline{Z}_0(\omega) }{(\overline{Z}_\mathrm{TL}(\omega)/\overline{Z}_0(\omega))^2+G^2\overline{Z}_\mathrm{TL}^2(\omega)}\boldsymbol{1} - \frac{i G \overline{Z}_\mathrm{TL}(\omega) }{(\overline{Z}_\mathrm{TL}(\omega)/\overline{Z}_0(\omega))^2+G^2\overline{Z}_\mathrm{TL}^2(\omega)}\boldsymbol{\sigma_y}.
\end{equation}
By putting the last two equations equal we find the effective load impedance due to the coupling inductance
\begin{equation}
    \overline{Z}_0(\omega) = \frac{Z_0(\omega)}{1 + (Z_c(\omega)/Z_0(\omega)) (1+G^2Z_0^2(\omega))},
\end{equation}
and the effective frequency-dependent characteristic impedance of the lines
\begin{equation}\label{eq: overline ZTL(omega)}
    \overline{Z}_\mathrm{TL}(\omega)  = \frac{Z_\mathrm{TL}}{\left(1 +  Z_c(\omega)/Z_0(\omega)\right)^2 +G^2 Z_c^2(\omega)}.
\end{equation}
With these definitions we find that 
\begin{equation}
    \left(\frac{\boldsymbol{Z}(\omega)}{Z_\mathrm{TL}}-1\right)^{-1}\cdot  \left(\frac{\boldsymbol{Z}(\omega)}{Z_\mathrm{TL}}+1\right) = \frac{((a-1) \boldsymbol{1} - i b \boldsymbol{\sigma_y})\cdot ((a+1) \boldsymbol{1} + i b \boldsymbol{\sigma_y})}{(a-1)^2+b^2} = \frac{(a^2+b^2-1)\boldsymbol{1}-i 2 b \boldsymbol{\sigma_y}}{(a-1)^2+b^2},
\end{equation}
where~$\boldsymbol{Z}(\omega) = a \boldsymbol{1}+i b \boldsymbol{\sigma_y}$ with 
\begin{align}
    & a = \frac{\overline{Z}_\mathrm{TL}(\omega)/\overline{Z}_0(\omega) }{(\overline{Z}_\mathrm{TL}(\omega)/\overline{Z}_0(\omega))^2+G^2\overline{Z}_\mathrm{TL}^2(\omega)}, \\ 
    & b = - \frac{G \overline{Z}_\mathrm{TL}(\omega) }{(\overline{Z}_\mathrm{TL}(\omega)/\overline{Z}_0(\omega))^2+G^2\overline{Z}_\mathrm{TL}^2(\omega)}.
\end{align}
We therefore find that the scattering matrix reduces to 
\begin{equation}
    \boldsymbol{S}(\omega) = \cos(2\theta_\omega)\boldsymbol{1} +i \sin(2\theta_\omega)\boldsymbol{\sigma_y}
\end{equation}
where we defined the angle~$\theta_\omega$ via
\begin{equation}
    \tan(2\theta_\omega) = \frac{2G\overline{Z}_\mathrm{TL}(\omega)}{1-\overline{Z}_\mathrm{TL}^2(\omega)/\overline{Z}_0^2(\omega)-G^2\overline{Z}_\mathrm{TL}^2(\omega)}. \label{eq:tan 2 theta}
\end{equation}

\paragraph{Central frequency.} 

The central frequency~$\omega_0'$ of the device corresponds to the frequency for which the denominator in~\cref{eq:tan 2 theta} vanishes, i.e. 
\begin{equation}
    G^2 = \overline{Z}_\mathrm{TL}^{-2}(\omega_0') - \overline{Z}_0^{-2}(\omega_0'). \label{eq:G^2 condition}
\end{equation}
Moreover we wish for the optimal conductance~$G$ to be minimized, i.e. $\overline{Z}_\mathrm{TL}^2(\omega_0') / \overline{Z}_0^2(\omega_0') \to 0$ and~$G = 1/\overline{Z}_\mathrm{TL}(\omega_0')$. Naturally we wish for both~$\overline{Z}_\mathrm{TL}(\omega_0')$ and~$\overline{Z}_0(\omega_0')$ to be large. $\omega_0'$ must therefore be the frequency at which~$\overline{Z}_\mathrm{TL}(\omega_0')$ peaks. Using~$G = 1/\overline{Z}_\mathrm{TL}(\omega_0')$ we find that~\cref{eq: overline ZTL(omega)} becomes
\begin{equation}
    0=Z_c^2(\omega_0') \overline{Z}_\mathrm{TL}^{-2}(\omega_0')-Z_\mathrm{TL}\overline{Z}_\mathrm{TL}^{-1}(\omega_0') + \left(1+Z_c(\omega_0')/Z_0(\omega_0')\right)^2,
\end{equation}
which reduces to 
\begin{equation} \label{eq:overline Z inverse}
    \overline{Z}_\mathrm{TL}^{-1}(\omega_0') = \frac{Z_\mathrm{TL}}{2Z_c^2(\omega_0')}\left(1 - \sqrt{1 - \frac{4Z_c^2(\omega_0')}{Z_\mathrm{TL}^2}\left(1+\frac{Z_c(\omega_0')}{Z_0(\omega_0')}\right)^2}\right),
\end{equation}
where~$Z_c^2(\omega_0')<0$. 

We would like to minimize~$|\overline{Z}_\mathrm{TL}^{-1}(\omega_0')|$ such as to maximize~$|\overline{Z}_\mathrm{TL}(\omega_0')|$. We find a root at~$\omega_0'/\omega_0 = \sqrt{1+Z_0/L_c\omega_0}$. However this is not sufficient to define the central frequency. Indeed, we notice that for~$L_c\to 0$,~$|\overline{Z}_\mathrm{TL}^{-1}(\omega_0')|= Z_\mathrm{TL}^{-1}$ is flat meaning that it cannot be minimized. This leads us to also consider the second condition that~$\overline{Z}_0^{-1}(\omega_0')=0$ to satisfy~\cref{eq:G^2 condition}:
\begin{equation} \label{eq:central_freq_condition}
    0=Z_0^{-1}(\omega_0') + Z_c(\omega_0') Z_0^{-2}(\omega_0') + Z_c(\omega_0') \overline{Z}_\mathrm{TL}^{-2}(\omega_0').
\end{equation}
We quickly observe that for~$L_c =0$, thus~$Z_c(\omega_0')=0$, we exactly find~$\omega_0' = \omega_0$. For large~$L_c$, such that~$\overline{Z}_\mathrm{TL}^{-1}(\omega_0')\approx -i Z_0^{-1}(\omega_0')$, we still find~$\omega_0' \approx \omega_0$.

In what follows we therefore approximate~$\omega_0' \approx \omega_0$ to compute optimal parameters for gyration such as the conductance~$G  \approx 1/\overline{Z}_\mathrm{TL}(\omega_0)$. Because of this approximation we emphasize that in the end the central frequency will be found by numerically solving~\cref{eq:G^2 condition}. 

However for an arbitrary $G$, solving $\overline{Z}_\mathrm{TL}^{-2}(\omega)-\overline{Z}_0^{-2}(\omega) - G^2 =0$ reveals that  
\begin{align}
    \frac{\omega_c}{\omega_0}-1 \approx \frac{\overline{Z}_\mathrm{TL}^{-2}(\omega_0)-G^2 + G^4L_c^2\omega_0^2}{2G^2(2L_c\omega_0/Z_0-G^2 L_c^2\omega_0^2)+4Z_\mathrm{TL}^{-2}(1-G^2L_c^2\omega_0^2) (2L_c\omega_0/Z_0+G^2L_c^2\omega_0^2)}. 
\end{align}

\begin{figure}[h!]
    \centering
    \includegraphics[width=.5\textwidth]{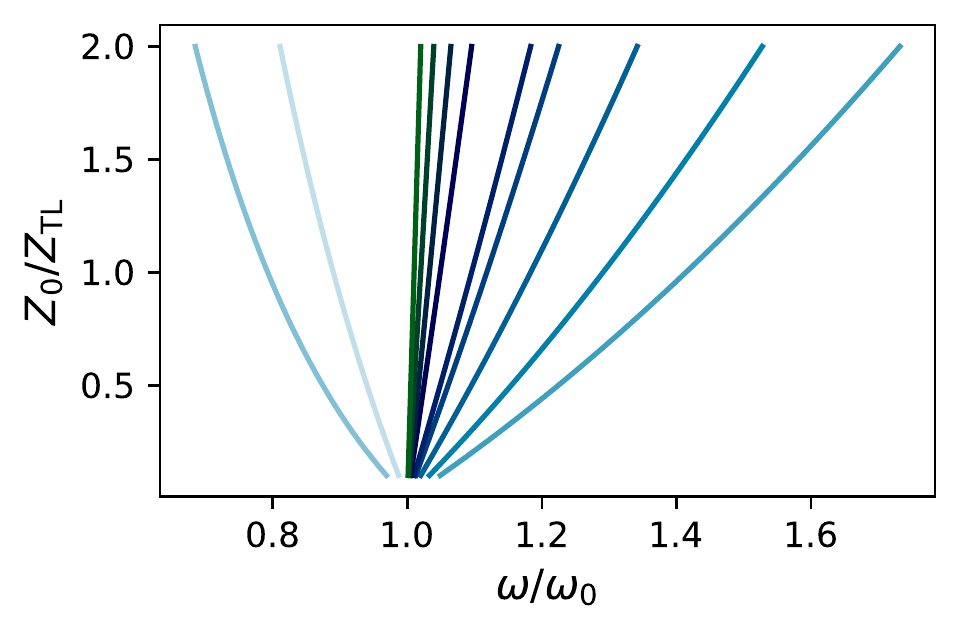}
    \caption{Central frequency obtained by numerically solving for the root of~\cref{eq:central_freq_condition} along with~\cref{eq:overline Z inverse} from~$L_c\omega_0/Z_\mathrm{TL}=0.05$ (dark blue) to~$L_c\omega_0/Z_\mathrm{TL}=50.00$ (dark green).}
    \label{fig:central freq deviation}
\end{figure}

\paragraph{Optimal conductance.} 

$|\tan(2\theta_\omega)| \to \infty$ corresponds to perfect gyration where~$\boldsymbol{S}$ resembles~\cref{eq:S gyr ideal}. This occurs at central frequency, where the denominator of~\cref{eq:tan 2 theta} vanishes, which we found to be 
\begin{equation}
    \omega_0' \approx \omega_0 = 1/\sqrt{L_0 C_0}.
\end{equation}
This is approximately where~$G$ can take on a minimal value (assuming~$\overline{Z}_\mathrm{TL}^2(\omega_0)/\overline{Z}_0^2(\omega_0)\approx 0$)
\begin{equation}
    G_0=  Z_\mathrm{TL}^{-1} \left(\sqrt{1+2 x^2}-1\right)/x^2, \quad x=\sqrt{2} \omega_0' L_c/Z_\mathrm{TL}, \label{eq:G_0_app}
\end{equation}
such~$G_0= \overline{Z}_\mathrm{TL}(\omega_0)^{-1}$ and therefore~$|\tan(2\theta_\omega)| \to \infty$. The approximation we make in this work is~$\omega_0'\approx \omega_0$ in~\cref{eq:G_0_app}. 

\paragraph{Frequency bandwidth.} 

We define the frequency bandwidth~$\Delta = \omega_+-\omega_-$ for gyration with the cut-off frequencies~$\omega_\pm$ for which reflection equals transmission, i.e. when~$|\tan(2\theta_\omega)| = 1$. We consider two limiting cases:~$L_c=0$ and~$L_c\gg 0$.

\begin{itemize}
\item[]~$L_c=0$. For~$\theta_\omega = \pm \pi/8$ solve the equation  
\begin{equation}
    1 + |Z_\mathrm{TL}/Z_0(\omega)|^2 - G^2 Z_\mathrm{TL}^2 \mp 2 G Z_\mathrm{TL} = 0
\end{equation}
according to~\cref{eq:tan 2 theta}. Here we observe that this equation can only be valid if~$\mp G <0$ at perfect impedance matching where~$G Z_\mathrm{TL}=1$. We therefore simplify the equation to 
\begin{equation}
    1 + |Z_\mathrm{TL}/Z_0(\omega)|^2 - G^2 Z_\mathrm{TL}^2 - 2|G| Z_\mathrm{TL} = 0 \ \rightarrow \ Z_\mathrm{TL}/Z_0(\omega) = \pm i\sqrt{G^2Z_\mathrm{TL}^2 + 2|G|Z_\mathrm{TL} -1} = \pm i 2 \epsilon .
\end{equation}
We finally arrive at the explicit constraint 
\begin{equation}
    (\omega/\omega_0)^2 \pm (2\epsilon Z_0/Z_\mathrm{TL}) (\omega/\omega_0) - 1 = 0,
\end{equation}
where~$Z_0=\sqrt{L_0/C_0}$ is the impedance of the load. We find the solutions 
\begin{equation}
    \frac{\omega_{\pm}}{\omega_0} = \frac{\epsilon Z_0}{Z_\mathrm{TL}}\pm \sqrt{1+\left(\frac{\epsilon Z_0}{Z_\mathrm{TL}}\right)^2}.
\end{equation}
We finally find the frequency bandwidth 
\begin{equation}
    \frac{\omega_+-\omega_-}{\omega_0} = 2\sqrt{1+\left(\frac{\epsilon Z_0}{Z_\mathrm{TL}}\right)^2}.
\end{equation}
\item[]~$L_c\gg 0$. As a simplification we focus on perfect impedance matching, i.e. we choose~$G$ such that~$\abs{G \overline{Z}_\mathrm{TL}(\omega_0)} = 1$, which corresponds to~$G=1/L_c\omega_0$ according to~\cref{eq:G_0_app} in the large~$L_c$ limit. As seen in~\cref{fig:freq bandwidth app}a)-b) we find that~$\overline{Z}_0(\omega)\approx \left(G^2Z_c(\omega_0)Z_\mathrm{TL}\right)^{-1}$ and~$\overline{Z}_\mathrm{TL}(\omega)\approx Z_0(\omega)Z_\mathrm{TL}/2Z_c(\omega_0)$ in the large~$Z_0(\omega)$ limit and for~$G^2Z_c^2(\omega_0)\approx -1$. We also observe that~$\overline{Z}_\mathrm{TL}^2(\omega)/\overline{Z}_0^2(\omega)\approx -1$. This fact allows us to approximate
\begin{equation}
    \tan(2\theta_\omega) \approx  2G \overline{Z}_\mathrm{TL}(\omega), \label{eq: effective tan 2 theta}
\end{equation}
which shows near perfect agreement in~\cref{fig:freq bandwidth app}c) near~$\omega=\omega_0$. We also emphasize that the~$|\tan(2\theta_\omega)| = 1$ condition occurs in a range smaller than the frequency range plotted here. We are ultimately interested in finding frequencies for which~$|\tan(2\theta_\omega)|=1$ closest to~$\omega_0$. \cref{eq: effective tan 2 theta} indicates that this occurs for 
\begin{equation}
    \abs{2G \overline{Z}_\mathrm{TL}(\omega)} = 1.
\end{equation}
Finally we find the approximate constraint 
\begin{equation}
    \frac{\omega_0}{\omega} - \frac{\omega}{\omega_0}=\pm\frac{Z_0Z_\mathrm{TL}}{L_c^2\omega_0^2},
\end{equation}
where we used~$Z_0(\omega) = iZ_0 \left(\omega_0/\omega-\omega/\omega_0\right)^{-1} \approx (i Z_0/2)(1-\omega/\omega_0)^{-1}$ to leading order in~$\omega-\omega_0$. This leads to the cut-off frequencies
\begin{equation}
    \frac{\omega_\pm}{\omega_0} =  \sqrt{\left(\frac{Z_0Z_\mathrm{TL}}{2L_c^2\omega_0^2}\right)^2+1}\pm\frac{Z_0Z_\mathrm{TL}}{2L_c^2\omega_0^2} \approx 1 \pm \frac{Z_0Z_\mathrm{TL}}{2L_c^2\omega_0^2}.
\end{equation}
Finally we find the frequency bandwidth 
\begin{equation}
    \frac{\omega_+-\omega_-}{\omega_0} \approx \frac{Z_0Z_\mathrm{TL}}{L_c^2\omega_0^2}.\label{eq:freq_bandwidth_estimate}
\end{equation}

\begin{figure}[h!]
    \centering
    \includegraphics[width=\textwidth]{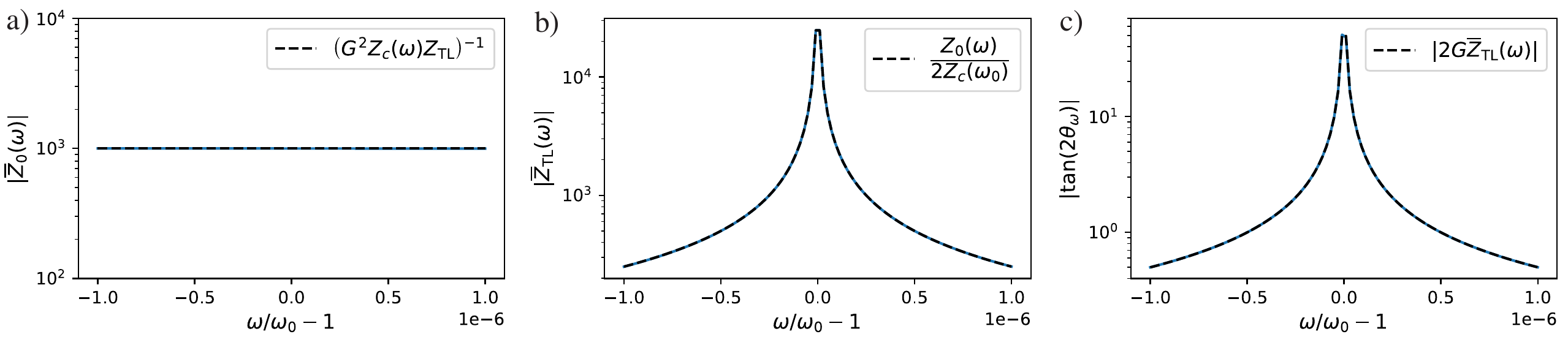}
    \caption{Analytical estimates in presence of a coupling inductance.}
    \label{fig:freq bandwidth app}
\end{figure}

\end{itemize}

\paragraph{Compression level.} 

At central frequency we find that~$|\tan (2\theta_{\omega_0})| \approx 2 (1-x)/(1 - (1-x)^2)$ where~$x=\pi Z_0 N/2R_Q$ and~$N = \zeta_1 N_1 + \zeta_2 N_2$ given that~$|G_0\overline{Z}_\mathrm{TL}(\omega_0)|=1$ for perfect impedance matching at~$N_1=N_2=0$ and~$G = G_0(1-x)$, and where~$\zeta_{1(2)}$ are determined from~\cref{eq:G_mf}. For identical junctions and flux biases satisfying~$\sin(\varphi_1^\mathrm{ex})=-\sin(\varphi_2^\mathrm{ex})$ we find that~$\zeta_1=\zeta_2=1/2$. When transmission drops by 1~dB such that~$|\sin(2\theta_{\omega_0})| = 10^{-0.1}$ and therefore reflection is~$|\cos(2\theta_{\omega_0})| = \sqrt{1-10^{-0.2}}$, we find that~$\abs{\tan(2\theta_\omega)}=10^{-0.1}/\sqrt{1-10^{-0.2}}$. We therefore find the constraint 
\begin{equation}
     \frac{10^{-0.1}}{\sqrt{1-10^{-0.2}}}= \frac{2 (1-x)}{1 - (1-x)^2}, \quad x=\pi Z_0 N/2R_Q. 
\end{equation}
We obtain the maximum average photon number
\begin{equation}
\label{eqn:maxphotnumber}
    N_\mathrm{max} = \frac{R_Q}{\pi Z_0}.
\end{equation}

\subsubsection{Numerics}
\label{app:scattering_matrix_numerics}

\paragraph{Dimensionless parameters.}

It is useful to define the following dimensionless quantities: 
\begin{itemize}
    \item Renormalized frequency: \[\omega' = \omega/\omega_0\] where~$\omega_0 = 1/\sqrt{L_0C_0}$ is the central frequency.
    \item Renormalized coupling inductance: \[L_c' = L_c \omega_0/Z_\mathrm{TL}\] where~$Z_\mathrm{TL}$ is the characteristic impedance of the transmission lines. 
    \item Renormalized characteristic impedance of the load: \[Z_0' = Z_0/Z_\mathrm{TL}\] where~$Z_0 = \sqrt{L_0/C_0}$ is the characteristic impedance of the load. 
    \item Renormalized conductance:~$G' = G Z_\mathrm{TL}$ where~$G$ is the conductance of the gyrator.
\end{itemize}
The system can therefore be entirely characterized by three parameters~$L_C'$,~$Z_0'$ and~$G'$ to be optimized. Here~$\omega_0$ and~$Z_\mathrm{TL}$ are parameters to be defined. 

\paragraph{Frequency bandwidth}

\begin{figure}[h!]
    \centering
    \includegraphics[width=\textwidth]{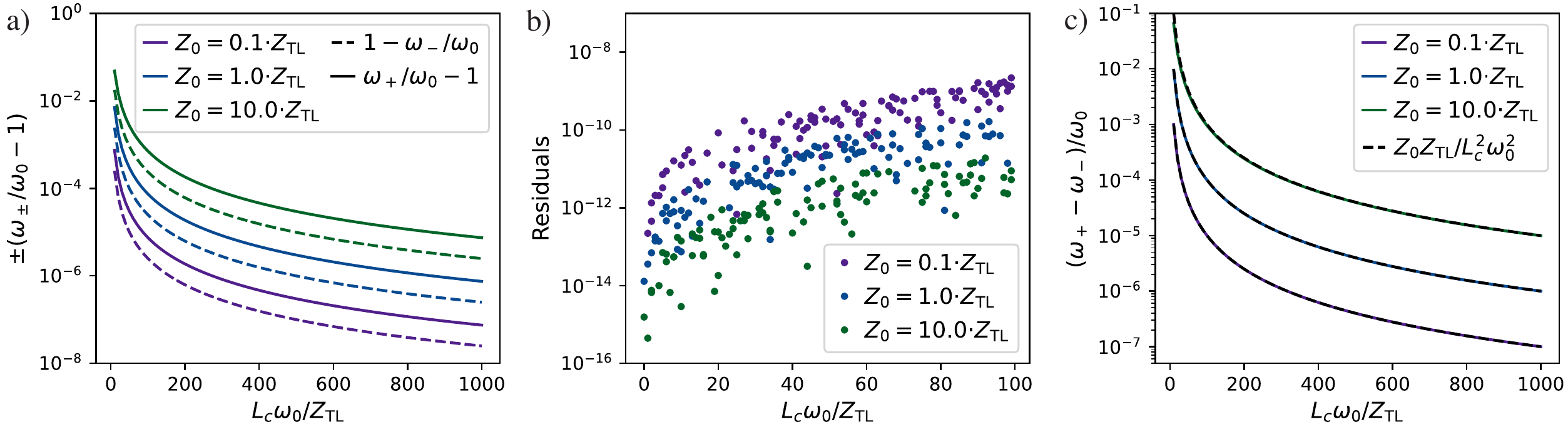}
    \caption{Numerical computation of the frequency bandwidth. a) Cut-off frequencies~$\omega_\pm$ for which~$|\tan(2\theta_\omega)|=\pi/8$ for different~$L_c$'s, closest to~$\omega_0$. b) Residuals of the \emph{least\_squares} algorithm in \texttt{scipy}. c) Frequency bandwidth obtained from a) and compared against~\cref{eq:freq_bandwidth_estimate}.}
    \label{fig:residuals_app}
\end{figure}

We numerically solve for when~\cref{eq:tan 2 theta} is~$\pm \pi/8$ using the \emph{least\_squares} algorithm in \emph{Scipy}. The distance between the two solutions closest to~$\omega_0$ is used to defined the bandwidth. The solutions are shown in~\cref{fig:residuals_app}a) for different~$L_c$'s with corresponding residuals plotted in b). The frequency bandwidth is then shown in c) and compared against the analytical estimate in~\cref{eq:freq_bandwidth_estimate} (black lines). We observe near perfect quantitative agreement in the large~$L_c$ limit. 

\section{System Hamiltonian and Effective Lindblad Master equation}

\label{app:Lindblad Master equation}

\subsection{Canonical quantization}

We consider the gyrator Lagrangian in~\cref{eq: app full gyrator Lagrangian} and add transmission lines interacting with internal gyrator modes via a coupling inductance as in~\cref{eq: app mean field full gyrator Lagrangian}. We stop at the first derivative of the junctions' transmission coefficients.

The canonical charge fields are
\begin{align}
    \tilde{q}_1(x,t) = \frac{\partial \mathcal{L}}{\partial (\partial_t \tilde{\Phi}_1(x,t))} = c \partial_ t \Phi_1(x,t),\quad \tilde{q}_2(x,t) = \frac{\partial \mathcal{L}}{\partial (\partial_t \tilde{\Phi}_1(x,t))} = c \partial_ t \Phi_2(x,t),
\end{align}
while the canonical charges are 
\begin{align}
    q_1 = C_0 \dot{\Phi}_1 - \frac{\Delta}{4} \frac{\partial T}{\partial V}\Big|_{V_0} \sin(\varphi_2), \quad q_2 = C_0 \dot{\Phi}_2 + \frac{\Delta}{4} \frac{\partial T}{\partial V}\Big|_{V_0} \sin(\varphi_1).
\end{align}
The full system Hamiltonian is~$\mathcal{H} = \sum_i \int_{-\infty}^0 \tilde{q}_i(x,t) \partial_t \tilde{\Phi}_i(x,t) + q_i\dot{\Phi}_i - \mathcal{L}$, 
\begin{equation}
\begin{split}
    \mathcal H &= \sum_{i=1}^2\int_{-\infty}^0 dx \left[\frac{(\tilde{q}_i(x,t))^2}{2c}+\frac{1}{2\ell}\left(\partial_x \tilde{\Phi}_i(x,t)\right)^2\right]+ \sum_{i=1}^2 \frac{\left(\tilde{\Phi}_i(0,t)-\Phi_i\right)^2}{2L}\\
    & + \frac{1}{2C_0}\left(q_1+\frac{C_0 g}{2e} \sin( \varphi_2)\right)^2 + \frac{\Phi_1^2}{2  L_0}+ \frac{1}{2C_0}\left(q_2 -\frac{C_0 g}{2e}\sin( \varphi_1)\right)^2 + \frac{\Phi_2^2}{2  L_0}.
\end{split}
\end{equation}
The quantized fields in the transmission line~$j$ have the form
\begin{align}
    & \hat \Phi_j(x,t) =  \sqrt{\frac{\hbar}{4\pi c}}\int_0^\infty \frac{d\omega}{\sqrt{\omega}}\left(e^{ik_\omega x}\hat a_{j,\omega}^+ + e^{- ik_\omega x}\hat a_{j,\omega}^-+\mathrm{h.c.}\right), \\
    & \hat q_j(x,t) =  -i\sqrt{\frac{\hbar c}{4\pi }}\int_0^\infty d\omega \sqrt{\omega} \left(e^{ik_\omega x}\hat a_{j,\omega}^+ + e^{- ik_\omega x}\hat a_{j,\omega}^--\mathrm{h.c.}\right)
\end{align}
with the dispersion relation~$k_\omega = \omega\sqrt{c \hspace{0.1em} \ell}$. Here~$\comm{\hat a_{j,\omega}}{\hat a_{j',\omega'}^\dagger}=\delta_{ij}\delta(\omega-\omega')$.
The quantized Hamiltonian then takes the form~$\hat H  \approx \hat{H}_g + \hat{H}_t + \hat{H}_{gt}$, where
\begin{equation}
\begin{split}
    \hat{H}_g  &= \sum_{i=1}^2\sum_{j\neq i} \frac{1}{2C_0}\left(\hat q_i- \frac{(-1)^iC_0 g}{2e} \sin( \hat \varphi_j)\right)^2 + \frac{\hat \Phi_0^2\hat{\varphi}_i^2}{8\pi^2L_0} \\
    &=  \sum_{i=1}^2\sum_{j\neq i} 4 E_C  \left(\hat n_i- \frac{(-1)^i g}{8E_C} \sin \hat \varphi_j\right)^2 + \frac{E_L\hat \varphi_i^2}{2},\\
    \hat{H}_t &= \sum_{i=1}^2 \int_0^\infty d\omega \hbar \omega \hat a_{i,\omega}^\dagger \hat a_{i,\omega},\\
    \hat{H}_{gt} &=  - \frac{\Phi_0}{2\pi L}\sqrt{\frac{\hbar}{4\pi c}} \sum_{i=1}^2\int_0^\infty \frac{d\omega }{\sqrt{\omega }}\left(\hat a_{i,\omega}^+ + \hat a_{i,\omega}^-+\mathrm{h.c.}\right)\hat \varphi_i,
\end{split}
\end{equation}
where~$E_C = e^2/2C_0$ and~$E_L = (\Phi_0/2\pi)^2/L_0$. Here we assume small zero-point fluctuations~$\sqrt[4]{2 E_C/E_L}\ll 1$ and small gyrator strength~$g\ll\sqrt{8 E_C E_L}$. Thus, we can expand the gyrator modes in the Fock basis: 
\begin{equation}
    \hat{n}_i = \frac{1}{2i\eta}(b_i - b_i^\dagger), \  \hat{\varphi}_i = \eta(b_i + b_i^\dagger), \textrm{with} \ \eta = \sqrt[4]{2E_C(E_L + (g^2/16E_C))^{-1}},
\end{equation}
and truncate $\textrm{sin}(\hat{\varphi_i})$ to~$5$-th order in~$\hat{\varphi_i}$. 
    
\subsection{Time evolution} 

Following a similar process to that in~\cite{Muller2018}, we can derive a master equation for the gyrator. Assuming a single-mode, coherent-field input in the transmission lines, we find
\begin{equation}
\label{eqn:MeInputOutput}
     \dot{\hat \rho} = \mathcal{L}(t)\hat \rho = \-i [\hat H'(t),\hat \rho]/\hbar + \sum_j \kappa \left[\hat b_j \hat \rho \hat b_j^\dagger - \acomm{\hat b_j^\dagger \hat b_j}{\hat \rho}/2\right],    
\end{equation}
where we defined the reduced system Hamiltonian
\begin{equation}
    \hat H'(t)/\hbar=\hat H_g/\hbar - \sum_{i=1}^2 \frac{i\sqrt{\kappa}}{2}\left(\beta_i e^{-i\omega_s t}\hat b_i^{\dagger}-\mathrm{h.c.}\right),
\end{equation}
where we included an incoming photon fluxes with coherent amplitude~$\beta_i$ and frequency~$\omega_s$ in each of the lines, with decay rate~$\kappa = \hbar/(2 c L^2 \omega_0)$.

The Master equation is used to compute the average outgoing fields in the steady limit, 
\begin{equation}
    \alpha_j=\lim_{t\to \infty} \textrm{Tr} \left\lbrace e^{i\omega_s t}\hat b_j \hat \rho(t) \right\rbrace,
\end{equation}
where~$\hat b_j$ is the annihilation operator of the~$j$th gyrator mode, and reserving contributions from~$\sin(\hat{\varphi}_i)$ to~$5$-th order. The components of the scattering matrix are then given by~$\mathcal{S}_{ij} = \alpha_i/\beta_j-\delta_{ij}$. 

The scattering matrix is extracted from the time-ordered integral of the evolution operator over a period of the drive~$T=2\pi/\omega_s$,

\begin{equation}
    \mathcal{V}(T) = \mathcal{T}\exp \left(\int_0^T\mathcal{L}(t) dt\right),
\end{equation}
which is calculated using an exponential integrator~\cite{Shillito2021}. The evolution is performed in the diagonalized basis of~$\hat{H}_g$ and truncated to the first~$21$ states, allowing for a total of~$5$ excitations in the composite system.  The steady state is subsequently found by renormalizing the right eigenvector of~$\mathcal{V}$ with eigenvalue norm~$1$. These results were corroborated by numerical integration of~\cref{eqn:MeInputOutput} using the \texttt{mesolve} function of QuTiP~\cite{JOHANSSON2013Qutip}.

\subsection{Comparison with mean-field theory}

\begin{figure}[h!]
    \centering
    \includegraphics[width=.4\textwidth]{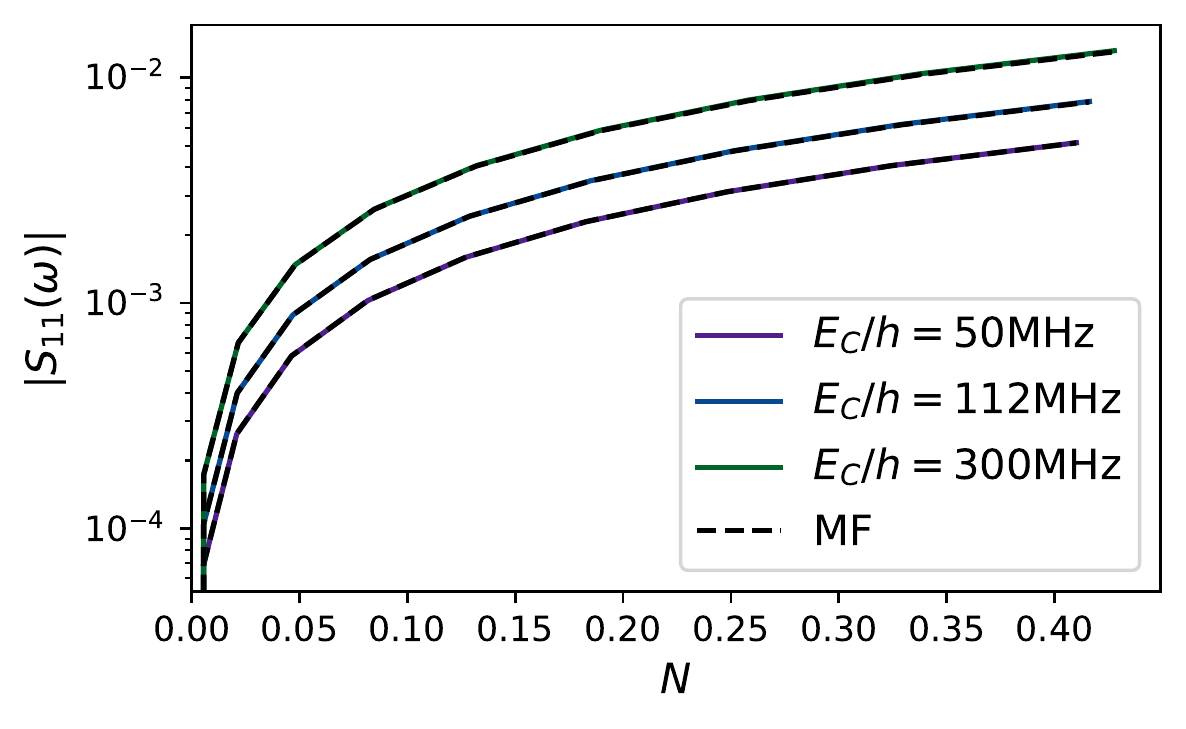}
    \caption{Numerical agreement between mean-field theory (black lines) and full-circuit time-evolution for different charging energies (i.e. different load impedances~$Z_0$) against the average photon number in the gyrator.}
    \label{fig:mean field agreement app}
\end{figure}

Numerical results are shown in~\cref{fig:mean field agreement app}.

\section{Circuit nonidealities}
\label{app:Circuit disorder}

We start from the effective linear response theory proposed in~\cref{app:Gyrator Effective linear response theory}: 
\begin{equation}
    \boldsymbol{b_\omega} \simeq \boldsymbol{S}(\omega) \cdot \boldsymbol{a_\omega} =\left(\frac{\boldsymbol{Z}(\omega)}{Z_\mathrm{TL}}-1\right)^{-1}\cdot\left(\frac{\boldsymbol{Z}(\omega)}{Z_\mathrm{TL}}+1\right)\cdot\boldsymbol{a_\omega},
\end{equation}
where 
\begin{align}
    \boldsymbol{Z}(\omega) = i \omega \boldsymbol{L_c} + \left(i\omega \boldsymbol{C}  + (i\omega\boldsymbol{L})^{-1}  + i  G\boldsymbol{\sigma_y}\right)^{-1}.
\end{align}
We now consider the general case~$\boldsymbol{L_c}=L_c \boldsymbol{1} + d L_c \boldsymbol{\sigma_z}$,~$\boldsymbol{C}=C_0 \boldsymbol{1} + d C_0 \boldsymbol{\sigma_z} - C_{12}\boldsymbol{\sigma_x}$ and~$\boldsymbol{L}=L_0 \boldsymbol{1} + d L_0 \boldsymbol{\sigma_z}-L_{12}\boldsymbol{\sigma_x}$. Here~$C_{12}$ and~$L_{12}$ are the stray capacitive and inductive couplings respectively between the two internal gyrator modes. $d L_c$,~$d C_0$ and~$d L_0$ are due to asymmetries in the circuit. We highlight that~$G$ is the average of the FENNEC interaction strength on both sides of the gyrator and is therefore insensitive to asymmetries -- we only care about impedance matching with the characteristic impedance of the transmission lines. 

We assume that~$d L_c$,~$d C_0$,~$d L_0$,~$C_{12}$ and~$L_{12}$ are much smaller than~$G$ and Taylor expand the perturbed impedance and scattering matrix~$\boldsymbol{Z}'(\omega)$ and~$\boldsymbol{S}'(\omega)$ to leading order in those quantities, i.e. $\boldsymbol{Z}'(\omega) = \boldsymbol{Z}(\omega) + d \boldsymbol{Z}(\omega)$ and~$\boldsymbol{S}'(\omega) = \boldsymbol{S}(\omega) + d \boldsymbol{S}(\omega)$ where 
\begin{equation}
    d \boldsymbol{Z}(\omega) = i\omega d L_c \boldsymbol{\sigma_z} - \frac{i\omega d C_0  - d L_0/(iL_0^2\omega)}{Z_0^{-2}(\omega)+G^2}\boldsymbol{\sigma_z}  +  \frac{i\omega C_{12} - L_{12}/(iL_0^2\omega)}{Z_0^{-2}(\omega)+G^2}\boldsymbol{\sigma_x},
\end{equation}
and therefore
\begin{equation}
    d \boldsymbol{S}(\omega) =  -\left(\boldsymbol{1}- \boldsymbol{S}(\omega)\right)\cdot\frac{d \boldsymbol{Z}(\omega)}{2Z_\mathrm{TL}} \cdot \left(\boldsymbol{1}- \boldsymbol{S}(\omega)\right).  
\end{equation} 
Recall that 
\begin{equation}
    \boldsymbol{S}(\omega) = \cos(2\theta_\omega)\boldsymbol{1} +i \sin(2\theta_\omega)\boldsymbol{\sigma_y},
\end{equation}
and therefore we find that 
\begin{equation}
\begin{split}
    d \boldsymbol{S}(\omega) =  \frac{1-\cos(2\theta_\omega)}{Z_\mathrm{TL}} \left(i\omega d L_c \boldsymbol{\sigma_z} - \frac{i\omega d C_0  - d L_0/(iL_0^2\omega)}{Z_0^{-2}(\omega)+G^2}\boldsymbol{\sigma_z}  +  \frac{i\omega C_{12} - L_{12}/(iL_0^2\omega)}{Z_0^{-2}(\omega)+G^2}\boldsymbol{\sigma_x}\right). 
\end{split}
\end{equation}
At central frequency~$\omega_0=1/\sqrt{L_0C_0}$, where~$Z_0^{-1}(\omega_0) = 0$, and at perfect impedance matching~$\theta_\omega =\pi/4$ we observe that
\begin{equation}
\begin{split}
    d \boldsymbol{S}(\omega) =  \frac{1}{Z_\mathrm{TL}} \left(i\omega d L_c \boldsymbol{\sigma_z} - \frac{i\omega_0 d C_0  - d L_0/(iL_0^2\omega_0)}{G^2}\boldsymbol{\sigma_z}  +  \frac{i\omega_0 C_{12} - L_{12}/(iL_0^2\omega_0)}{G^2}\boldsymbol{\sigma_x}\right). 
\end{split}
\end{equation}
Frequency mismatches, due to disorder in the circuit design, yield a~$\boldsymbol{\sigma_z}$ error in~$\boldsymbol{S}(\omega)$, which mostly affects reflection, whereas stray couplings result in a~$\boldsymbol{\sigma_x}$ error, which instead mostly impacts transmission. We get the constraints 
\begin{align}
    |\omega_0 d L_c/Z_\mathrm{TL}|, \quad |\omega_0d C_0/Z_\mathrm{TL}G^2|, \quad |d L_0/(Z_\mathrm{TL}L_0^2\omega_0 G^2)|, \quad |\omega_0 C_{12}/Z_\mathrm{TL}G^2|, \quad |L_{12}/(Z_\mathrm{TL}L_0^2\omega_0G^2)| \ll 1.
\end{align}
These constraints come without surprise: Gyration is known to be fragile to circuit disorder and parasitic couplings. Another error specific to our circuit is due to deviations in the areas of the loops. This can in principle yield residual~$\sin \hat \varphi_k$ potential terms. This is not captured by the calculation above. The leading order effect of this term is the renormalization of the photon number in the internal gyrator modes, or in another words, more compression and frequency mixing. 

\begin{figure}[h!]
    \centering
    \includegraphics[width=.7\textwidth]{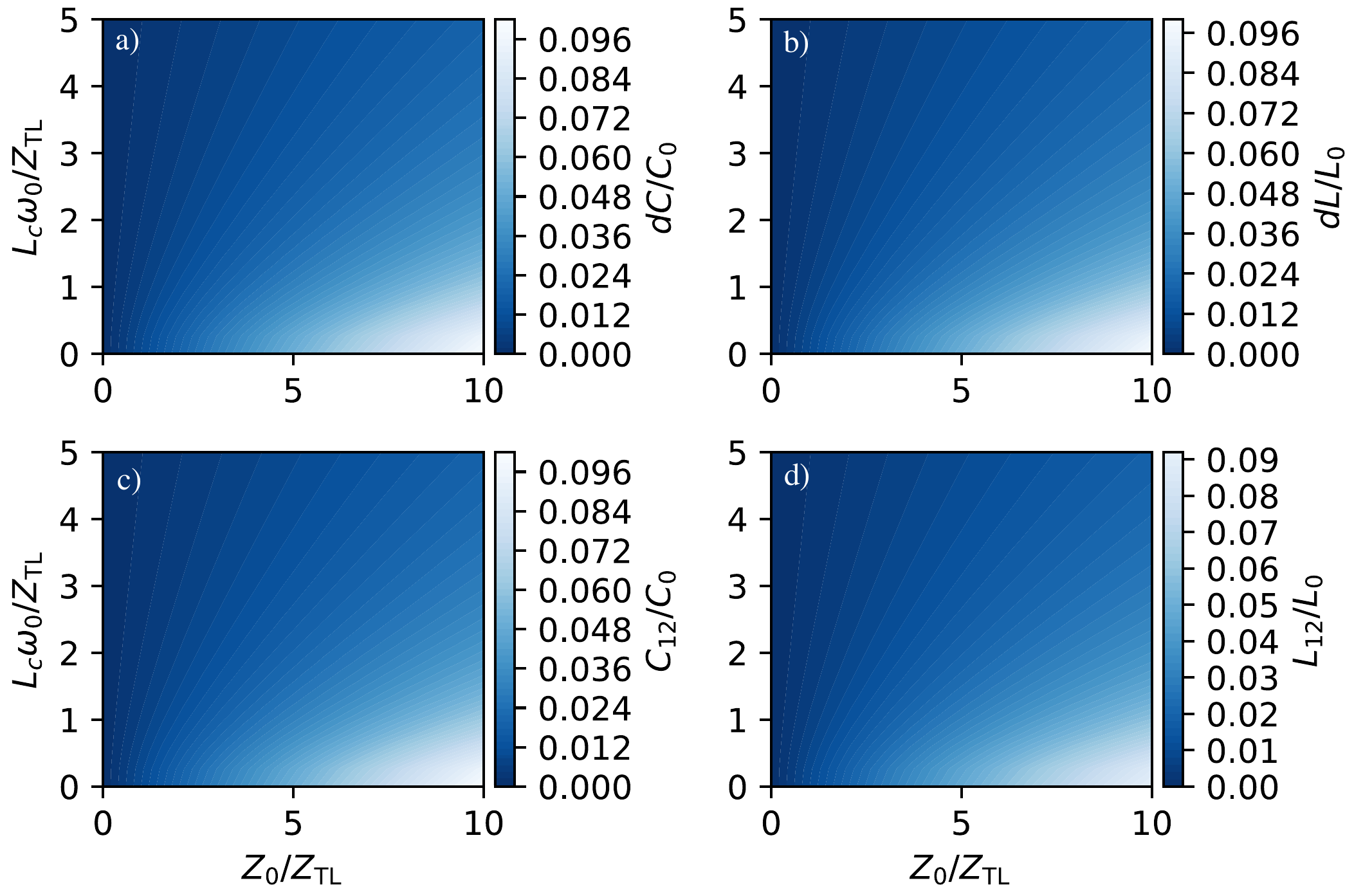}
    \caption{Magnitude of~$dC$,~$dL$,~$dC_{12}$ and~$dL_{12}$ resulting in a 1\% error in the scattering matrix for a), b), c) and d)  respectively.}
    \label{fig:circuit disorder in gyrator app}
\end{figure}

\begin{figure}[h!]
    \centering
    \includegraphics[width=0.35\textwidth]{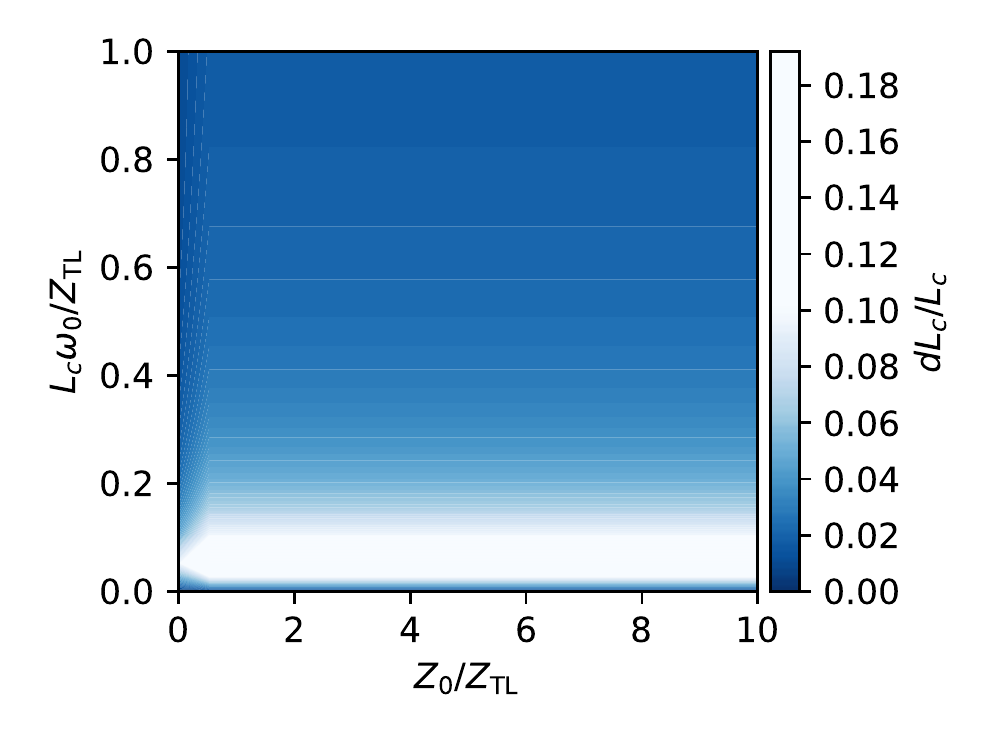}
    \caption{Magnitude of~$dL_c$ resulting in a 1\% error in the scattering matrix.}
    \label{fig:circuit disorder in gyrator coupling app}
\end{figure}

In~\cref{fig:circuit disorder in gyrator app} we plot the magnitude of circuit disorder resulting in a 1\% error in the scattering matrix, found numerically using the \emph{least\_squares} algorithm in \texttt{scipy}. We indeed observe that larger~$Z_0$ (i.e. $C_0\omega_0 = 1/Z_0$ and~$L_0 \omega_0 = Z_0$) and smaller~$L_c$ (i.e. larger~$G$) yield larger tolerances. Similarly in~\cref{fig:circuit disorder in gyrator coupling app} we observe that disorder in~$L_c$ is not limited by~$Z_0$ it is however impacted by~$L_c$ itself-- a smaller coupling inductance allows for stronger disorder.

\section{Generic Lagrangian}
\label{app: generic lagrangian}

We consider a generic circuit with total Lagrangian of the circuit is~$\mathcal L = \mathcal L_0 + \mathcal L_{\mathrm cp}$, where 
\begin{equation}
    \mathcal L_0 =  \dot{\boldsymbol{\Phi}}^T\cdot \frac{\boldsymbol{C_0}}{2}\cdot \dot{\boldsymbol{\Phi}} +\dot{\boldsymbol{\Phi}}^T\cdot \frac{\boldsymbol{C_c}}{2}\cdot \dot{\boldsymbol{\Phi}} + \left(\dot{\boldsymbol{\Phi}}-\boldsymbol{V_J}\right)^T\cdot \frac{\boldsymbol{C_J}}{2}\cdot \left(\dot{\boldsymbol{\Phi}}-\boldsymbol{V_J}\right)  - U\left(\boldsymbol{\varphi}\right),
\end{equation}
is the Lagrangian all the capacitive and inductive contributions commonly found in superconducting circuits and 
\begin{equation} \label{eq:Lint}
\begin{split}
    \mathcal L_{\mathrm int} = -\varepsilon_{J,1}(\dot{\Phi}_2,\varphi_1) - \varepsilon_{J,2}(\dot{\Phi}_1,\varphi_2), \quad  \varepsilon_{J,k}(\dot{\Phi}_\ell,\varphi_k) = -\Delta_k \sqrt{1-T_k(V_{J,k},\dot{\Phi}_\ell)\sin^2\left(\frac{ \varphi_k-\varphi_{\mathrm ex,k}}{2}\right)},
\end{split}
\end{equation}
results from the FENNEC interaction alone. Here~$\boldsymbol{\Phi} = \left(\Phi_1, \ \Phi_2\right)$ is a vector comprising the branch flux~$\Phi_1$ ($\Phi_2$) of the first (second) mode,~$\boldsymbol{\varphi} = 2\pi\boldsymbol{\Phi}/\Phi_0$ are the associated branch phases,~$\boldsymbol{C_0}$ and~$\boldsymbol{C_c}$ are capacitance matrices due to the shunt capacitors and the coupling capacitors respectively,~$\boldsymbol{C_J}$ is the capacitance matrix associated with the coupling to the control voltage lines~$\boldsymbol{V_J}$,~$U(\boldsymbol{\varphi})$ is the potential energy of the two modes defined by the shaded regions in the circuit,~$\Delta_1$ ($\Delta_2$) is the ABS energy of the small junction of mode 1 (mode 2),~$T_1$ ($T_2$) is the transmission probability of the small junction of mode 1 (mode 2),~$\Phi_{\mathrm ex,1}$ ($\Phi_{\mathrm ex,2}$) is an external flux threading the loop enclosing the small junction and the shaded region of mode 1 (mode 2). 

\subsection{Taylor-expanded form}

We will now simplify the interaction Lagrangian. First we do a Taylor expansion in~$\dot{\Phi}_k$,
\begin{equation} \label{eq:Lint}
\begin{split}
    \mathcal L_{\mathrm int} =-\sum_{n=0}^\infty \frac{\dot{\Phi}_2^n}{n!} \frac{\partial^n \varepsilon_{J,1}(\dot{\Phi}_2,\varphi_1)}{\partial \dot{\Phi}_2^n}\Big|_{\dot{\Phi}_2=0} - \sum_{n=0}^\infty \frac{\dot{\Phi}_1^n}{n!} \frac{\partial^n \varepsilon_{J,2}(\dot{\Phi}_1,\varphi_2)}{\partial \dot{\Phi}_1^n}\Big|_{\dot{\Phi}_1=0}
\end{split}
\end{equation}
Second we also do a Taylor expansion in~$T_k$,
\begin{equation} \label{eq:Lint_expanded}
\begin{split}
    \mathcal L_{\mathrm int} = \sum_{n,m=0}^\infty g_{1,n,m} \frac{\dot{\Phi}_2^n}{n!}\sin^{2m}\left(\frac{ \varphi_1-\varphi_{\mathrm ex,1}}{2}\right) + \sum_{n,m=0}^\infty g_{2,n,m}\frac{\dot{\Phi}_1^n}{n!}\sin^{2m}\left(\frac{ \varphi_2-\varphi_{\mathrm ex,2}}{2}\right),
\end{split}
\end{equation}
where we defined the couplings
\begin{align}
    g_{1,n,m} = (-1)^m\begin{pmatrix}
    1/2\\ m
    \end{pmatrix}\Delta_1\left(\frac{\partial^n T_1^m(V_{J,1},\dot{\Phi}_2)}{\partial \dot{\Phi}_2^n}\Big|_{\dot{\Phi}_2=0}\right), \quad g_{2,n,m} = (-1)^m\begin{pmatrix}
    1/2\\ m
    \end{pmatrix}\Delta_2\left(\frac{\partial^n T_2^m(V_{J,2},\dot{\Phi}_1)}{\partial \dot{\Phi}_1^n}\Big|_{\dot{\Phi}_1=0}\right).
\end{align}
In vectorized form the Lagrangian then takes the form 
\begin{equation}
    \mathcal{L} = \dot{\boldsymbol{\Phi}}^T \cdot\frac{\boldsymbol{C}(\boldsymbol{\varphi})}{2}\cdot \dot{ \boldsymbol{\Phi}} + \left[\frac{\boldsymbol{q_0}^T(\boldsymbol{\varphi})}{2} \cdot  \dot{\boldsymbol{\Phi}}+\dot{\boldsymbol{\Phi}}^T\cdot\frac{\boldsymbol{q_0}(\boldsymbol{\varphi})}{2}   \right] + \sum_{n=3}^\infty \left[\boldsymbol{1_{1\times 2}} \cdot \frac{\boldsymbol{g_n}(\boldsymbol{\varphi})}{2} \cdot \frac{\dot{\boldsymbol{\Phi}}^{\circ n}}{n!}+\frac{\dot{\boldsymbol{\Phi}}^{T \circ n}}{n!}\cdot \frac{\boldsymbol{g_n}(\boldsymbol{\varphi})}{2} \cdot  \boldsymbol{1_{2\times 1}} \right]- U\left(\boldsymbol{\varphi}\right),
\end{equation}
where we defined the charge offset
\begin{equation}
    \boldsymbol{q_0}(\boldsymbol{\varphi}) = - \boldsymbol{C_J}\cdot  \boldsymbol{V_J} +  \boldsymbol{g_1}(\boldsymbol{\varphi})\cdot \boldsymbol{1_{2\times 1}},
\end{equation}
the total capacitance matrix
\begin{equation}
    \boldsymbol{C}(\boldsymbol{\varphi})=\boldsymbol{C_0}+\boldsymbol{C_c}+\boldsymbol{C_J} + \boldsymbol{g_2}(\boldsymbol{\varphi}),
\end{equation}
the diagonal matrices
\begin{equation}
    \boldsymbol{g_n}(\boldsymbol{\varphi}) = \sum_{m=0}^\infty\mathrm{diag}\left(
    g_{2,n,m}\sin^{2m}\left(\dfrac{\varphi_2-\varphi_{\mathrm ex,2}}{2}\right), g_{1,n,m} \sin^{2m}\left(\dfrac{ \varphi_1-\varphi_{\mathrm ex,1}}{2}\right)\right),
\end{equation}
and where~$\circ$ is the Hadamard product. We highlight the identity 
\begin{equation}
    \sin^{2m}(x/2) = -\frac{(-1)^m}{4^m} \left(2\sum_{k=0}^{m-1}(-1)^k \begin{pmatrix}
    2m \\ k 
    \end{pmatrix} \cos((m-k)x) - (-1)^m \begin{pmatrix}
    2m \\ m 
    \end{pmatrix}\right)
\end{equation}
obtained with the help of the binomial theorem. 

\subsection{Canonical quantization}

In virtue of Hamilton's principle, the canonical charges,~$\boldsymbol{q}$, associated with ~$\boldsymbol{\Phi}$ are given by 
\begin{equation}
\begin{split}
    \boldsymbol{q} =\frac{\partial \mathcal L}{\partial \dot{ \boldsymbol{\Phi}}} = \boldsymbol{q_0}(\boldsymbol{\varphi}) +  \boldsymbol{C}(\boldsymbol{\varphi})\cdot \dot{ \boldsymbol{\Phi}} + \sum_{n=2}^\infty \boldsymbol{g_{n+1}}(\boldsymbol{\varphi})   \cdot \frac{\dot{\boldsymbol{\Phi}}^{\circ n}}{n!},
\end{split}
\end{equation}
where we observe that~$\boldsymbol{q}$ is nonlinear in~$\dot{\boldsymbol{\Phi}}$. 

Let's consider~$\boldsymbol{g_{n+1}}\to \lambda \boldsymbol{g_{n+1}}$ to be a perturbation for~$n\geq 2$. We will now write~$\dot{\boldsymbol{\Phi}}$ using a perturbative expansion,
\begin{equation}
    \dot{\boldsymbol{\Phi}} =  \boldsymbol{C}^{-1}(\boldsymbol{\varphi})\cdot\left(\boldsymbol{q} -\boldsymbol{q_0}(\boldsymbol{\varphi})\right)  + \sum_{k=1}^\infty \lambda^k \boldsymbol{X_k}(\boldsymbol{q}) =  \sum_{k=0}^\infty \lambda^k \boldsymbol{X_k}(\boldsymbol{q}), \quad \boldsymbol{X_0}(\boldsymbol{q}) = \boldsymbol{C}^{-1}(\boldsymbol{\varphi})\cdot\left(\boldsymbol{q} -\boldsymbol{q_0}(\boldsymbol{\varphi})\right),
\end{equation}
where~$\lambda$ is used to define the order of the expansion. Plugging this expansion in the definition of the canonical charges yields
\begin{equation}
\begin{split}
    0 = \sum_{k=1}^\infty \lambda^{k-1} \boldsymbol{C}(\boldsymbol{\varphi})\cdot \boldsymbol{X_k} + \sum_{n=2}^\infty \boldsymbol{g_{n+1}}(\boldsymbol{\varphi})  \cdot \frac{1}{n!}\left(\sum_{k=0}^\infty \lambda^k \boldsymbol{X_k}\right)^{\circ n}
\end{split}
\end{equation}
where we can then use the binomial theorem to obtain 
\begin{equation}
\begin{split}
    \frac{1}{n!}\left(\sum_{k=0}^\infty \lambda^k \boldsymbol{X_k}\right)^{\circ n}  =\lim_{K\to \infty} \sum_{m_0,m_1,\cdots ,m_K=0}^n \delta_{n,\sum_{j=0}^K m_j}\frac{\left(\lambda^0 \boldsymbol{X_0}\right)^{\circ m_0}}{m_0!}\circ \frac{\left(\lambda^1 \boldsymbol{X_1}\right)^{\circ m_1}}{m_1!} \circ \cdots \circ \frac{\left(\lambda^K \boldsymbol{X_K}\right)^{\circ m_K}}{m_K!},
\end{split}
\end{equation}
where~$\delta_{ij}=\begin{cases} 1, & i=j \\ 0, & i\neq j \end{cases}$ is the discrete Dirac delta function. By grouping terms of same order in~$\lambda$ we find that
\begin{equation}
\begin{split}
    \boldsymbol{X_k} = - \sum_{n=2}^\infty \boldsymbol{C}^{-1}(\boldsymbol{\varphi})\cdot \boldsymbol{g_{n+1}}(\boldsymbol{\varphi})  \cdot \sum_{m_1=0}^{(k-1)/1}\cdots \sum_{m_{k-1}=0}^{(k-1)/(k-1)}\delta_{k-1,\sum_{j=1}^{k-1} j m_j} \frac{\left(\lambda^0 \boldsymbol{X_0}\right)^{\circ \left(n-\sum_{j=1}^{k-1} m_j\right)}}{\left(n-\sum_{j=1}^{k-1} m_j\right)!}\circ \frac{\left(\lambda^1 \boldsymbol{X_1}\right)^{\circ m_1}}{m_1!} \\
    \circ \cdots \circ \frac{\left(\lambda^{k-1} \boldsymbol{X_{k-1}}\right)^{\circ m_{k-1}}}{m_{k-1}!}.
\end{split}
\end{equation}
The first correction terms are explicitly 
\begin{align}
    & \boldsymbol{X_0} = \boldsymbol{C}^{-1}(\boldsymbol{\varphi})\cdot\left(\boldsymbol{q} -\boldsymbol{q_0}(\boldsymbol{\varphi})\right), \\
    & \boldsymbol{X_1} = - \sum_{n=2}^\infty \boldsymbol{C}^{-1}(\boldsymbol{\varphi})\cdot \boldsymbol{g_{n+1}}(\boldsymbol{\varphi})  \cdot \frac{\boldsymbol{X_0}^{\circ n}}{n!} , \\
    & \boldsymbol{X_2} = - \sum_{n=2}^\infty \boldsymbol{C}^{-1}(\boldsymbol{\varphi})\cdot \boldsymbol{g_{n+1}}(\boldsymbol{\varphi})  \cdot \frac{\boldsymbol{X_0}^{\circ (n-1)}}{(n-1)!} \circ \boldsymbol{X_1} \\
    & \boldsymbol{X_3} = - \sum_{n=2}^\infty \boldsymbol{C}^{-1}(\boldsymbol{\varphi})\cdot \boldsymbol{g_{n+1}}(\boldsymbol{\varphi})  \cdot\left( \frac{\boldsymbol{X_0}^{\circ (n-1)}}{(n-1)!} \circ \boldsymbol{X_2} + \frac{\boldsymbol{X_0}^{\circ (n-2)}}{(n-2)!} \circ \frac{\boldsymbol{X_1}^{\circ 2}}{2!}\right) \\
    & \cdots 
\end{align}
We emphasize that~$\comm{\boldsymbol{\Phi}}{\boldsymbol{q}} = i$ following the canonical quantization.

\subsection{Full system Hamiltonian}

Now that we have expressions for the canonical charges we can write the Hamiltonian. The total system Hamiltonian, given by~$\mathcal H =\dot{\boldsymbol{\Phi}}^T \cdot \boldsymbol{q}-\mathcal{L}$, is (for~$\lambda = 1$)
\begin{equation}
    \mathcal{H} = \dot{\boldsymbol{\Phi}}^T \cdot\frac{\boldsymbol{C}(\boldsymbol{\varphi})}{2}\cdot \dot{ \boldsymbol{\Phi}} + \sum_{n=3}^\infty \left[\boldsymbol{1_{1\times 2}} \cdot \frac{(n-1)\boldsymbol{ g_n}(\boldsymbol{\varphi})}{2} \cdot \frac{\dot{\boldsymbol{\Phi}}^{\circ n}}{n!}+\frac{\dot{\boldsymbol{\Phi}}^{T \circ n}}{n!}\cdot \frac{(n-1)\boldsymbol{g_n}(\boldsymbol{\varphi})}{2} \cdot  \boldsymbol{1_{2\times 1}} \right]+ U\left(\boldsymbol{\varphi}\right), \quad \dot{\boldsymbol{\Phi}} = \sum_{k=0}^\infty \boldsymbol{X_k}(\boldsymbol{q}). 
\end{equation}

We can also divide the system Hamiltonian~$\mathcal{H} = \mathcal{H}_{\mathrm{quad}} + \mathcal{H}_{\mathrm{nln}}$ into two parts: 
\begin{equation}
    \mathcal{H}_{\mathrm{quad}} = \left(\boldsymbol{q} -\boldsymbol{q_0}(\boldsymbol{\varphi})\right)^T \cdot \frac{\boldsymbol{C}^{-1}(\boldsymbol{\varphi})}{2}\cdot \left(\boldsymbol{q} -\boldsymbol{q_0}(\boldsymbol{\varphi})\right) + U\left(\boldsymbol{\varphi}\right)
\end{equation}
involves all the contributions up to quadratic order in the charge operators~$\boldsymbol{q}-\boldsymbol{q_0}$, and 
\begin{equation}
\begin{split}
    \mathcal{H}_{\mathrm{nln}} & = \left(\sum_{k=0}^\infty \boldsymbol{X_k}(\boldsymbol{q})\right)^T \cdot\frac{\boldsymbol{C}(\boldsymbol{\varphi})}{2}\cdot \left(\sum_{k=0}^\infty \boldsymbol{X_k}(\boldsymbol{q})\right) - \boldsymbol{X_0}^T \cdot\frac{\boldsymbol{C}(\boldsymbol{\varphi})}{2}\cdot \boldsymbol{X_0} \\
    & + \sum_{n=3}^\infty \left[\boldsymbol{1_{1\times 2}} \cdot \frac{(n-1)\boldsymbol{ g_n}(\boldsymbol{\varphi})}{2} \cdot \frac{\left(\sum_{k=0}^\infty \boldsymbol{X_k}(\boldsymbol{q})\right)^{\circ n}}{n!}+\frac{\left(\sum_{k=0}^\infty \boldsymbol{X_k}(\boldsymbol{q})\right)^{T \circ n}}{n!}\cdot \frac{(n-1)\boldsymbol{g_n}(\boldsymbol{\varphi})}{2} \cdot  \boldsymbol{1_{2\times 1}} \right]
\end{split}
\end{equation}
comprises all remaining terms that are nonlinear in the charge operators, due to higher derivatives of the transmission coefficients. 

\subsubsection{$q$-quadratic Hamiltonian in expanded form}

Let us consider~$g_{j,1,1}$ to be the largest components by design and any other~$g_{j,n,m}\to \lambda g_{j,n,m}$ be an error term. Moreover we consider the coupling capacitance~$C_c\to \lambda C_c$ to be an error on the same order. We want to write the q-quadratic Hamiltonian to leading order in~$\lambda$. The capacitance matrix to have the form 
\begin{equation}
    \boldsymbol{C}(\boldsymbol{\varphi})= \begin{pmatrix}
   C_1 + \sum_{\ell=1}^\infty \lambda \Lambda_{1,2,\ell} \cos(\ell (\varphi_2-\varphi_{\mathrm ex,2})) & -\lambda C_c \\ -\lambda C_c & C_2 + \sum_{\ell=1}^\infty \lambda \Lambda_{2,2,\ell} \cos(\ell (\varphi_1-\varphi_{\mathrm ex,1}))
    \end{pmatrix},
\end{equation}
where~$C_j$ are the total capacitances and where we defined the coefficients
\begin{equation}
    \Lambda_{1,n,\ell} = -\sum_{m=1}^\infty\sum_{k=0}^{m-1} \dfrac{2(-1)^{m+k} g_{2,n,m}}{4^m} \begin{pmatrix}
    2m \\ k 
    \end{pmatrix} \delta_{m-k,\ell}, \quad \Lambda_{2,n,\ell} =- \sum_{m=1}^\infty\sum_{k=0}^{m-1} \dfrac{2(-1)^{m+k} g_{1,n,m}}{4^m} \begin{pmatrix}
    2m \\ k 
    \end{pmatrix} \delta_{m-k,\ell}.
\end{equation}
To leading order in~$\lambda$ we find that
\begin{equation}
    \boldsymbol{C}^{-1}(\boldsymbol{\varphi}) \approx \begin{pmatrix}
  \left(1 -  \sum_{\ell=1}^\infty \lambda \Lambda_{1,2,\ell}C_1^{-1}\cos(\ell (\varphi_2-\varphi_{\mathrm ex,2}))\right) C_1^{-1} & \lambda C_cC_1^{-1}C_2^{-1} \\ \lambda C_cC_1^{-1}C_2^{-1} & \left(1 -  \sum_{\ell=1}^\infty \lambda \Lambda_{2,2,\ell}C_2^{-1}\cos(\ell (\varphi_1-\varphi_{\mathrm ex,1}))\right)C_2^{-1}
    \end{pmatrix}.
\end{equation}
The charge offsets can be written as 
\begin{equation}
    \boldsymbol{q_0}(\boldsymbol{\varphi}) = \begin{pmatrix}
   q_{01}-g_1\cos\left(\varphi_2-\varphi_{\mathrm ex,2}\right) + \sum_{\ell=2}^\infty \lambda \Lambda_{1,1,\ell} \cos(\ell (\varphi_2-\varphi_{\mathrm ex,2})) \\
    q_{02}-g_2 \cos\left( \varphi_1-\varphi_{\mathrm ex,1}\right)+\sum_{\ell=2}^\infty \lambda \Lambda_{2,1,\ell} \cos(\ell (\varphi_1-\varphi_{\mathrm ex,1})),
    \end{pmatrix},
\end{equation}
where~$q_{0j}$ are some scalars and where we defined the couplings 
\begin{equation}
    g_1 = -g_{2,1,1}/2, \quad g_2 = -g_{1,1,1}/2.
\end{equation} 

The~$q$-quadratic Hamiltonian to leading order in~$\lambda$ then approximately takes the form 
\begin{equation}
\begin{split}
    \mathcal{H}_{\mathrm{quad}} = \frac{\left(q_1-q_{01}+g_1\cos\left(\varphi_2-\varphi_{\mathrm ex,2}\right)\right)^2}{2C_1} + \frac{\left(q_2-q_{02}+g_2 \cos\left( \varphi_1-\varphi_{\mathrm ex,1}\right)\right)^2}{2C_2} + U\left(\varphi_1,\varphi_2\right) + \lambda \mathcal{H}_{\mathrm{quad}}^\lambda +\mathcal{O}(\lambda^2),
\end{split}
\end{equation}

and the error Hamiltonian 
\begin{equation}
    \begin{split}
    \mathcal{H}_{\mathrm{quad}}^\lambda = &- \frac{\left(q_1-q_{01}\right)}{C_1}\sum_{\ell=2}^\infty \Lambda_{1,1,\ell} \cos(\ell (\varphi_2-\varphi_{\mathrm ex,2})) - \frac{\left(q_2-q_{02}\right)}{C_2}\sum_{\ell=2}^\infty \Lambda_{2,1,\ell} \cos(\ell (\varphi_1-\varphi_{\mathrm ex,1})) \\
    &- \frac{\left(q_1-q_{01}\right)^2}{2C_1^2}\sum_{\ell=1}^\infty \Lambda_{1,2,\ell} \cos(\ell (\varphi_2-\varphi_{\mathrm ex,2})) - \frac{\left(q_2-q_{02}\right)^2}{2C_2^2}\sum_{\ell=1}^\infty \Lambda_{2,2,\ell} \cos(\ell (\varphi_1-\varphi_{\mathrm ex,1})) \\
    &+ \frac{C_c}{C_1C_2}\left(q_1-q_{01}\right)\left(q_2-q_{02}\right).
    \end{split}
\end{equation}

\subsubsection{q-nonlinear Hamiltonian in expanded form}

To leading order in~$\boldsymbol{g_n}$ for~$n\geq 3$ we can approximate 
\begin{equation}
    \mathcal{H}_{\mathrm{nln}} \approx -\sum_{n=3}^\infty \left[\boldsymbol{1_{1\times 2}} \cdot \frac{\lambda \boldsymbol{ g_n}(\boldsymbol{\varphi})}{2} \cdot \frac{\left(\boldsymbol{C}^{-1}(\boldsymbol{\varphi})\cdot\left(\boldsymbol{q} -\boldsymbol{q_0}(\boldsymbol{\varphi})\right)\right)^{\circ n}}{n!}+\frac{\left(\boldsymbol{C}^{-1}(\boldsymbol{\varphi})\cdot\left(\boldsymbol{q} -\boldsymbol{q_0}(\boldsymbol{\varphi})\right)\right)^{T \circ n}}{n!}\cdot \frac{\lambda \boldsymbol{g_n}(\boldsymbol{\varphi})}{2} \cdot  \boldsymbol{1_{2\times 1}} \right].
\end{equation}
To leading order in~$\lambda$ we find that~$\mathcal{H}_{\mathrm{nln}} \approx \lambda \mathcal{H}_{\mathrm{nln}}^\lambda + \mathcal{O}\left(\lambda^2\right)$ with 
\begin{equation}
\begin{split}
    \mathcal{H}_{\mathrm{nln}}^\lambda = -\sum_{n=3}^\infty \frac{\left(q_1-q_{01}\right)^n}{n!C_1^n}\sum_{\ell=1}^\infty \Lambda_{1,n,\ell} \cos(\ell (\varphi_2-\varphi_{\mathrm ex,2})) -\sum_{n=3}^\infty \frac{\left(q_2-q_{02}\right)^n}{n!C_2^n}\sum_{\ell=1}^\infty \Lambda_{2,n,\ell} \cos(\ell (\varphi_1-\varphi_{\mathrm ex,1})) \\
    - \sum_{n=3}^\infty \frac{\left(q_1-q_{01}\right)^n}{n!C_1^n}\xi_{1,n} - \sum_{n=3}^\infty \frac{\left(q_2-q_{02}\right)^n}{n!C_2^n}\xi_{2,n},
\end{split}
\end{equation}
where we defined the coefficients 
\begin{equation}
    \xi_{1,n} = \sum_{m=0}^\infty \dfrac{g_{2,n,m}}{4^m} \begin{pmatrix}
    2m \\ m 
    \end{pmatrix}, \quad \xi_{2,n} = \sum_{m=0}^\infty \dfrac{g_{1,n,m}}{4^m} \begin{pmatrix}
    2m \\ m 
    \end{pmatrix} .
\end{equation}

\subsubsection{Approximate form}

Combining the results of the previous sections with~$\lambda=1$ we therefore find the approximate Hamiltonian 
\begin{equation}
\begin{split}
    \mathcal{H} \simeq  \frac{\left(q_1-q_{01}+g_1\cos\left(\varphi_2-\varphi_{\mathrm ex,2}\right)\right)^2}{2C_1} + \frac{\left(q_2-q_{02}+g_2 \cos\left( \varphi_1-\varphi_{\mathrm ex,1}\right)\right)^2}{2C_2} + U\left(\varphi_1,\varphi_2\right) + \mathcal{H}_{\mathrm{err}},
\end{split}
\end{equation}
where we defined the error Hamiltonian 
\begin{equation}
    \begin{split}
    \mathcal{H}_{\mathrm{err}} =& - \frac{\left(q_1-q_{01}\right)}{C_1}\sum_{\ell=2}^\infty \Lambda_{1,1,\ell} \cos(\ell (\varphi_2-\varphi_{\mathrm ex,2})) - \frac{\left(q_2-q_{02}\right)}{C_2}\sum_{\ell=2}^\infty \Lambda_{2,1,\ell} \cos(\ell (\varphi_1-\varphi_{\mathrm ex,1})) \\
    &-\sum_{n=2}^\infty \frac{\left(q_1-q_{01}\right)^n}{n!C_1^n}\sum_{\ell=1}^\infty \Lambda_{1,n,\ell} \cos(\ell (\varphi_2-\varphi_{\mathrm ex,2})) -\sum_{n=2}^\infty \frac{\left(q_2-q_{02}\right)^n}{n!C_2^n}\sum_{\ell=1}^\infty \Lambda_{2,n,\ell} \cos(\ell (\varphi_1-\varphi_{\mathrm ex,1})) \\
    &- \sum_{n=3}^\infty \frac{\left(q_1-q_{01}\right)^n}{n!C_1^n} \xi_{1,n} - \sum_{n=3}^\infty \frac{\left(q_2-q_{02}\right)^n}{n!C_2^n} \xi_{2,n} + \frac{C_c}{C_1C_2}\left(q_1-q_{01}\right)\left(q_2-q_{02}\right).
    \end{split}
\end{equation}
Here the couplings are explicitly
\begin{align}
    & g_1 = \frac{\Delta_2}{4}\left(\frac{\partial T_2(V_{J,2},\dot{\Phi}_1)}{\partial \dot{\Phi}_1}\Big|_{\dot{\Phi}_1=0}\right),\\
    & g_2 = \frac{\Delta_1}{4}\left(\frac{\partial T_1(V_{J,1},\dot{\Phi}_2)}{\partial \dot{\Phi}_2}\Big|_{\dot{\Phi}_2=0}\right), \\
    & \Lambda_{1,n,\ell} = -\sum_{m=1}^\infty\sum_{k=0}^{m-1} \dfrac{2(-1)^{k}}{4^{m-1}}\begin{pmatrix}
    1/2\\ m
    \end{pmatrix}\begin{pmatrix}
    2m \\ k 
    \end{pmatrix} \delta_{m-k,\ell}\frac{\Delta_2}{4}\left(\frac{\partial^n T_2^m(V_{J,2},\dot{\Phi}_1)}{\partial \dot{\Phi}_1^n}\Big|_{\dot{\Phi}_1=0}\right), \\
    & \Lambda_{2,n,\ell} =  -\sum_{m=1}^\infty\sum_{k=0}^{m-1} \dfrac{2(-1)^{k}}{4^{m-1}}\begin{pmatrix}
    1/2\\ m
    \end{pmatrix}\begin{pmatrix}
    2m \\ k 
    \end{pmatrix} \delta_{m-k,\ell}\frac{\Delta_1}{4}\left(\frac{\partial^n T_1^m(V_{J,1},\dot{\Phi}_2)}{\partial \dot{\Phi}_2^n}\Big|_{\dot{\Phi}_2=0}\right), \\
    & \xi_{1,n} = \sum_{m=0}^\infty \dfrac{(-1)^m}{4^{m-1}} \begin{pmatrix}
    1/2\\ m
    \end{pmatrix}\begin{pmatrix}
    2m \\ m 
    \end{pmatrix}\frac{\Delta_2}{4}\left(\frac{\partial^n T_2^m(V_{J,2},\dot{\Phi}_1)}{\partial \dot{\Phi}_1^n}\Big|_{\dot{\Phi}_1=0}\right), \\ 
    & \xi_{2,n} = \sum_{m=0}^\infty \dfrac{(-1)^m}{4^{m-1}}\begin{pmatrix}
    1/2\\ m
    \end{pmatrix}\begin{pmatrix}
    2m \\ m 
    \end{pmatrix}\frac{\Delta_1}{4}\left(\frac{\partial^n T_1^m(V_{J,1},\dot{\Phi}_2)}{\partial \dot{\Phi}_2^n}\Big|_{\dot{\Phi}_2=0}\right). 
\end{align} 

Tolerance values of the higher derivatives in the error terms are ultimately determined by the mode impedances, i.e.
\begin{align}
    \left|\sum_{m=1}^\infty\sum_{k=0}^{m-1} \dfrac{2(-1)^{k}}{4^{m-1}}\begin{pmatrix}
    1/2\\ m
    \end{pmatrix}\begin{pmatrix}
    2m \\ k 
    \end{pmatrix} \delta_{m-k,\ell}\frac{(2e)^n\Delta_2}{4n!C_1^n}\left(\frac{\partial^n T_2^m(V_{J,2},\dot{\Phi}_1)}{\partial \dot{\Phi}_1^n}\Big|_{\dot{\Phi}_1=0}\right)\right|\ll \sqrt{\frac{4\pi Z_1}{R_Q}}^n, \\
    \left|\sum_{m=1}^\infty\sum_{k=0}^{m-1} \dfrac{2(-1)^{k}}{4^{m-1}}\begin{pmatrix}
    1/2\\ m
    \end{pmatrix}\begin{pmatrix}
    2m \\ k 
    \end{pmatrix} \delta_{m-k,\ell}\frac{(2e)^n\Delta_1}{4n!C_2^n}\left(\frac{\partial^n T_1^m(V_{J,1},\dot{\Phi}_2)}{\partial \dot{\Phi}_2^n}\Big|_{\dot{\Phi}_2=0}\right)\right| \ll \sqrt{\frac{4\pi Z_2}{R_Q}}^n,
\end{align}
where~$Z_j$ is the impedance of mode~$j$ and~$R_Q \simeq 6.5~$k$\Omega$ is the resistance quantum. The constraints are obtained by considering the smallest non-zero matrix elements of the charge quadrature~$q_j$ to be given by~$i2e\sqrt{R_Q/(4\pi Z_j)}$.

\section{Classical scattering matrix for a 3-port circulator}
\label{app:Circulator_from_gyrator}

Following standard circuit theory~\cite{Carlin:1964}, we can compute the classical (linear) scattering matrix response for a gyrator-based 3-port circulator with symmetric LC-resonator loads at their ports, see Fig.~\cref{fig:Circulator_sketch}b). Such matrix relates the amplitudes of single-tone input ($a_n(\omega)=(V_n+Z_{\mathrm{TL}} I_n)/\sqrt{Z_{\mathrm{TL}}}$) and output ($b_n(\omega)=(V_n-Z_{\mathrm{TL}} I_n)/\sqrt{Z_{\mathrm{TL}}}$) signals, where~$I_n$ ($V_n$) is the current (voltage) of port~$n$, such that~$\boldsymbol{b}=\boldsymbol{S}(\omega)\boldsymbol{a}$. Solving Kirchhoff's equations in the frequency domain, we obtain 
\begin{align}
    \boldsymbol{S}=&\frac{1}{r}\left(
\begin{array}{ccc}
 R^2 (Z_{\mathrm{TL}}-\tilde{Z}_0)^2-Z_{\mathrm{TL}}^2 \tilde{Z}_0^2 & 2 R \tilde{Z}_0 (\tilde{Z}_0 (R+Z_{\mathrm{TL}})-R Z_{\mathrm{TL}}) & 2 R \tilde{Z}_0 (R (Z_{\mathrm{TL}}+\tilde{Z}_0)-Z_{\mathrm{TL}} \tilde{Z}_0) \\
 2 R \tilde{Z}_0 (R (\tilde{Z}_0-Z_{\mathrm{TL}})-Z_{\mathrm{TL}} \tilde{Z}_0) & R^2 (Z_{\mathrm{TL}}-\tilde{Z}_0)^2-Z_{\mathrm{TL}}^2 \tilde{Z}_0^2 & -2 R \tilde{Z}_0 (R (Z_{\mathrm{TL}}+\tilde{Z}_0)+Z_{\mathrm{TL}} \tilde{Z}_0) \\
 2 R \tilde{Z}_0 (R (Z_{\mathrm{TL}}+\tilde{Z}_0)+Z_{\mathrm{TL}} \tilde{Z}_0) & -2 R \tilde{Z}_0 (R (Z_{\mathrm{TL}}+\tilde{Z}_0)-Z_{\mathrm{TL}} \tilde{Z}_0) & R^2 (Z_{\mathrm{TL}}+\tilde{Z}_0)^2-Z_{\mathrm{TL}}^2 \tilde{Z}_0^2 \\
\end{array}
\right).
\end{align}
Here, we have defined the parameters
\begin{align}
    r=&(Z_{\mathrm{TL}}^2 \tilde{Z}_0^2-R^2 (Z_{\mathrm{TL}}-3 \tilde{Z}_0) (Z_{\mathrm{TL}}+\tilde{Z}_0)),\qquad
\tilde{Z}_0(\omega)=Z_0 \frac{- i \omega \omega_r}{\omega^2-\omega_r^2},
\end{align}
and~$Z_0=\sqrt{L_0/C_0}$ and~$\omega_0=1/\sqrt{L_0 C_0}$. Impedance-matching the whole system to the reference transmission-lines ($R=Z_0=Z_{\mathrm{TL}}$), and working on resonance condition ($\omega=\omega_r$), the linear response becomes that ideal circulator 
\begin{align}
    \boldsymbol{S}\rightarrow\begin{pmatrix}
      0&1&0\\0&0&-1\\1&0&0
    \end{pmatrix}.
\end{align}

\end{document}